\documentclass[a4paper,11pt]{article}
\pdfoutput=1
\usepackage{jcappub}
\usepackage{slashed}
\usepackage[T1]{fontenc}
\usepackage{adjustbox}
\usepackage{multirow}
\usepackage{hhline}
\usepackage{tabularx,booktabs,caption}
\usepackage{float}
\usepackage{placeins}
\usepackage{caption}
\usepackage{verbatim}
\usepackage{soul}
\usepackage{color}
\usepackage[utf8]{inputenc}
\usepackage{xcolor}
\newcommand{\expt}[1]{\left\langle #1 \right\rangle}
\newcommand{\ie}{$i.e.,$}
\newcommand{\eg}{$e.g.,$}
\newcommand{\nn}{\nonumber{\nonumber}}
\newcommand{\abs}[1]{\left| #1 \right|}
\newcommand{\beq}{\begin{equation}}
\newcommand{\eeq}{\end{equation}}
\newcommand{\beqn}{\begin{eqnarray}}
\newcommand{\eeqn}{\end{eqnarray}}
\newcommand{\bea}{\begin{eqnarray}}
\newcommand{\eea}{\end{eqnarray}}

\title{\boldmath {Revisiting} Dark Matter Freeze-in and Freeze-out through Phase-Space Distribution}

\author[a]{Yong Du,}
\author[a,b]{Fei Huang,}
\author[a,g]{Hao-Lin Li,}
\author[a,c]{Yuan-Zhen Li}
\author[a,c,d,e,f]{and Jiang-Hao Yu}

\affiliation[a]{CAS Key Laboratory of Theoretical Physics, Institute of Theoretical Physics,\\
Chinese Academy of Sciences, Beijing 100190, China}
\affiliation[b]{Department of Physics and Astronomy, University of California, Irvine, CA 92697 USA}
\affiliation[c]{School of Physical Sciences, University of Chinese Academy of Sciences, Beijing 100049, P.\ R.\ China}
\affiliation[d]{Center for High Energy Physics, Peking University, Beijing 100871, China}
\affiliation[e]{School of Fundamental Physics and Mathematical Sciences, Hangzhou Institute for Advanced
Study, UCAS, Hangzhou 310024, China}
\affiliation[f]{International Centre for Theoretical Physics Asia-Pacific, Beijing/Hangzhou, China}
\affiliation[g]{Centre for Cosmology, Particle Physics and Phenomenology (CP3),
Universite catholiqu\'e de Louvain,
Chemin du Cyclotron, 2
B-1348 Louvain-la-Neuve,
Belgium}

\emailAdd{yongdu@itp.ac.cn}
\emailAdd{huangf4@uci.edu}
\emailAdd{haolin.li@uclouvain.be}
\emailAdd{liyuanzhen@itp.ac.cn}
\emailAdd{jhyu@itp.ac.cn}

\abstract{
We revisit dark-matter production through freeze-in and freeze-out by solving the Boltzmann equations at the level of the phase-space distribution $f(p,t)$.
Using the $2\to2$ annihilation and the $1\to2$ decay processes for illustration, we compare the resulting dark-matter relic abundance with that from the number-density approach.
In the transition regime between freeze-in and freeze-out, we find the difference can be 
quite significant, or even by orders of magnitude if the annihilation of dark-matter particles or the decaying mediator is neglected.
The freeze-in production in the $2\to2$ and the $1\to 2$ processes can also result in non-thermal phase-space distributions, or even multi-modal ones with out-of-equilibrium decay, which can potentially affect structure formation at late times.
We also investigate how elastic scatterings can distort such non-thermal distributions.
}

\begin{document} 
\captionsetup[figure]{labelfont={bf},labelformat={default},labelsep=period,name={FIG.}}
\maketitle
\flushbottom

%%%%%%%%%%%%%%%%%%%%
\section{Introduction}\label{sec:intro}
%%%%%%%%%%%%%%%%%%%%

The nature of dark matter remains one of the most challenging problems facing physics today.
However, despite ample evidence supporting its existence through its gravitational interaction and countless efforts in understanding it in both theoretical (see, for example \,\cite{Jungman:1995df,Bergstrom:2000pn,Bertone:2004pz,Bertone:2018krk,Bertone:2019irm}) and experimental frontiers\,\cite{Drukier:1986tm,Freese:1987wu,LUX:2016ggv,LUX:2017ree,Roszkowski:2017nbc,PandaX-II:2017hlx,XENON:2018voc,Montanari:2021yic}, our knowledge of dark matter is still very limited.
So far, years of attempt in direct and indirect searches still yield null result \cite{LUX:2016ggv,LUX:2017ree,PandaX-II:2017hlx,XENON:2018voc,Montanari:2021yic,Roszkowski:2017nbc}.\footnote{The annual modulation observed in Refs.\,\cite{Drukier:1986tm,Freese:1987wu} is still under debate and a discussion on this point is beyond the scope of this work. However, the corresponding parameter space will be scanned in future by detectors with higher sensitivities (see, for example, Refs.\,\cite{Angloher:2016ooq,SABRE:2018lfp,Coarasa:2018qzs,PICO-LON:2015rtu,DM-Ice:2016snk,Park:2017jvs,Adhikari:2017esn}).}
The only properties that we know about dark matter are that it does not interact appreciably with the standard model (SM) fields or with itself other than via gravity, it has a large relic abundance --- $\Omega_{\rm DM}\simeq 0.26$, it has to be stable on cosmological timescales, and it must be non-relativistic after matter-radiation equality in order to seed the structure formation.
On the other hand, we do not know its particle content, nor do we know its non-gravitational interaction with the SM or with itself, or even whether such interactions exist.
As a result, we do not know how dark matter is produced in the early universe.

Nevertheless, when studying the production of dark matter, it is often assumed that dark matter is produced by particle interactions in the early universe such as particle decays and scatterings.
In standard thermal freeze-out scenarios, dark matter is assumed to be a particle species in thermal equilibrium with the visible-sector thermal bath deep within the radiation dominated (RD) epoch,
and then decouples first chemically and then kinetically from the thermal bath as the interaction rates fall below the Hubble expansion rate.
For this type of scenarios, the phase-space distribution of dark matter $\chi$ is thermal before kinetic decoupling occurs, \ie~
\beq
f_\chi(p)=f_\chi^{\rm eq}(p)=\frac{{1}}{e^{\frac{E-\mu_\chi}{T}}\pm 1},
\eeq
in which \mbox{$E~=~\sqrt{p^2+m_\chi^2}$}, $\mu_\chi$ is the chemical potential, and $+/-$ stands for Fermi-Dirac/Bose-Einstein distribution, respectively.
After kinetic decoupling, the phase-space distribution is only subject to the expansion of the universe until being disrupted by structure formation.
{Since kinetic decoupling is often assumed to be much later than the chemical decoupling, for the epoch relevant for estimating the dark-matter relic abundance, it is often sufficient to work only at the level of the particle number density.
In this situation, the interaction rates can be evaluated in a closed form represented by the \emph{thermally averaged cross section} in which all relevant 
species are assumed to be in kinetic equilibrium with the same temperature \cite{Gondolo:1990dk}.}
{On the other hand, it is pointed out in Refs.~\cite{Binder:2017rgn,Binder:2021bmg} that if kinetic decoupling occurs very early such that it is close to the chemical decoupling, the number-density approach could be very inaccurate in estimating the dark-matter relic abundance. 
In this case, it is necessary to keep track of the evolution of relevant phase-space distributions.}

In another well-studied dark-matter production mechanism, the freeze-in mechanism (see \cite{Hall:2009bx,Elahi:2014fsa,Bernal:2017kxu} and references therein), the interaction between dark matter and the visible sector is so feeble that thermal equilibrium could never be established between them.
Consequently, the phase-space distribution of dark matter cannot be known \textit{a priori}.
Nevertheless, for the simplest freeze-in scenario --- the pure freeze-in scenario in which dark-matter particles barely annihilate, one can take advantage of the fact that $f_\chi\ll f_\chi^{\rm eq}$ and estimate the dark-matter relic abundance without even including processes that decrease dark-matter particles. 
Therefore, the need to know the evolution of the dark-matter phase-space distribution is circumvented.
In more complicated scenarios where dark-matter particles could establish thermal equilibrium within the dark sector, (\eg~scenarios described in \cite{Cheung:2010gj,Cheung:2010gk,Chu:2011be,Bernal:2015ova,Bernal:2017kxu,Krnjaic:2017tio,Berger:2018xyd,Evans:2019vxr,Du:2020avz,Hryczuk:2021qtz}), the typical approach is to keep track of the energy density of the dark sector and solve for the temperature of the dark thermal bath such that the dark-matter phase-space distribution is simply a thermal distribution at that temperature.

However, between the freeze-out and the pure freeze-in, there still exists a vast intermediate regime in which back reactions against the particle injection from the visible sector, 
\eg~annihilations of dark-matter particles into visible-sector particles,
are not entirely negligible even though thermal equilibrium between the dark and the visible sector is never established.
In this regime, $f_\chi$ can be both non-thermal and non-negligible.
It is not clear quantitatively at which point such back reactions can no longer be safely neglected.
It is also not clear how good the approximation is if one only solves Boltzmann equations of number densities.
Therefore, in this regime, one in principle needs to work at the level of the phase-space distribution to correctly account for the back reactions when simulating the evolution of the dark-matter number density.

Besides the estimation of dark-matter relic abundance, the phase-space distribution of dark matter can also have important phenomenological implications on the formation of the cosmic structure --- no matter being produced from freeze-out or freeze-in, structure formation would be suppressed as long as dark-matter particles have non-negligible velocities.
{To study the effects on structure formation, the traditional approach for freeze-out scenarios is to find the temperature at which kinetic decoupling occurs \cite{Bringmann:2006mu,vandenAarssen:2012ag}.
For freeze-in scenarios, since dark matter is never in thermal equilibrium with the thermal bath, one in general needs to keep track of the evolution of the phase-space distribution \cite{Bae:2017dpt,DEramo:2020gpr,Hufnagel:2021pso}.}
{In either scenarios, } for dark matter with a relatively simple phase-space distribution, \eg~thermal distribution, or a distribution concentrated at one particular velocity, the effect on structure formation can be characterized by the 
\emph{free-streaming} horizon \cite{Kolb:1990vq}
\beq
\lambda_{\rm FSH}=\int dt \frac{\expt{v(t)}}{a(t)},
\eeq
which is the maximum average distance that dark-matter particles can travel.
Structure formation on a scale below the free-streaming horizon
are suppressed because dark-matter particles can free-stream out of overdense regions of (comoving) size $\sim \lambda_{\rm FSH}$ and delay the gravitational collapse.

However, the average velocity only represents the zeroth-order behavior of dark matter.
In general, the distribution function can be more complicated than being characterized simply by a single quantity.
In scenarios in which the phase-space distribution is highly non-thermal, not concentrated around a particular velocity, or multi-modal (consisting of more than one single packet), a single free-streaming scale would fail to capture all the information in the distribution.
In fact, the features in the phase-space distribution can leave identifiable imprints in the structure of the universe
which can be seen from physical quantities of structure formation such as the matter power spectrum and the halo mass function \cite{Boyarsky:2008mt,Lovell:2011rd,Konig:2016dzg,Murgia:2017lwo, Dienes:2020bmn, Dienes:2021itb},
and can be constrained by observations such as the Lyman-$\alpha$ forest and the Milky-Way satellite count\cite{Schneider:2016uqi,Konig:2016dzg,Murgia:2017lwo,Murgia:2018now}.
These imprints can even be exploited to reconstruct the primordial phase-space distribution of dark matter \cite{Dienes:2020bmn, Dienes:2021itb}.
Since structure formation depends on gravitational interaction only,
the imprints and constraints on structure formation provide a way of learning about the early-universe dynamics in the dark sector without relying on non-gravitational interactions of dark matter.
Such methods therefore provide further motivation for studying the dark-matter phase-space distribution itself, especially in freeze-in scenarios in which interactions between dark matter and the SM particles are usually very feeble.

In this paper, we revisit the freeze-out and freeze-in\footnote{To be specific, we only study the IR freeze-in instead of the UV freeze-in in this paper.} mechanisms, and provide a systematic study by solving the Boltzmann equations at the level of the phase-space distribution $f(p,t)$.
We consider two types of production processes in which dark matter is produced via $2\to2$ annihilation and $1\to2$ decay, respectively.
For the first type, we study both freeze-out and freeze-in, whereas, for the second type, we focus on the freeze-in production of dark matter but classify different scenarios according to the thermal history of the decaying particle --- it could have a short lifetime and decay while in thermal equilibrium with the visible sector, or have a relatively long lifetime and decay after it is produced via freeze-in or freeze-out.
In all cases, we calculate the relic abundance of dark matter and compare it with the estimates from different approaches --- methods by solving the Boltzmann equations of the phase-space distribution and by solving that of the number density only.
Since, in freeze-in scenarios, one often consider the particle injection from the visible sector as dominant and thus neglect any back reaction to it,
we also investigate the extent to which potential back reactions can be safely neglected as one dials the interaction strength, and point out when a full solution using the phase-space distribution is necessary.

For the distribution function itself, we examine not only the evolution of the average momentum/velocity but also how much it can deviate from the thermal distribution in different situations.
For this part, we shall focus on the freeze-in mechanism since it is generically nonthermal.
We are especially interested in scenarios in which the final phase-space distribution possesses highly nonthermal or multi-modal features.
We shall see that such distributions can be easily realized in the $1\to2$ scenario if the fraction of dark matter produced via freeze-in can be accompanied by a comparable fraction via a late decay.

Moreover, the phase-space distribution is also affected by elastic scatterings which in principle are always present.
The elastic scattering is also important because its rate is crucial in determining the temperature of kinetic-decoupling, which naturally sets a small-scale cut-off in the matter power spectrum \cite{Bringmann:2006mu}.
We therefore also study the effects of elastic scatterings by incorporating them in the Boltzmann equation in a model-independent way.
In particular, we investigate how much the overall momentum of dark-matter particles can be affected by varying the elastic-scattering rate.
We also assess the potential of elastic scatterings in distorting or even erasing nonthermal features in the phase-space distribution.

\begin{figure}
	\centering
	\includegraphics[width=0.8\textwidth]{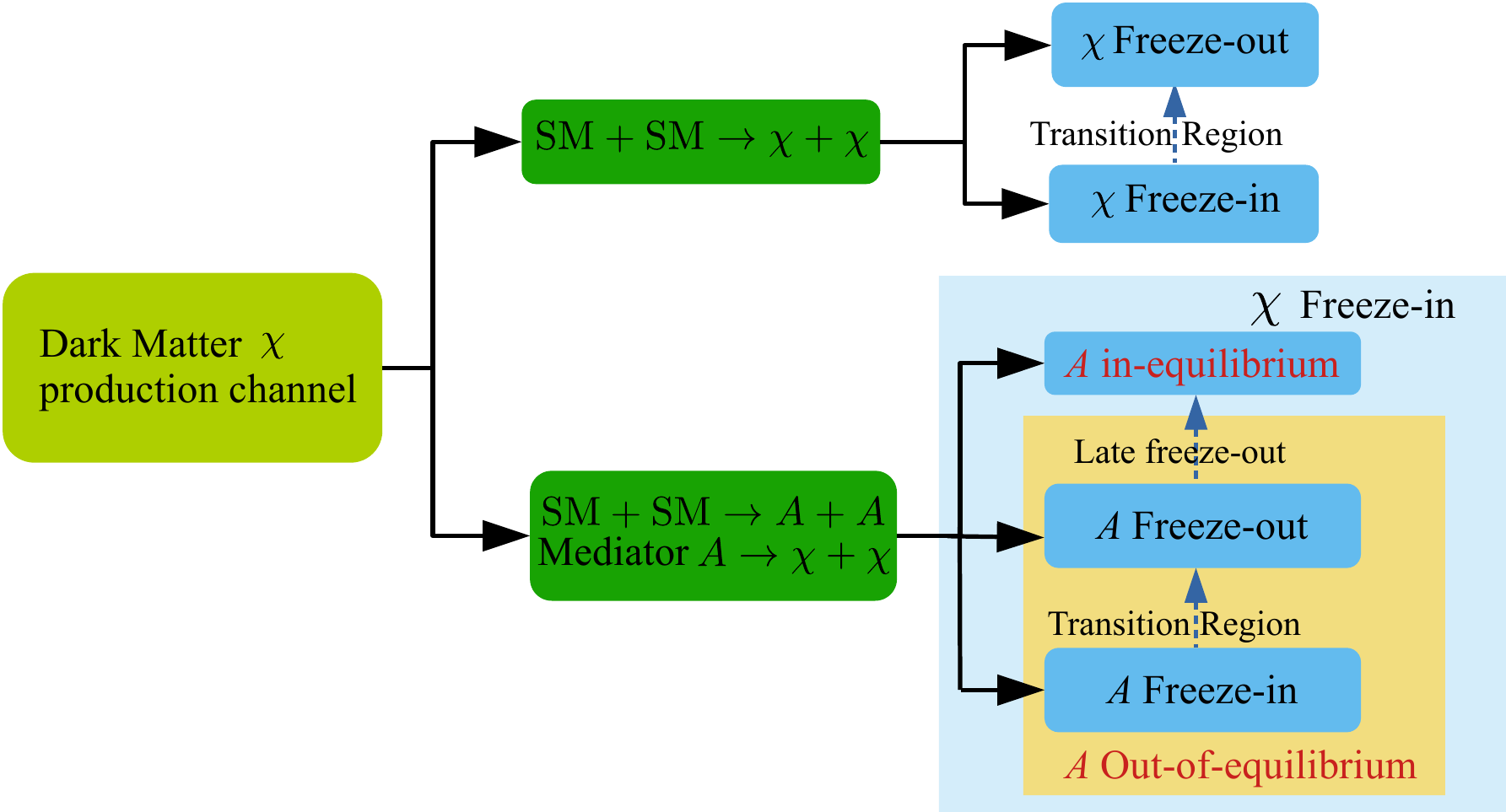}
	\caption{All scenarios of dark-matter production studied in this work. 
	The scenarios are classified by their dark-matter production channels, \ie~$2\to 2$ annihilation or $1\to 2$ decay. 
	In the $2\to 2$ case, we revisit the vanilla freeze-in and freeze-out mechanisms.
	In the $1\to 2$ case, we focus on the freeze-in production of $\chi$ and classify scenarios according to 
	the thermal history of the decaying mediator $A$ --- it could undergo freeze-in/out, or be always in thermal equilibrium with the visible sector.
	The dashed vertical lines between the blue boxes suggests the transition regime between different types of scenarios with the arrows indicating the direction of increasing dark-matter or mediator annihilation rate in corresponding scenarios.
	}
	\label{fg:guide}
\end{figure}

This paper is organized as follows.
In Sec.~\ref{sec:Collision}, we provide the general setup of our analysis and describe the numerical procedure.
In Sec.~\ref{sec:tworegimes}, we consider the production of dark matter through the $2\to2$ annihilation channel and revisit the freeze-in and freeze-out scenarios using information obtained from the evolution of the phase-space distribution.
In Sec.~\ref{sec:parent_decay}, we study another production channel, \ie~production via $1\to2$ decay, focusing on several freeze-in scenarios classified by the thermal history of the decaying particle.
Finally, we conclude our study in Sec.~\ref{sec:conclusion}.
To guide the readers, all scenarios studied in this paper are summarized in FIG.~\ref{fg:guide}.

%%%%%%%%%%%%%%%%%%%%
\section{Phase-space distribution and its evolution}\label{sec:Collision}
\subsection{General setup}
For a particular particle species, its phase-space distribution $f(\vec{x},\vec{p}, t)$ describes its distribution in both the position space and the momentum space.
Since, at the zeroth order, the universe is spatially homogeneous and isotropic as described by the Friedmann-Lema{\^i}tre-Robertson-Walker (FLRW) cosmology, one often ignores the dependence on spatial coordinates and the direction of the momentum.
Therefore, the distribution function can be approximated as a function that depends only on time and the magnitude of momentum:
\beq
f(\vec{x},\vec{p}, t)\approx f(p,t)\,.
\eeq
With the phase-space distribution, quantities such as the number density, energy density, pressure can be obtained straightforwardly:
\beqn
n(t)&=&{g}\int \frac{d^3p}{(2\pi)^3} f(p,t)\,,\\
\rho(t)&=&{g}\int \frac{d^3p}{(2\pi)^3} f(p,t)E\,,\\
P(t)&=&{g} \int \frac{d^3p}{(2\pi)^3} f(p,t)\frac{p^2}{3E}\,,
\eeqn
{in which $g$ is the internal degrees of freedom of the corresponding particle species.}

The evolution of $f(p,t)$ is essentially governed by the interplay between particle interactions and the expansion of the universe. 
For particles with a particular momentum $p$, on one hand, interactions such as decays, annihilations and elastic scatterings would produce or deplete them, 
or scatter them to other momenta.
On the other hand, the expansion of the universe would redshift those particles to a lower momentum at $p(t')=p(t)a(t)/a(t')$, where $a$ is the scale factor.
These two effects are collectively described by the Boltzmann equation:
\beq
\frac{\partial f}{\partial t} - Hp\frac{\partial f}{\partial p}=C[f]\,,\label{eq:Boltzmann_f}
\eeq
where the term with the Hubble parameter, $H\equiv\dot{a}/{a}$, describes the redshift effect due to the expansion of the universe,
and the term on the right-hand side is the collision term which describes particle interactions.
Integrating both sides over the phase space, we then have the Boltzmann equation for the number density:
\beq
\frac{dn}{dt}+3Hn={g}\int\frac{d^3 p}{(2\pi)^3}C[f]\,.\label{eq:Boltzmann_num}
\eeq
Note that, the integration of the collision term requires full information of the phase-space distributions of all participating particles, which in some cases are not all known \textit{a priori}.

Considering a particle species labeled by $a$, for a specific process $a+b+\dots\leftrightarrow i+j+\dots$, the collision term in Eq.~\eqref{eq:Boltzmann_f} takes the following general form
\beqn
C[f]&=&-\frac{1}{2E_{a}}\int d\pi_b\dots d\pi_i d\pi_j\dots (2\pi)^4 \delta^{(4)}(p_a+p_b+\dots-p_i-p_j-\dots)\nn\\
&~&\times\bigg[\abs{\mathcal{M}_{a+b\dots\rightarrow i+j+\dots}}^2 f_a f_b\dots(1\pm f_i)(1\pm f_j)\dots\nn\\
&~&~~~~ - \abs{\mathcal{M}_{i+j+\dots\rightarrow a+b+\dots}}^2f_i f_j\dots (1\pm f_a)(1\pm f_b)\dots \bigg]\,,\label{eq:collision_general}
\eeqn
where {$d\pi_i=\frac{g_i}{(2\pi)^3}\frac{d^3p_{i}}{2E_{i}}$}, and ``$\pm$'' describes the effects from quantum statistics, \ie~the Bose-enhancement/Pauli-blocking effects, and the two amplitudes correspond to the forward and inverse processes, respectively.
{Note that, we are assuming here that indices $a,~b,\dots~i,~j\dots$ label different particles.
In general, this need not be true as the same species can appear multiple times on both sides of the process.
In those cases, appropriate multiplicity factors must be taken in to account.}

In principle, the collision term in Eq.~\eqref{eq:Boltzmann_f} contains all relevant processes that the corresponding particle participates such as decays, inverse decays, annihilations and scatterings, etc.
For the study of dark matter, it is often assumed that one or a few processes dominates the dynamical evolution.
In what follows, we shall focus on studying the phase-space distribution of dark matter resulting from different production mechanisms.
We shall also compare the estimate on the dark-matter relic abundance with the approach using the Boltzmann equation of the number density only. 
Before that, let us briefly review the numerical procedures in our study.

%%%%%%%%%%
\subsection{Numerical strategy}\label{sec:numerical}
%%%%%%%%%%
To solve for the phase-space distribution $f$ numerically, one first realizes that Eq.~\eqref{eq:Boltzmann_f} in principle describes a system of an infinite number of coupled differential equations --- one for each particle species involved in the problem at each continuous momentum value which extends from zero to infinity.
In practice, one needs to choose a finite range of momentum which covers all the relevant regions for the physical processes under study, and then discretize this finite momentum space. 
The final result, however, shall not depend on the details of the discretization.
In our work, we set the range and discretization based on the comoving momentum in order to conveniently accommodate the effects of the cosmological redshift.
The range of the comoving momentum should cover 
1) the entire range where the phase-space distribution is non-negligible;
2) the minimum and maximum comoving momenta at which a non-negligible number of particles in consideration can be produced or scattered into.
Once the range is fixed, we then discretize the comoving momentum space into $\mathcal{N}_p$ slices labeled by its comoving momentum $\tilde{p}_\ell$ with $\ell=1,2,\dots,\mathcal{N}_p$.
To further reduce the size of the system, notice that not all the distribution functions in the problem are unknown, \ie~particle species 
in thermal equilibrium with the visible-sector thermal bath follow thermal distributions which are completely specified by the temperature and their masses.
Therefore, we only need to solve for relevant particle species which are out of thermal equilibrium with the visible sector. 
Assuming $\mathcal{N}_s$ such species,
the entire Boltzmann-equation system then consists of $\mathcal{N}_p\times \mathcal{N}_s$ coupled differential equations.

We choose to discretize the comoving momentum space \emph{logarithmically} since it is naturally compatible with the (comoving) phase-space distribution $a^3p^3f(p,t)/(2\pi^2)$ defined with respect to $\log p$.
Such a choice has the advantage that shape of the distribution is invariant under the cosmological redshift, and that patterns in the distribution function can be conveniently mapped with patterns in observables of structure formation which are usually presented on logarithmic mass/length scales \cite{Dienes:2020bmn,Dienes:2021itb}.
In fact, from hereon , we will be referring to $a^3p^3f(p)/(2\pi^2)$ when presenting the phase-space distribution.
With this choice, the comoving number density associated with each momentum bin $\ell$ for a particle species $\alpha$ is therefore $a^3p_\ell^3f_{\alpha,\ell}\times \Delta\log p_\ell$, in which the width of the bin $\Delta \log p_\ell=\log(p_{\ell+1}/p_{\ell})$ is fixed since the physical momentum $p_\ell=\tilde{p}_\ell/a(t)$.

We then focus on the time evolution of each $f_{\alpha,\ell}$.
Let us first deal with the cosmological redshift and ignore the collision term for the moment.
In the RD or matter dominated (MD) epoch, the scale factor $a\sim t^{\kappa/3}$ with $\kappa=3/2$ for RD and $2$ for MD.
As a result, particles with momentum $p(t)$ at time $t$ redshift to a smaller momentum $p(t')$ at a later time $t'$ with 
\beq
p(t')=p(t)\times\frac{a(t)}{a(t')}=p(t)\times\left(\frac{t}{t'}\right)^{\kappa/3}\,.
\eeq
In numerical treatment, this redshift effect can be simply incorporated by redefining the physical momentum associated with each comoving momentum slice from $p_\ell(t)$ to $p_\ell(t')$ when evolving the system from $t$ to $t'$, which amounts to taking $f_\alpha(p_\ell(t),t)$ to $f_\alpha(p_\ell(t'),t')$.
The value $f_{\alpha,\ell}=f_{\alpha}(\tilde{p}_\ell/a(t),t)$ associated with each comoving momentum bin is in fact invariant under redshift.
Therefore, once particle interactions are taken into account, it turns out that the evolution of $f_{\alpha,\ell}$ only depends on the collision term:
\beq
\frac{df_{\alpha,\ell}}{dt}=C[f]\,.\label{eq:dfdt_discrete}
\eeq

The collision term involves integration over phase-space distributions of all relevant particles.
For the species whose phase-space distribution we seek to solve, their distribution functions are just the solutions of the Boltzmann-equation system at each time step.
On the other hand, for species in thermal equilibrium with the visible-sector thermal bath, their distribution functions $f_\alpha=f_\alpha^{\rm eq}$ are only functions of their masses and the thermal-bath temperature $T$.
Therefore, a conversion between the temperature and the cosmological time $t$ is needed.
For simplicity, we shall assume that the universe is dominated by the SM radiation bath during the entire simulation such that the Hubble parameter
\beq
H~\simeq~\sqrt{\frac{\rho_R}{3M_P^2}}=\sqrt{\frac{\pi^2}{90}}~g_{\star}^{1/2}(T)\frac{T^2}{M_P}\,\,\,,\label{eq:H_RD} 
\eeq
where $\rho_R$ is the energy density of the SM thermal bath, $M_P$ is the reduced Planck mass and $g_\star(T)$ is the effective number of relativistic degrees of freedom.
Comparing with $H=1/(2t)$ during RD epoch, the time-temperature relation is
\beq
t=\sqrt{\frac{45}{2\pi^2}}g_\star^{-1/2}(T)\frac{M_P}{T^2}\,.
\eeq
With this relation, all distribution functions in the phase-space integrals in the collision term are then completely specified.
After some simplifications on explicit formulas (see Appendix~\ref{sec:app}), numerical integration can then be performed straightforwardly at each time step.

Finally, after solving Eq.~\eqref{eq:dfdt_discrete}, the solution $f_{\alpha,\ell}(t)$ at time $t$ is then identified with $f_{\alpha}(\tilde{p}_\ell/a(t),t)$.
The entire Boltzmann system is thus solved with this procedure. 

\section{Freeze-out vs freeze-in: number density and beyond  }\label{sec:tworegimes}

\subsection{\texorpdfstring{A fresh look at a vanilla 2$\to$2 example}{}}

The production of dark matter can be realized in many different ways.
In this section, we focus on two most popular production mechanisms --- the \emph{freeze-out} and the \emph{freeze-in} mechanisms.

The freeze-out mechanism is a thermal production mechanism, in which dark matter is assumed to be in thermal equilibrium with the visible sector at early times, and decouples later as the expansion of the universe decreases the interaction rate such that the chemical equilibrium can no longer be maintained.
On the other hand, the freeze-in mechanism is non-thermal in the sense that thermal equilibrium is never established between dark matter and the visible sector.

To understand how the phase-space distribution of dark matter evolves in these two production mechanisms,
we take a simple example in which the dark-matter particle $\chi$ is produced only through a $2\to2$ annihilation 
process $\psi+\psi\leftrightarrow \chi+\bar\chi$.
We shall assume that $\psi$ is a species in the SM thermal bath with a mass smaller than that of $\chi$.
For simplicity, we shall neglect elastic-scattering processes like $\psi+\chi\leftrightarrow \psi+\chi$ and postpone the discussion of their effects to the next subsection.
Applying the general formula in Eq.~\eqref{eq:collision_general} to this example and simplifying the expressions,
the collision term for this $2\to2$ process can be rewritten in the following form (See Appendix \ref{sec:col_2to2_ann} for a detailed derivation):
\beqn
C_{\rm ann}(t,p_\chi)&=&\frac{{g_{\bar\chi}g_\psi^2}}{512\pi^3 E_\chi p_\chi}\int_{4m_\chi^2}^{\infty} ds \int_{{E_{\bar\chi}}^{\rm min}(s)}^{{E_{\bar\chi}}^{\rm max}(s)} \frac{dE_{\bar\chi}}{\sqrt{s}p_{\chi}^*(s)}\nn\\
&~&
\times \int^{t^{\rm max}}_{t^{\rm min}} dt~\Big[f^{\rm eq}_\chi(p_\chi) f^{\rm eq}_\chi(p_{\bar\chi})-f_\chi(p_\chi) f_{\chi}(p_{\bar\chi})\Big]\overline{\abs{\mathcal{M}_{\chi\bar\chi\to\psi\psi }}^2},\label{eq:col_ann_balance}
\eeqn
in which $s$ is the Mandelstam variable, and we have implicitly assumed that $f_\chi(p)=f_{\bar\chi}(p)$.
Obviously, the terms in the squared bracket suggest that the collision term always tends to bring $f_\chi$ to its equilibrium distribution $f_\chi^{\rm eq}$.

For freeze-out scenarios, the two terms in the squared brackets are balanced against each other at earlier times when $\psi$ and $\chi$ are in thermal equilibrium, \ie~$f_{\chi}=f_{\chi}^{\rm eq}$.
As the temperature drops below $m_\chi$, the creation of $\chi$ will be exponentially suppressed due to the Boltzmann suppression from the $f_{\chi}^{\rm eq}$ term --- the visible sector is suffering from a lack of kinetic energy in producing $\chi$.
However, the comoving number density will not immediately freeze, as $\chi\bar\chi$  annihilations tend to lower $f_\chi$ (and thus the comoving number density $a^3n_\chi$) to match $f_{\chi}^{\rm eq}$.
As the universe continues to expand and the temperature continues to drop, the interaction rate in both the forward and the inverse direction decreases and eventually becomes negligible compared with the redshift effect.
At this point, the dark-matter particle $\chi$ decouples chemically from the thermal bath, and the production stops.

For the simplest freeze-in scenarios, the factor $f_\chi(p_\chi) f_\chi(p_{\bar\chi})$ which controls the dark-matter annihilation rate is much smaller than the product $f_\chi^{\rm eq}(p_\chi) f_\chi^{\rm eq}(p_{\bar\chi})$ at all times --- the collision term is never strong enough to bring dark-matter particles $\chi$ into thermal equilibrium.
The dynamics is therefore predominantly affected by the production from the thermal bath
and the term corresponding to the $\chi\bar\chi$ annihilation in Eq.~\eqref{eq:col_ann_balance} can often be neglected in the calculation.
Just like the freeze-out scenario, the forward process in freeze-in scenario will be Boltzmann suppressed when the temperature falls below the mass of $\chi$.
However, since the inverse process is now completely negligible, the production stops as soon as $T\lesssim m_\chi$.

To obtain the dark-matter number density, the exact approach is to numerically solve Eq.~\eqref{eq:Boltzmann_f} and then integrate $f_{\chi}(p,t)$ at all times.
Another approach is to use the Boltzmann equation for number density directly, which, for the type of collision terms in Eq.~\eqref{eq:col_ann_balance}, is often written as
\beq
\frac{dn_\chi}{dt}+3Hn_\chi\simeq ({n_\chi^{\rm eq}}^2-n_\chi^2)\expt{\sigma_{\rm ann}v}_T\,,\label{eq:Boltzmann_n_chi}
\eeq
where $\expt{\sigma_{\rm ann} v}_T$ is the thermally averaged cross section for $\chi+\bar\chi\to \psi+\psi$.
Notice that this equation is exact only when $f_\chi\propto f_\chi^{\rm eq}$ , which is only possible when dark-matter particles follow a Boltzmann distribution and are in kinetic equilibrium with the thermal bath \ie~$f_\chi\sim e^{-(E-\mu_\chi)/T}$.
Indeed, Eq.~\eqref{eq:Boltzmann_n_chi} actually amounts to assuming that there exist interactions sufficiently rapid to maintain kinetic equilibrium at all times, even when dark matter is chemically decoupled.
For freeze-in scenarios, one often neglects the annihilation of $\chi$ and writes Eq.~\eqref{eq:Boltzmann_n_chi} as
\beq
\frac{dn_\chi}{dt}+3Hn_\chi\simeq ({n_\chi^{\rm eq}}^2)\expt{\sigma_{\rm ann}v}_T\,.\label{eq:Boltzmann_n_chi_fi}
\eeq

\begin{figure}
	\centering
	\includegraphics[width=0.7\textwidth]{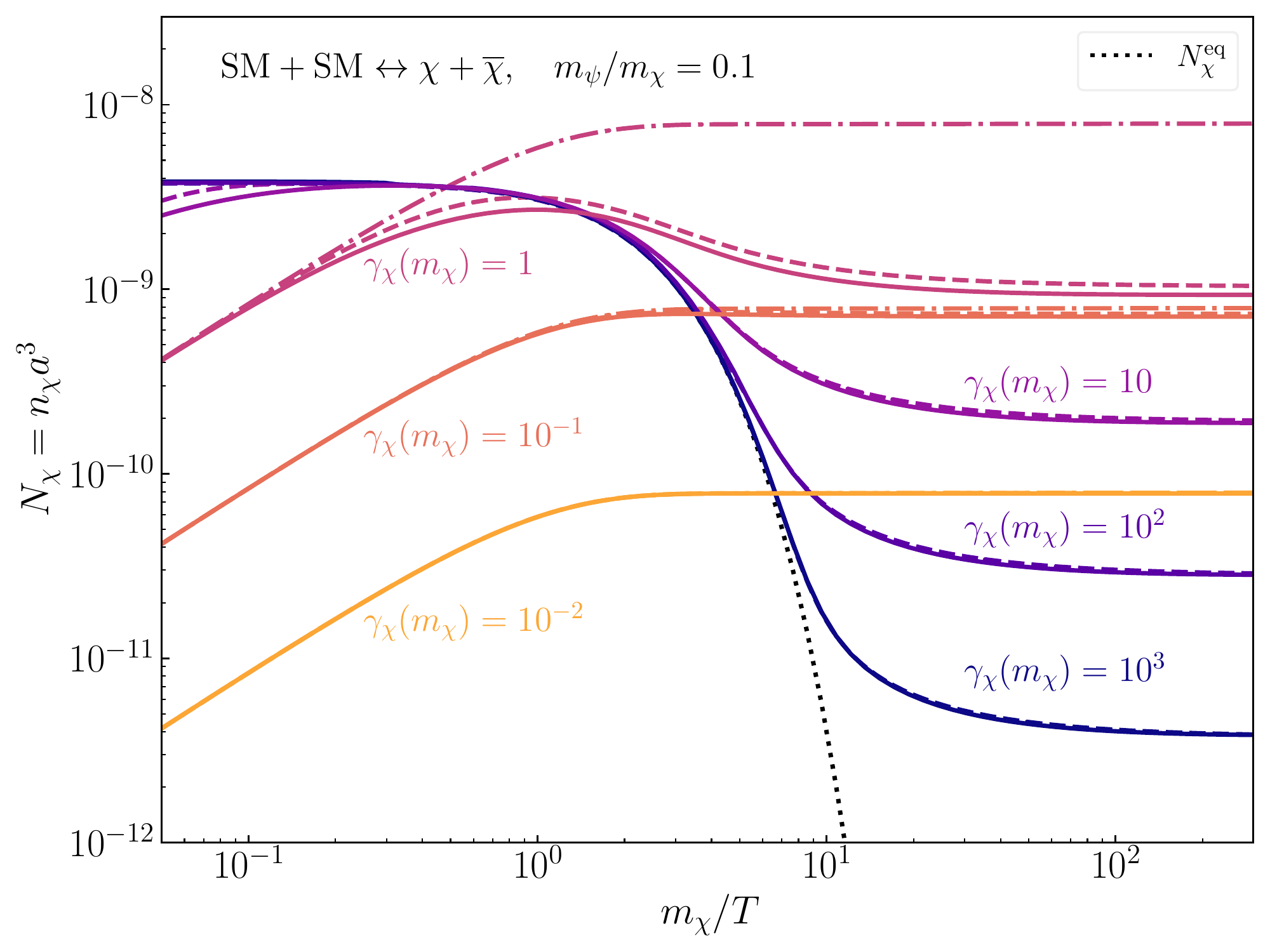}
	\caption{The evolution of the comoving number density for freeze-out and freeze-in scenarios governed by the process $\psi+\psi\leftrightarrow \chi+\bar\chi$ in which the masses are chosen such that $m_\psi/m_\chi=0.1$.
	The scale factor is normalized such that $a=1$ when $m_\chi/T=300$ (which is the boundary of the horizontal axis on the right).
	For each case, the Interaction strength is characterized by 
	$\gamma(m_\chi)$.
	The solid curves correspond to comoving number density obtained by integrating the solution of Eq.~\eqref{eq:Boltzmann_f}, whereas the dashed curves correspond to the solution of Eq.~\eqref{eq:Boltzmann_n_chi}.
	For freeze-in cases, the additional dash-dotted curves represent the solution of Eq.~\eqref{eq:Boltzmann_n_chi_fi}.
	Note that for $\gamma(m_\chi)=10^{-2}$ the three curves overlap as the three solutions agrees almost exactly.
	The dotted curve shows the case if $\chi$ is kept in thermal equilibrium with the radiation bath.}
	\label{fg:com_num_density}
\end{figure}

{Assuming $\chi$ and $\bar\chi$ are distinct particle species and setting $g_\chi=g_{\bar\chi}=g_\psi=1$ for simplicity, we present several examples of the comoving number density $N_\chi\equiv n_\chi a^3$} with gradually increasing interaction rate in FIG.~\ref{fg:com_num_density} in which solutions of Eq.~\eqref{eq:Boltzmann_f}, \eqref{eq:Boltzmann_n_chi} and \eqref{eq:Boltzmann_n_chi_fi} (for freeze-in only) are represented by solid, dashed and dash-dotted curves, respectively.
For simplicity, the amplitude $\abs{\mathcal{M_{\chi\bar\chi\to\psi\psi}}}^2$ is set to constant in all cases.\footnote{Of course, the form of the amplitude is model-dependent, which could be constant or dependent on the momenta of particles involved in the processes. 
For example, for fermionic dark matter annihilating into a pair of SM fermions through a light scalar or vector mediator, the leading amplitude-squared is approximately constant. Processes like this cover a lot of scenarios relevant for IR freeze-in.}
The interaction rate in each example is conveniently specified by the quantity
\beq
\gamma(T)\equiv\frac{n_\chi^{\rm eq}(T)\expt{\sigma_{\rm ann} v}_T}{H(T)}\,
\eeq
evaluated at $T=m_\chi$.
As such, the cases with $\gamma(m_\chi)=\{10^{-2},~10^{-1},~1\}$ correspond to freeze-in as the thermal equilibrium with the radiation bath is never reached.
On the contrary, the cases with $\gamma(m_\chi)=\{10,~10^{2},~10^{3}\}$ correspond to freeze-out since thermal equilibrium is reached at least for a short period as one can clearly see from the overlap of $N_\chi$ with $N_\chi^{\rm eq}$, which is the comoving number density of $\chi$ if it is kept in thermal equilibrium with the thermal bath.
Just as discussed before, in examples of freeze-out, the curves peel off from $N_{\chi}^{\rm eq}$ first before reaching their asymptotic values around $m_\chi/T\sim \mathcal{O}(10)$.
In examples of freeze-in, the curves stop increasing almost as soon as $m_\chi/T\sim 1$.
Indeed, we see that the comoving number density always freezes earlier in freeze-in cases than in freeze-out cases.

Comparing the solid curves with the dashed curves, we find excellent agreement between the two when $\gamma(m_\chi)$ is either much larger or much smaller than one. 
Nevertheless, we see that in general the latter tend to overestimate the total number of dark matter particles, and the largest discrepancy is seen when $\gamma(m_\chi)\sim \mathcal{O}(1)$, which is around 10\%.
{This discrepancy is determined by how important the inverse process (dark-matter annihilation) in Eq.~\eqref{eq:col_ann_balance} is and how well it is approximated by the $n_\chi^2\expt{\sigma_{\rm ann}v}$ after integrating the phase space.
For all the cases shown in FIG.~\ref{fg:com_num_density}, the number-density approach assumes kinetic equilibrium throughout, \ie~using $f_\chi\propto f_\chi^{\rm eq}$ in Eq.~\eqref{eq:col_ann_balance}.
As we shall see later, this means that, when out of thermal equilibrium, dark matter is assumed to have overall a momentum larger than what it actually is.
Meanwhile, in our example, $\expt{\sigma_{\rm ann}v}$ decreases with the center-of-mass energy in the relativistic regime and tends towards a constant in the non-relativistic regime.
Together, the above reasons suggest that the rate of dark-matter annihilations is underestimated in the number-density approach which explains the overestimate in the relic abundance.}

Besides comparing the results from solving Eq.~\eqref{eq:Boltzmann_f} and Eq.~\eqref{eq:Boltzmann_n_chi}, for freeze-in cases, solutions to Eq.~\eqref{eq:Boltzmann_n_chi_fi} (dash-dotted curves), which neglect dark-matter annihilations, are a good approximation only when $\gamma(m_\chi) \lesssim10^{-1}$.
In general, the agreement becomes worse as the interaction rate increases.
For example, in the case with $\gamma(m_\chi)=1$, dark-matter particles are very close to establishing thermal equilibrium with the thermal bath.
Consequently, annihilations of $\chi$ are no longer negligible, and the error can be of $\mathcal{O}(10)$.
This error can even be as large as $\mathcal{O}(100)$ with a slightly larger $\gamma(m_\chi)$ which still does not permit thermal equilibrium. 

{Based on the above observations, we conclude that for the vanilla $2\to2$ freeze-out and freeze-in, the number-density approach is in general in good agreement with the phase-space-distribution approach in estimating the dark-matter relic abundance even in the transition regime.
However, the annihilation of dark-matter particles cannot always be neglected --- neglecting such process would lead to a large error in the transition regime, even if the thermal equilibrium with the SM thermal bath cannot be established.
As we shall see in Sec.~\ref{sec:parent_decay}, similar behaviors also occur for the $1\to2$ process.}

\subsection{{Phase-space distribution from the \texorpdfstring{$2\to2$}{2->2} process}}

\begin{figure}
	\centering
	\includegraphics[width=0.49\textwidth]{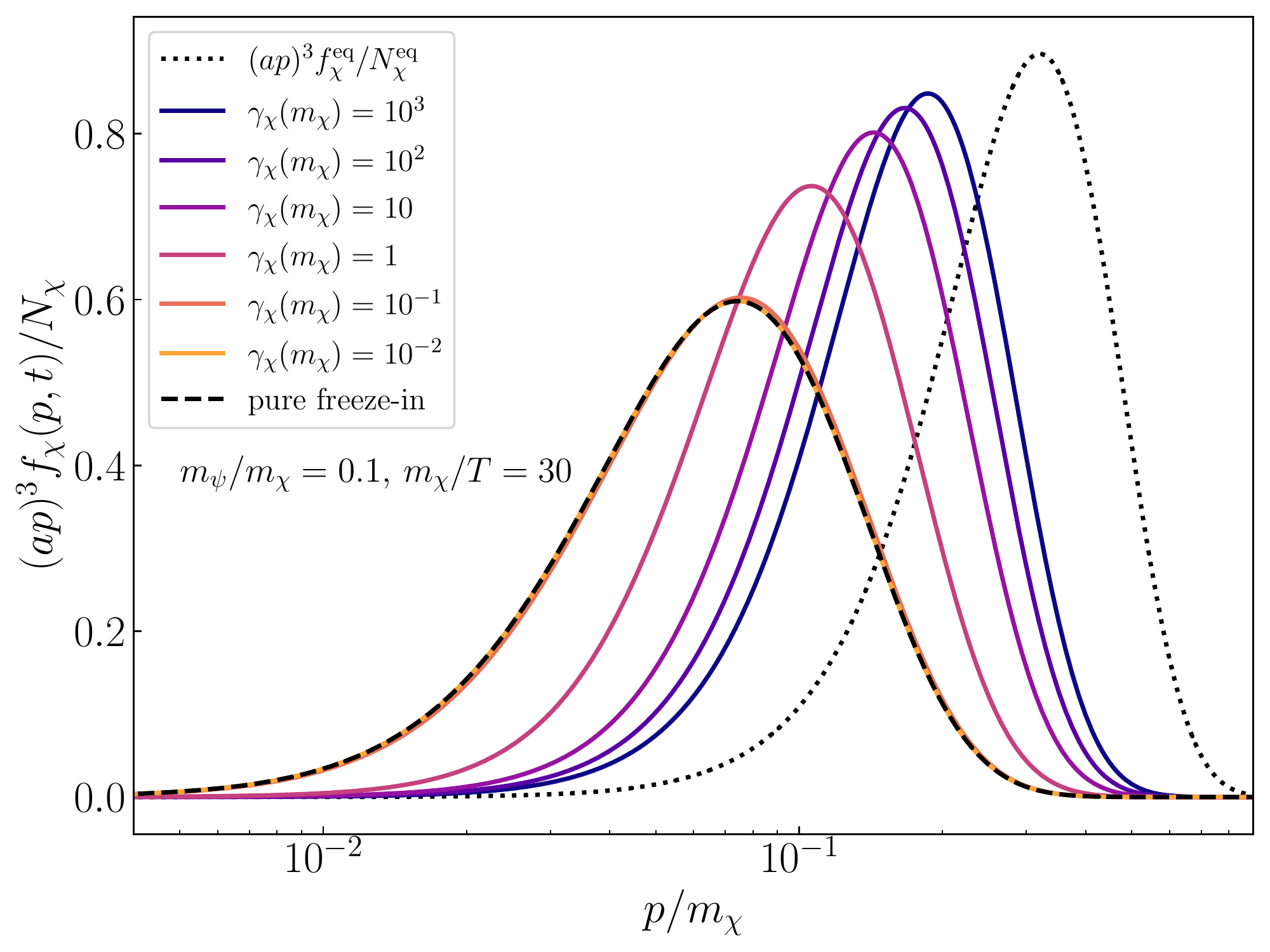}
	\includegraphics[width=0.49\textwidth]{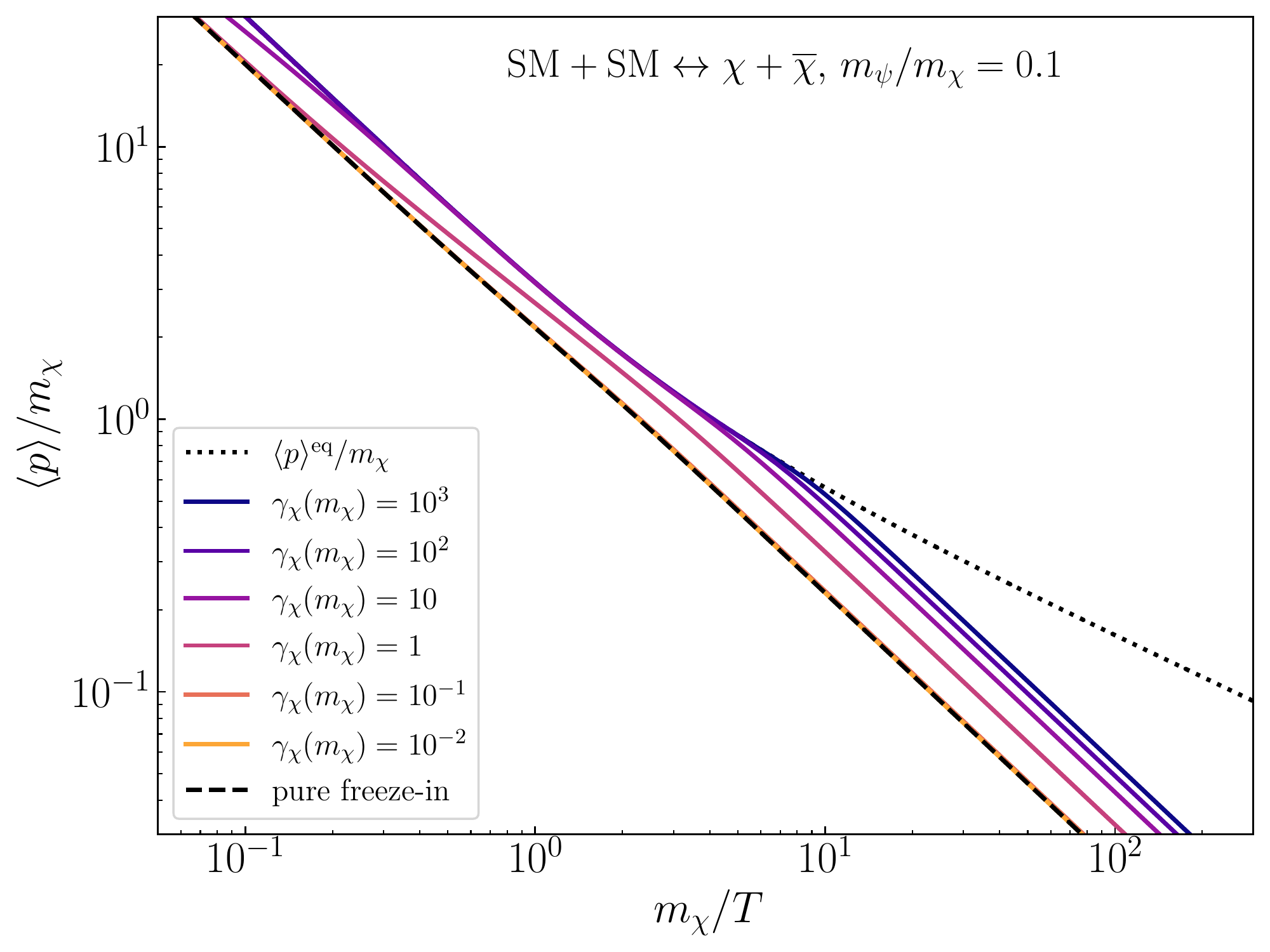}
	\caption{
	\textbf{Left panel}: Phase-space distribution of dark matter at $m_\chi/T=30$ normalized by the final comoving number density $N_\chi$.
	The MB thermal distribution at the same temperature is also shown by the dotted curve for comparison.
	\textbf{Right panel}: The evolution of the average momentum normalized by the dark-matter mass $m_\chi$.
	The dotted curve shows the case when dark matter is in equilibrium with the thermal bath.
	}
	\label{fg:2to2_avep}
\end{figure}

Beyond the number density, we now compare the momentum of dark-matter particles in freeze-in and freeze-out scenarios.
In the left panel of FIG.~\ref{fg:2to2_avep}, we show the phase-space distributions associated with the examples in FIG.~\ref{fg:com_num_density} together with the thermal distribution at $m_\chi/T=30$ when the production is well finished (when the comoving number density is fixed).
As already mentioned in Sec.~\ref{sec:numerical}, instead of showing $f_\chi(p)$ directly, we plot $(ap)^3f_\chi(p)$ against $p/m_\chi$ on log scale.
All distributions are normalized by the final comoving number density $N_\chi$ such that the areas under the curves are the same.
A few noticeable features can be seen immediately:
as one dials down the interaction strength continuously such that the production transitions from freeze-out to freeze-in, the shape of the distribution function after production also changes continuously --- it shifts towards the lower momentum, becomes lower and wider.
Interestingly, the change in the shape of the distribution is bounded.
While increasing the interaction strength makes the shape of the distribution more and more similar to that of the thermal distribution, in the limit where the interaction strength is very small, the distribution also deforms towards a fixed limiting shape which can be clearly seen as the curves with $\gamma(m_\chi)=10^{-1}$ and $10^{-2}$ almost overlap.
In fact, this limiting shape as $\gamma(m_\chi)\rightarrow 0$ can be obtained by completely neglecting the back reaction from dark matter, as is indicated by the dashed black curve. 
This is simply due to the fact that,
relative to the dark-matter production part in Eq.~\eqref{eq:col_ann_balance} which depends on the thermal distribution $f_\chi^{\rm eq}$, the annihilation part differs from the production part in that it depends on the actual distribution $f_\chi$.
As a result, as $\gamma(m_\chi)\to 0$, the annihilation part would vanish faster than the production part since $f_\chi\to 0$ while $f_\chi^{\rm eq}$ is not changed, making the $\gamma(m_\chi)\to 0$ limit indeed a limit in which there is no back reaction from dark matter.

In the right panel, we show the evolution of the average momentum.
In freeze-out cases, the average momentum of dark matter follows $\expt{p}^{\rm eq}$ before it decouples from the thermal bath, and its value after decoupling depends on when the decoupling occurs --- the later it decouples, the larger the final momentum is.
In freeze-in cases, similarly, more kinetic energy could be drained from the thermal bath when the interaction is stronger.
However, if the interaction rate is sufficiently small, the evolution of the average momentum approaches an asymptotic boundary (the dashed black curve) determined by the dashed limiting distribution in the left panel.
Therefore, consistent to what we have seen from the left panel, $\expt{p}$ is confined between $\expt{p}^{\rm eq}$ and this boundary, 
and, within the boundaries, the average momentum in general gets larger as the interaction strength increases.

\FloatBarrier

%%%%%%%%%%%%%%%%%%%%%%%%%%%%%%%%
\subsubsection{Effects of elastic scatterings on unimodal distribution}\label{subsec:elastic}
%%%%%%%%%%%%%%%%%%%%%%%%%%%%%%%%
So far, we have only considered $2\to2$ annihilation $\psi+\psi\leftrightarrow\chi+\bar\chi$ and have completely ignored the effect from elastic scattering. 
Under this simplification, the condition for establishing and maintaining the chemical and kinetic equilibrium is solely controlled by the annihilation rate.
However, this needs not be true.
The $2\to2$ annihilation naturally implies the elastic-scattering process $\chi+\psi\leftrightarrow \chi+\psi$.
Moreover, dark-matter particles can also scatter with other particles in the thermal bath.
{In some cases, the such processes might even be more rapid than the scatterings between $\chi$ and $\psi$ since the couplings that govern the dominant annihilation process and the dominant elastic-scattering process need not be the same.
For example, $\psi$ could be a scalar that has a small coupling with $\chi$ and $\bar\chi$ while having a large coupling with a pair of heavy SM fields. In this situation, the elastic-scattering with this heavy SM particle might have a much larger cross section than that with $\psi$.}

Since the elastic scattering is not a number-changing process,
at first glance, it seems that the evolution of the comoving number density should not depend on such processes at all.
However, the elastic scattering enables momentum exchange between dark-matter particles and the radiation bath even when they are not in thermal equilibrium, which means the dark-matter phase-space distribution is indeed expected to change if the elastic scattering is present.
Subsequently, the change in the phase-space distribution would in principle affect the evaluation of the collision term.
In fact, FIG.~\ref{fg:com_num_density} has already shown to what extent elastic scatterings could influence the relic abundance as the results from Eq.~\eqref{eq:Boltzmann_n_chi} are obtained by assuming kinetic equilibrium with the thermal bath at all times.

To see how elastic scatterings modify the phase-space distribution,
we take two benchmark points from FIG.~\ref{fg:com_num_density} with $\gamma(m_\chi)=10^{-2}$ and $10^{2}$ which correspond to freeze-in and freeze-out, respectively, 
{and then add an elastic-scattering channel.
Although the $\chi+\psi\leftrightarrow \chi+\psi$ channel is not necessarily the dominant process for elastic scattering, 
we shall use this channel as a proxy by simply including it in the collision term and adjusting its amplitude freely.
Notice that, this means that we shall not assume any theoretical connection between the annihilation amplitude $\abs{\mathcal{M}_{\chi\bar\chi\to\psi\psi}}^2$ and the elastic-scattering amplitude $\abs{\mathcal{M}_{\rm el}}^2\equiv \abs{\mathcal{M}_{\chi\psi\to\chi\psi}}^2$, other than that they are both constants.
Such choice allows us to study the effects of the elastic scattering independently and avoids the need of introducing additional particle species.
}

\begin{figure}
    \centering
    \includegraphics[width=\textwidth]{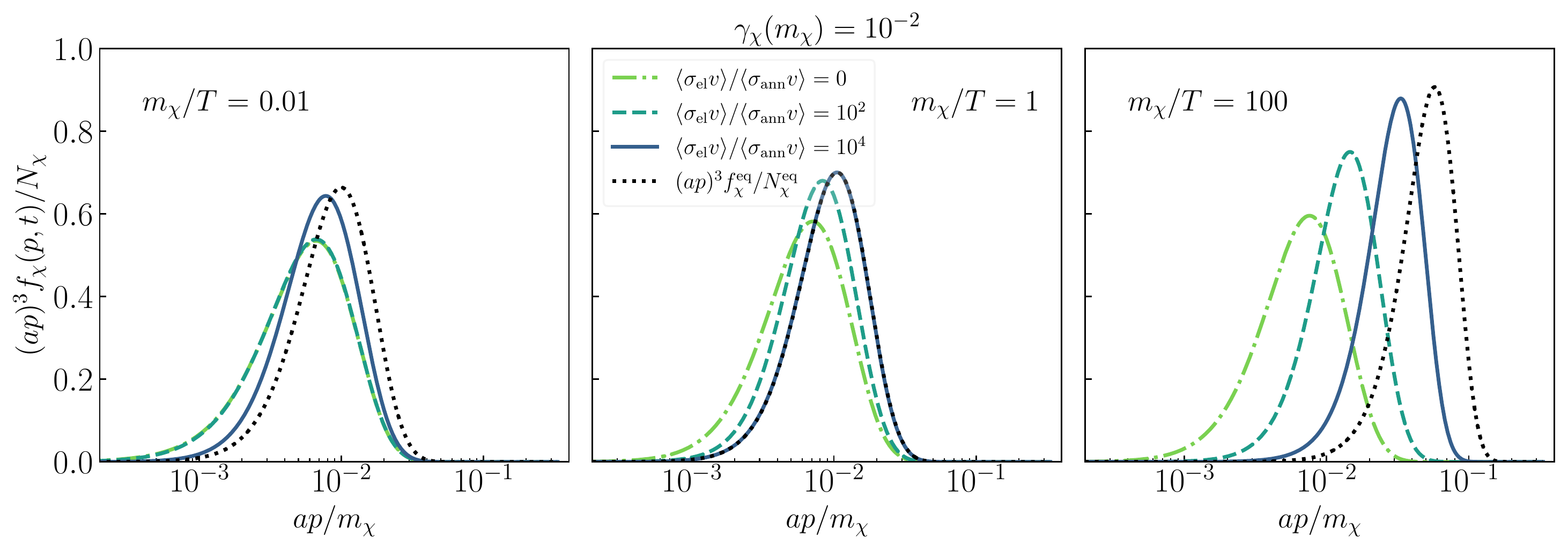}\\
    \includegraphics[width=\textwidth]{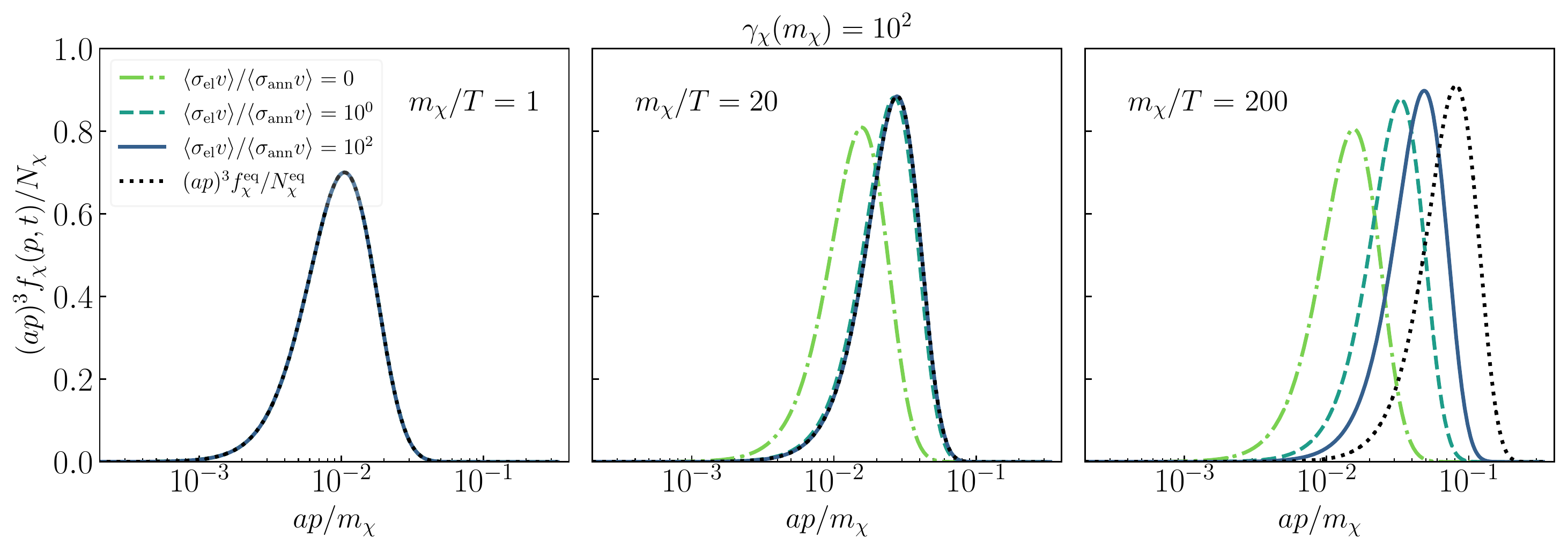}
    \caption{Snapshots of the phase-space distribution during freeze-in (upper panels) and freeze-out (lower panels) production {with different elastic-scattering rates. The ratios of the thermally averaged cross sections are evaluated at $T=m_\chi$.}
     The normalization of the scale factor is the same as in FIG.~\ref{fg:com_num_density}.}
    \label{fg:elastic_psd}
\end{figure}

We show the effects of elastic scatterings on the evolution of phase-space distribution in FIG.\,\ref{fg:elastic_psd}, where the first row corresponds to freeze-in and the second row to freeze-out.
In each row, we show cases with increasing elastic-scattering amplitudes {(characterized by the ratios between the thermally averaged cross sections $\expt{\sigma_{\rm el}v}/\expt{\sigma_{\rm ann}v}$ evaluated at $T=m_\chi$)} and take three snapshots during the production process.
We also show the thermal distribution \mbox{$(ap)^3f_\chi^{\rm eq}/N_\chi^{\rm eq}$} in all panels at the corresponding temperature.
Note that, all distributions are normalized so that they have the same area under the curve even though the number densities are not necessarily the same.
\bigskip

\noindent\textbullet~Freeze-in

In the first row, the left panel corresponds to the time when freeze-in just starts ($m_\chi/T=0.01$) and the kinetic equilibrium is not yet established.
Clearly, other than the case with the largest elastic scattering rate, the shapes of the distributions at that time all deviate significantly from the thermal distribution at the corresponding temperature.
As one increases the elastic-scattering rate, we see that the distribution deforms in a way that it becomes more and more similar to the thermal distribution.
Indeed, the case with the largest elastic scattering rate, \ie~ $\expt{\sigma_{\rm el}v}/\expt{\sigma_{\rm ann}v}=10^4$, has a shape that almost resembles the thermal distribution, suggesting the kinetic equilibrium {with the thermal bath} is almost established.
In general, a distribution associated with a larger elastic-scattering rate tends to have an overall larger momentum and a smaller width, and is more concentrated at the higher-momentum end.
This suggests that, in freeze-in scenarios where dark-matter particles produced are usually ``colder'' than the thermal bath, elastic scatterings always facilitate the extraction of kinetic energy from the thermal bath.
We shall see later by the end of Sec.~\ref{subsec:out_of_eq_dec} that elastic scatterings can also consume the overall kinetic energy of dark matter if a non-negligible portion of dark-matter particles are ``warmer'' than the thermal bath.\footnote{The wording ``colder'' and ``warmer'' here refers to the typical momentum of the particles under consideration and is not necessarily associated with a temperature.}

As dark-matter particles keep populating, at later times (\eg~$m_\chi/T=1$ in the middle panel), the blue curve merges with the thermal distribution which indicates $f_\chi\propto f_\chi^{\rm eq}$.
This is nothing but the evidence that the kinetic equilibrium with the thermal bath is established.
Indeed, in kinetic equilibrium, $f_\chi\sim e^{-(E-\mu_\chi)/T}\sim e^{\mu_\chi/T} f_\chi^{\rm eq}$,
the proportionality $e^{\mu_\chi/T}$ is exactly the ratio between the normalization factors $N_\chi/N_\chi^{\rm eq}$.
The other two curves, however, are more separated from the thermal distribution.
The one with a larger elastic-scattering rate tends to be closer to the thermal distribution and has more kinetic energy,
which is consistent with our observation in the left panel.

At even later times after the entire freeze-in process is well completed (\eg~$m_\chi/T=100$ in the right panel), the blue curve is peeled off from the thermal one suggesting that kinetic decoupling has already occurred.
Nevertheless, cases with larger elastic-scattering rates still show an overall larger momentum, and at the same time a smaller width.
\bigskip

\noindent\textbullet~Freeze-out

In the bottom row of FIG.~\ref{fg:elastic_psd}, we show the influence of elastic scatterings during the freeze-out production.
Unlike the freeze-in scenario, at high temperatures (\eg~$m_\chi/T= 1$ in the left panel), all the phase-space distributions in the freeze-out scenario are the same as the thermal one, despite how rapid the elastic scattering is.
Indeed, before the chemical decoupling, thermal equilibrium, which implies both kinetic equilibrium and chemical equilibrium, can be maintained solely by the annihilation process $\psi+\psi\leftrightarrow\chi+\bar\chi$.

Around the freezes-out temperature(\eg~$m_\chi/T= 20$ in the middle panel), the annihilation process decouples,
and only the case with the largest elastic-scattering rate can efficiently exchange momentum with the thermal bath and maintain a thermal shape in its distribution, which suggests that it is still in kinetic equilibrium.
As one dials down the elastic-scattering rate, the distribution in general starts to peel off from the thermal distribution, shifts towards lower momentum and becomes wider.

Well after freeze-out (\eg~$m_\chi/T= 200$ in the right panel), dark matter is kinetically decoupled from the thermal bath even in the case with the largest elastic-scattering rate.
Nevertheless, the effect of elastic scatterings can still manifest itself in the relative shift of the momentum distribution --- 
when the distribution under consideration is colder than the thermal bath, elastic scatterings always facilitate the extraction of kinetic energy from the thermal bath.

\begin{figure}
    \centering
    \includegraphics[width=0.49\textwidth]{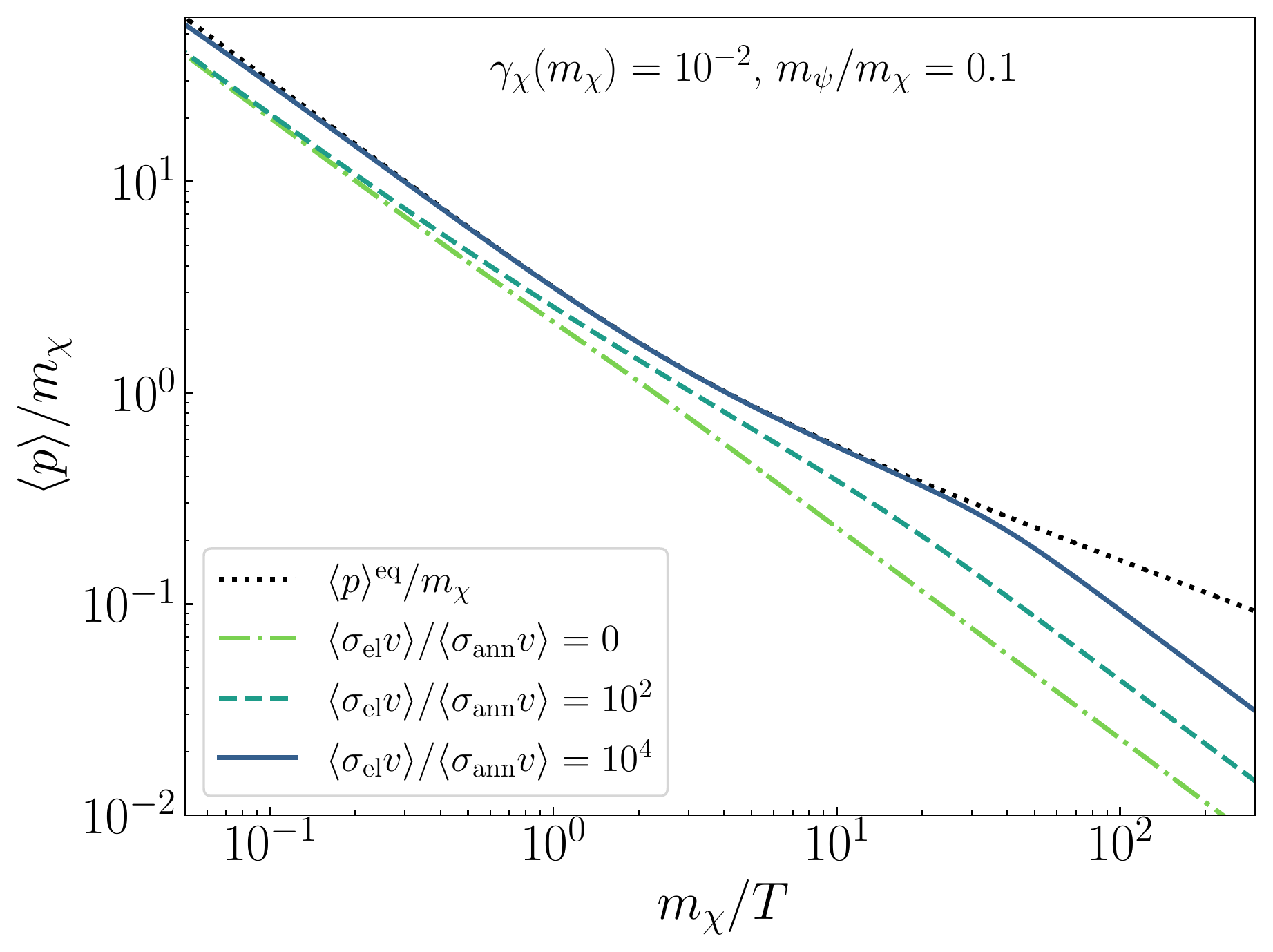}
    \includegraphics[width=0.49\textwidth]{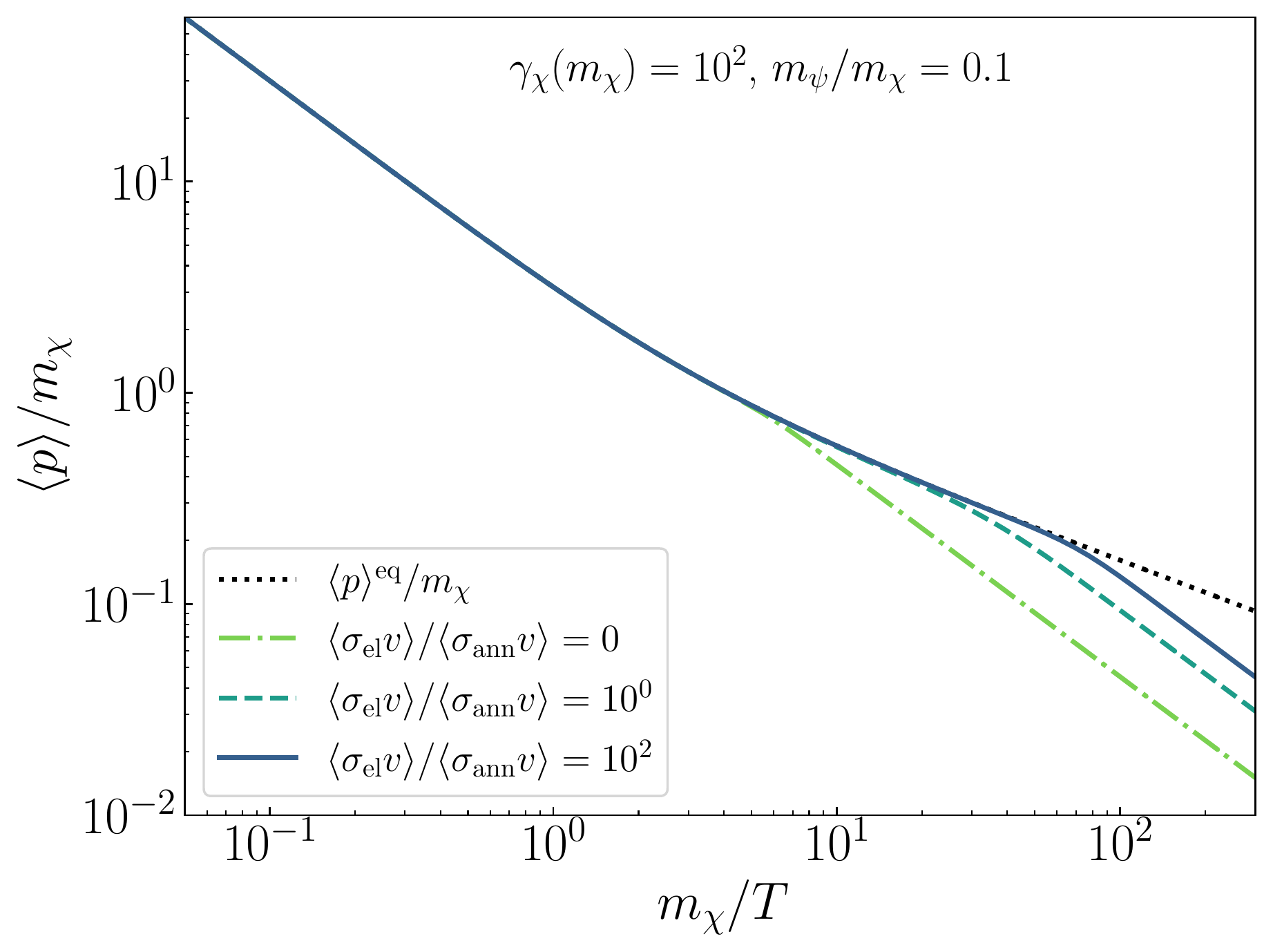}
    \caption{Evolution of average momentum with different elastic-scattering rates.}
    \label{fg:elastic_p}
\end{figure}

\bigskip
Finally, we show in FIG.~\ref{fg:elastic_p} the evolution of average momentum (normalized by $m_\chi$) for all the cases discussed in this subsection.
In both panels, the evolution of average momentum only varies from the curve with zero elastic-scattering rate to $\expt{p}_\chi^{\rm eq}$ as the elastic-scattering rate increases.
For freeze-in cases in the left panel, kinetic equilibrium can only be established when elastic scatterings are rapid enough -- when the corresponding curve merges with the black dotted one.
On the other hand, for freeze-out cases in the right panel, all of them show kinetic equilibrium at earlier times, and then the elastic-scattering rate determines when they peel off from $\expt{p}_\chi^{\rm eq}$, which subsequently determines the final average momentum.

\FloatBarrier

%%%%%%%%%%%%%%%%%%%%%%%%%%%%%%%%
\section{Freeze-in distributions from \texorpdfstring{$1\to2$}{1->2} decay}\label{sec:parent_decay}
%%%%%%%%%%%%%%%%%%%%%%%%%%%%%%%%
Besides the vanilla $2\to2$ annihilation process, the freeze-in mechanism can also be realized via other processes which might have different implications in the relic-abundance estimation when comparing the result from the phase-space distribution with that from the number density.
From the perspective of the dark-matter phase-space distribution itself, we are interested in exploring the possibility of generating distributions that are very different from what we have seen in Sec.~\ref{sec:tworegimes}.
As was shown in \cite{Konig:2016dzg,Heeck:2017xbu,Dienes:2020bmn}, the dark-matter phase-space distribution could have complicated, multi-modal structure if non-negligible, separate fractions of dark matter are produced at different times.
For this sake, we consider freeze-in production of dark matter through the $1\to2$ decay in this section, as the freeze-in production can naturally be accompanied with an additional contribution from the late-time decay of the parent particle.

To be concrete, we label the decay process as $A\leftrightarrow\chi+\bar{\chi}$, in which $\chi$ represents the dark-matter particle,
and the mediator $A$ is a particle species produced from the annihilation of a pair of SM particles: $\psi+\psi\leftrightarrow A+A$.
We consider two scenarios here in which $A$ decays either in or out of thermal equilibrium with $A$ produced through either freeze-out or freeze-in mechanisms.
Before delving into the detailed results,
we first apply Eq.~\eqref{eq:Boltzmann_f} to these two scenarios here.
The two coupled Boltzmann equations for the particle $A$ and the dark matter $\chi$ are 
\begin{align}
	\frac{\partial f_A}{\partial t} - Hp\frac{\partial f_A}{\partial p} &=C^A_{\rm ann}[f_A] + C^{A}_{\rm dec}[f_A,f_\chi]  \,,\label{eq:Boltzmann_fA} \\
	\frac{\partial f_\chi}{\partial t} - Hp\frac{\partial f_\chi}{\partial p} &= C^{\chi}_{\rm dec}[f_A,f_\chi]  \,.\label{eq:Boltzmann_fx}
\end{align}
The collision term $C^A_{\rm ann}[f_A]$ is just Eq.~\eqref{eq:col_ann_balance},
by replacing $\chi$ with $A$.
The two remaining collision terms contain both the decay and the inverse-decay processes of $A$  (See the Appendix~\ref{sec:app} for details).
 Separating the production and depletion parts, the collision terms can be written as\footnote{We have neglected the Bose-enhancement/Pauli-blocking effects, such that on the right-hand side the collision terms are functions of a single phase-space distribution.}:
  \begin{eqnarray}
      C^{A}_{\rm dec}[f_A,f_\chi] &=&C^{A}_{\chi\bar{\chi}\to A}[f_\chi]-C^{A}_{A\to\chi\bar{\chi}}[f_A],\\
       C^{\chi}_{\rm dec}[f_A,f_\chi] &=&C^{\chi}_{A\to\chi\bar{\chi}}[f_A]-C^{\chi}_{\chi\bar{\chi}\to A}[f_\chi],\\
       C^{A}_{\rm ann}[f_A] &=&  C^{A}_{\psi\psi\to AA}[f^{\rm eq}_A]-C^{A}_{AA\to \psi\psi}[f_A],
 \end{eqnarray}
in which the subscripts explicitly denotes the processes with specific directions, and we call both $C^{A}_{\chi\bar{\chi}\to A}[f_\chi]$ and $C^{\chi}_{\chi\bar{\chi}\to A}[f_\chi]$ inverse decay terms.
All the terms on the right-hand side are positive, thus the sign conventions are manifest with respect to Eq.~\eqref{eq:collision_general}.
 
In the freeze-in production of $A$ or $\chi$, the inverse decay from $\chi\chi$ to $A$ or back reaction from $AA$ to $\psi\psi$ are sometimes neglected in the literature, as the number density of the produced particle are usually small. An interesting consequence of neglecting the inverse processes would be the simplification of the Boltzmann equations. 
For example, neglecting the inverse-decay process will decouple the Eq.~\eqref{eq:Boltzmann_fA} and \eqref{eq:Boltzmann_fx}, thus one can solve for $f_A$ with Eq.~\eqref{eq:Boltzmann_fA} and insert the solution in Eq.~\eqref{eq:Boltzmann_fx} to solve for $f_\chi$, rather than solving the coupled differential equations which would require more computing resources. 
However, as we will see in the following subsections, these simplifications by neglecting part of the inverse process are not always valid especially in the transition regime between freeze-in and freeze-out, which means solving the coupled equations is necessary in some cases. 
In the following, we present the results for the numerical solutions to different benchmarks with and without the inverse-decay process in the in-equilibrium scenario, and including/excluding back reaction from $AA$ to $\psi\psi$ in the out-of-equilibrium decays scenario, 
and discuss the effects and necessity for solving  Boltzmann equations of phase-space distributions.
Again, for simplicity, we set the decay amplitude $|{\cal M}_{A\to \chi\bar\chi}|^2$ to constant in following analysis.

For the convenience of the readers, we also list the number-density Boltzmann equations followed by Appendix.~\ref{sec:app2} for $n_\chi$ and $n_A$ as a comparison, in which the assumption that dark matter and $A$ are in kinetic equilibrium with thermal bath is used:
\begin{align}
      \frac{d n_A}{d t} + 3 H n_A &= \left\langle \sigma_{\psi\psi \rightarrow A A} v \right\rangle \left( n_A^{\rm eq^2} - n_A^2 \right) - \left\langle \Gamma_{A\rightarrow \chi\bar{\chi}} \right\rangle \left( n_A - n_A^{\rm eq}\frac{ n_\chi^2}{ n_\chi^{\rm eq^2}}  \right)   \, ,\label{eq:Boltzmann_n_A_dec}\\
	  \frac{d n_\chi}{d t} + 3 H n_\chi &=   \left\langle \Gamma_{A\rightarrow \chi\bar{\chi}} \right\rangle \left( n_A - n_A^{\rm eq}\frac{ n_\chi^2}{ n_\chi^{\rm eq^2}}  \right) \,\label{eq:Boltzmann_n_chi_dec}.
 \end{align}
The inverse decay terms refer to the second term within the parentheses next to $\left\langle \Gamma_{A\rightarrow \chi\bar{\chi}} \right\rangle$, and the back reaction term for $A+A\to\psi+\psi$ are the term proportional to $n_A^2$.
%%%%%%%%%%%%%%%%%%%%%%%%%%%%%%%%
\subsection{In-equilibrium decay}\label{subsec:in_eq_dec}
%%%%%%%%%%%%%%%%%%%%%%%%%%%%%%%%

\begin{figure}
\includegraphics[width=0.49\textwidth]{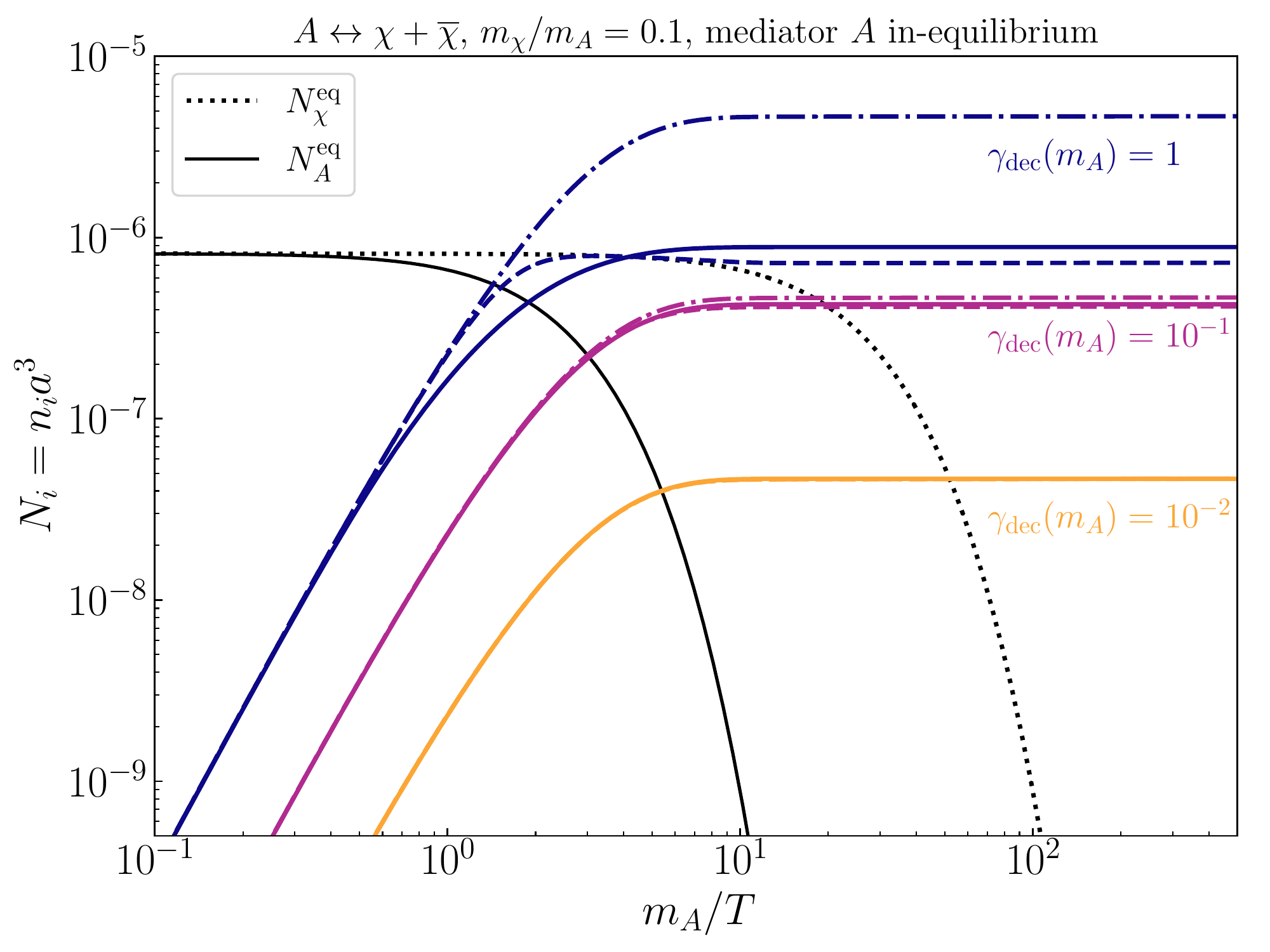}\includegraphics[width=0.49\textwidth]{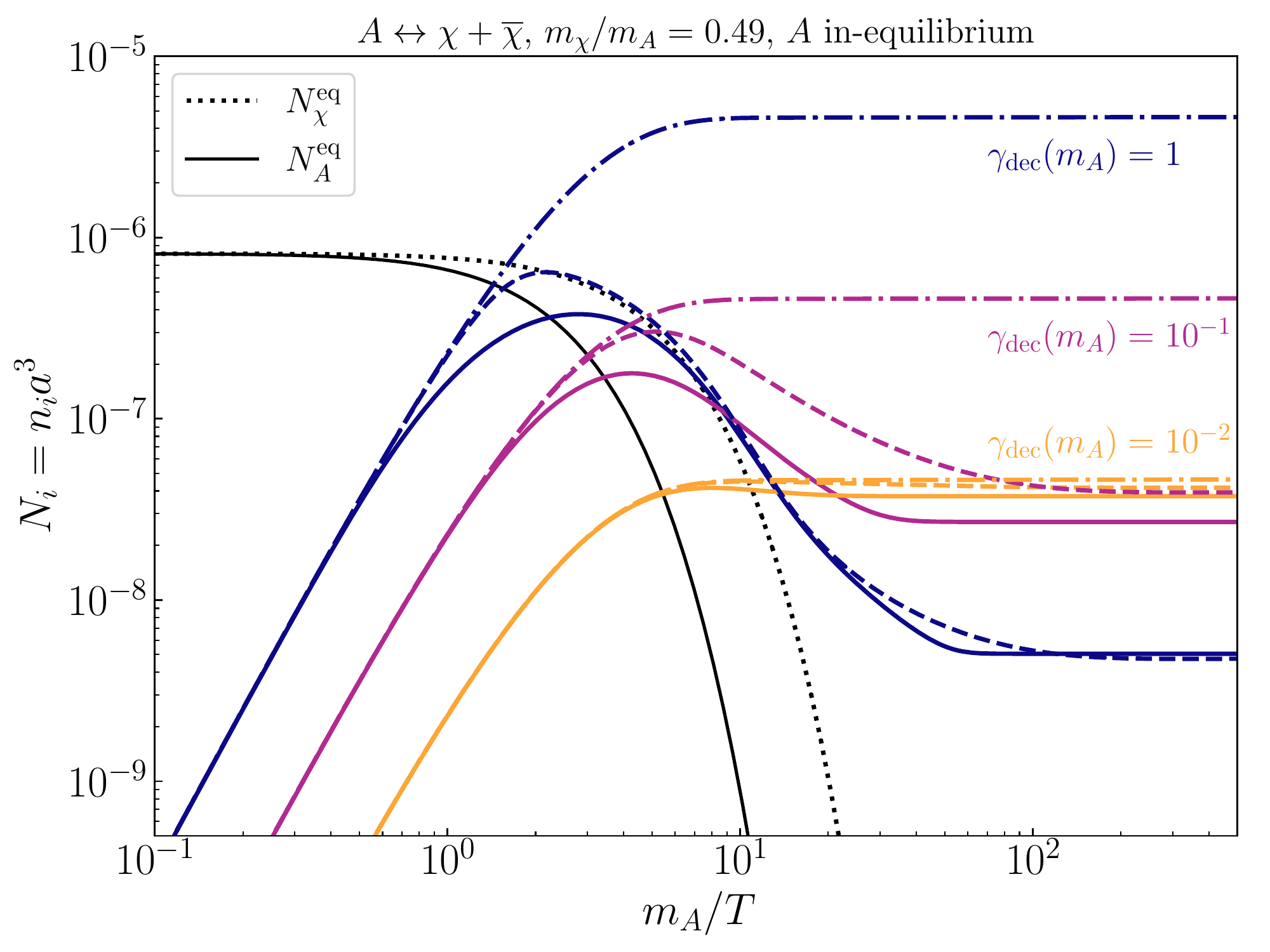} 
\caption{The freeze-in production of dark-matter particles $\chi$ through $A\leftrightarrow\chi+\bar{\chi}$ in which the mediator $A$ is assumed to be in thermal equilibrium with the SM thermal bath throughout the dark-matter production.
We choose two mass ratios $m_\chi/m_A=0.1$ (left panel) and $m_\chi/m_A=0.49$ (right panel).
The black curves stand for the comoving number density of $A$ (solid) and $\chi$ (dotted) if they are in thermal equilibrium.
The blue, purple and orange curves represent the comoving number density of $\chi$ associated with different decay widths of $A$, among which the three line styles, solid, dashed and dash-dotted, correspond to results obtained from solutions to Eq.~\eqref{eq:Boltzmann_fx}, Eq.~\eqref{eq:Boltzmann_n_chi_dec} and a simplified version of Eq.~\eqref{eq:Boltzmann_n_chi_dec} without including the inverse decay.
{Note that, the horizontal axis is normalized by $m_A$ which is fixed in all cases in this section. Likewise, the scale factor in this section is normalized such that $a=1$ when $m_A/T=500$.}
}\label{fg:numden}
\end{figure}

In this subsection, we focus on the freeze-in production of the dark matter $\chi$ via the decay of the particle $A$ when $A$ is in thermal equilibrium with the SM sector.
In other words, $A$ can be treated as a species in the thermal bath.
We summarize our results in the two plots in FIG.~\ref{fg:numden}.
Since $f_A=f^{\rm eq}_A$ in all these cases, we only need to solve for $f_\chi$ in Eq.~\eqref{eq:Boltzmann_fx}.
We choose two benchmarks for the mass ratio --- $m_\chi/m_A=0.1$ (left) for which the dark matter particles are likely to be boosted due to the release of the energy stored in the mass of $A$;  and $m_\chi/m_A=0.49$ (right) for which the mass of $A$ merely pass the threshold of producing a pair of $\chi$. 
In each plot, the solid curves represent the number density obtained by integrating the solutions to the Boltzmann equations Eq.~\eqref{eq:Boltzmann_fx}, the dashed curves represent the solution to Eq.~\eqref{eq:Boltzmann_n_chi_dec}, and the dash-dotted curves that to Eq.~\eqref{eq:Boltzmann_n_chi_dec} when the inverse decay is neglected. 
Three different colors---blue, purple and orange, represent three decay widths of $A\to\chi+\bar{\chi}$, measured by their ratios to the Hubble parameter $\gamma_{\rm dec}(m_A)\equiv \Gamma_A/H(m_A)=10^{-1},~10^{-2}$ and $10^{-3}$, respectively. 
Combining three decay widths and two mass ratios, we have in total six benchmarks.

From the two plots in FIG.~\ref{fg:numden}, we can find that the inclusion of the inverse decay $\chi+\bar\chi\to A$ has a large effect when $\gamma_{\rm dec}(m_A)$ approaches $\mathcal{O}(1)$ --- close to the critical point dividing the freeze-in and freeze-out regimes.
{When the inverse decay is not included, the relic abundance is overestimated by about one order of magnitude in the left panel and around three orders of magnitude in the right panel for $\gamma_{\rm dec}(m_A)=1$ (see the dash-dotted curves).}
Comparing the two plots, one can see that benchmarks with larger mass ratio $m_\chi/m_A$ are more sensitive to the process of inverse decay as the discrepancies between the dash-dotted and the solid curves are larger. {Even for $\gamma_{\rm dec}(m_A)=10^{-1}$, the error can still be as large as one order of magnitude}.
{The reason behind this is twofold --- a large mass ratio leads to a smaller average momentum of dark matter, and at the same time, for a constant decay amplitude, the inverse-decay rate increases as the momentum of dark matter decreases.
Therefore, at early times when dark matter is relativistic, there can be a significant inverse-decay rate if the mass ratio is large, which means a large error in the relic-abundance estimation could result if the inverse decay is not taken into account.}

Another feature that can be observed from the two plots is that, at early times, the dashed curves, which correspond to results from the number-density approach, are always above the solid ones (which is manifest for relatively large decay width).
This is because the number-density solution assumes the dark matter is in kinetic equilibrium with thermal bath, indicating a hotter phase-space distribution than the actual one obtained from the equation for the distribution. 
Given that the inverse-decay process is suppressed for dark matter with larger momentum in the relativistic regime, 
{this results in a smaller inverse-decay process in the result from the number-density approach and thus leads to a larger yield at early stage.}

One can also find that for a smaller mass ratio, the final number density obtained from the solution of the phase-space distribution is larger than that obtained directly from the solution of the number density equations,
this is because the inverse-decay process is stronger for dark matter with larger momentum in the non-relativistic regime, and the number density solution with assumption of a hotter distribution tends to deplete more dark matter at late time. 
On the contrary, for a larger mass ratio, the final comoving number density obtained by number-density equations is only smaller than that from distribution equations for relatively large decay rate, $\gamma_{\rm dec}(m_A)\sim 1$.
For smaller decay rates, since the discrepancy accumulated at the earlier stage is larger for larger mass ratios, there is not enough time to deplete the difference accumulated at early times.
{Thus, the results from the number-density approach remain larger than that from the phase-space-distribution approach, just like at the beginning of the freeze-in process.
The largest discrepancy we see in this case is $\sim$50\%, which occurs when $\gamma_{\rm dec}=10^{-1}$ in the right panel.}

\begin{figure}
\includegraphics[width=0.49\textwidth]{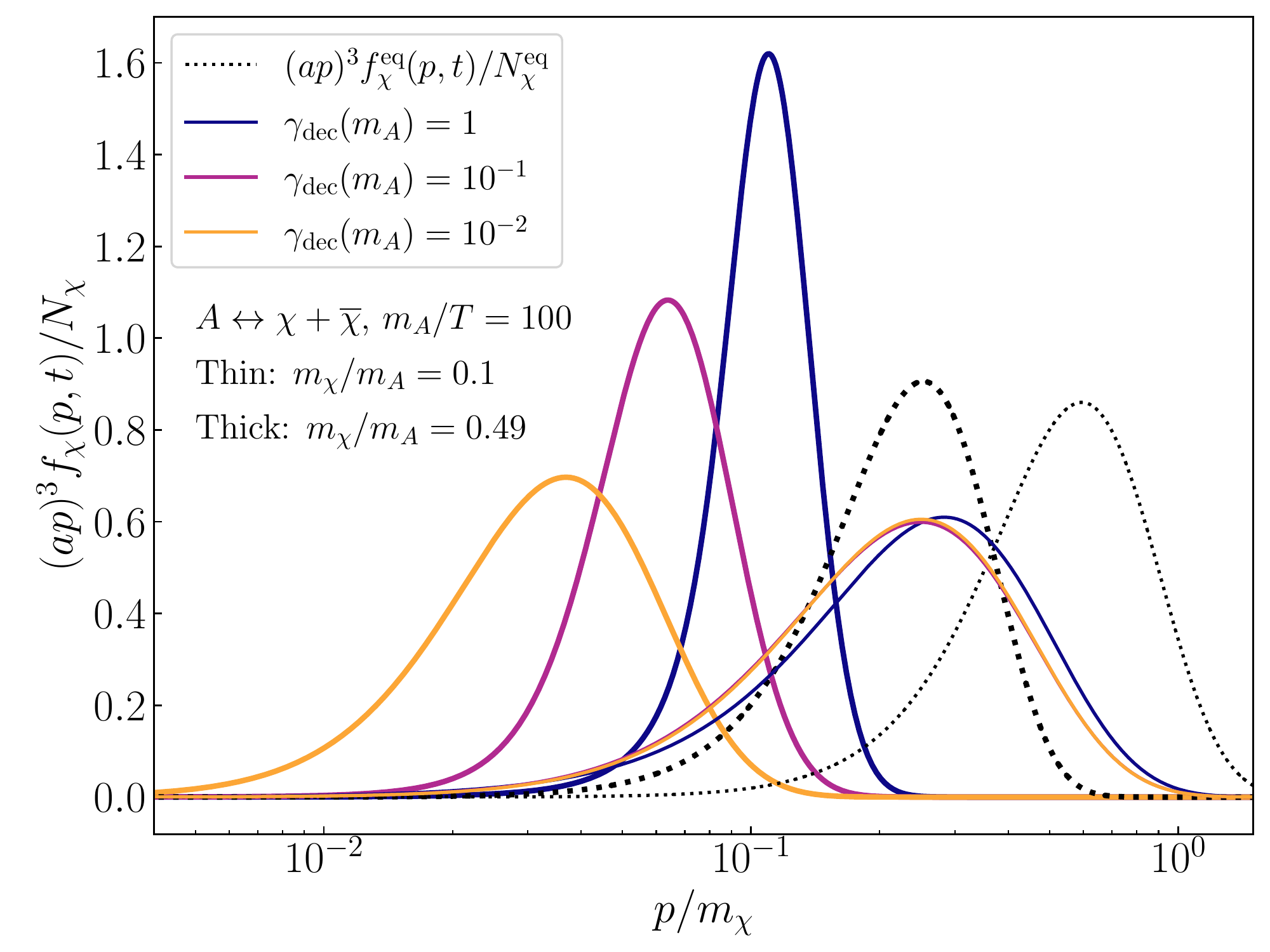}\includegraphics[width=0.49\textwidth]{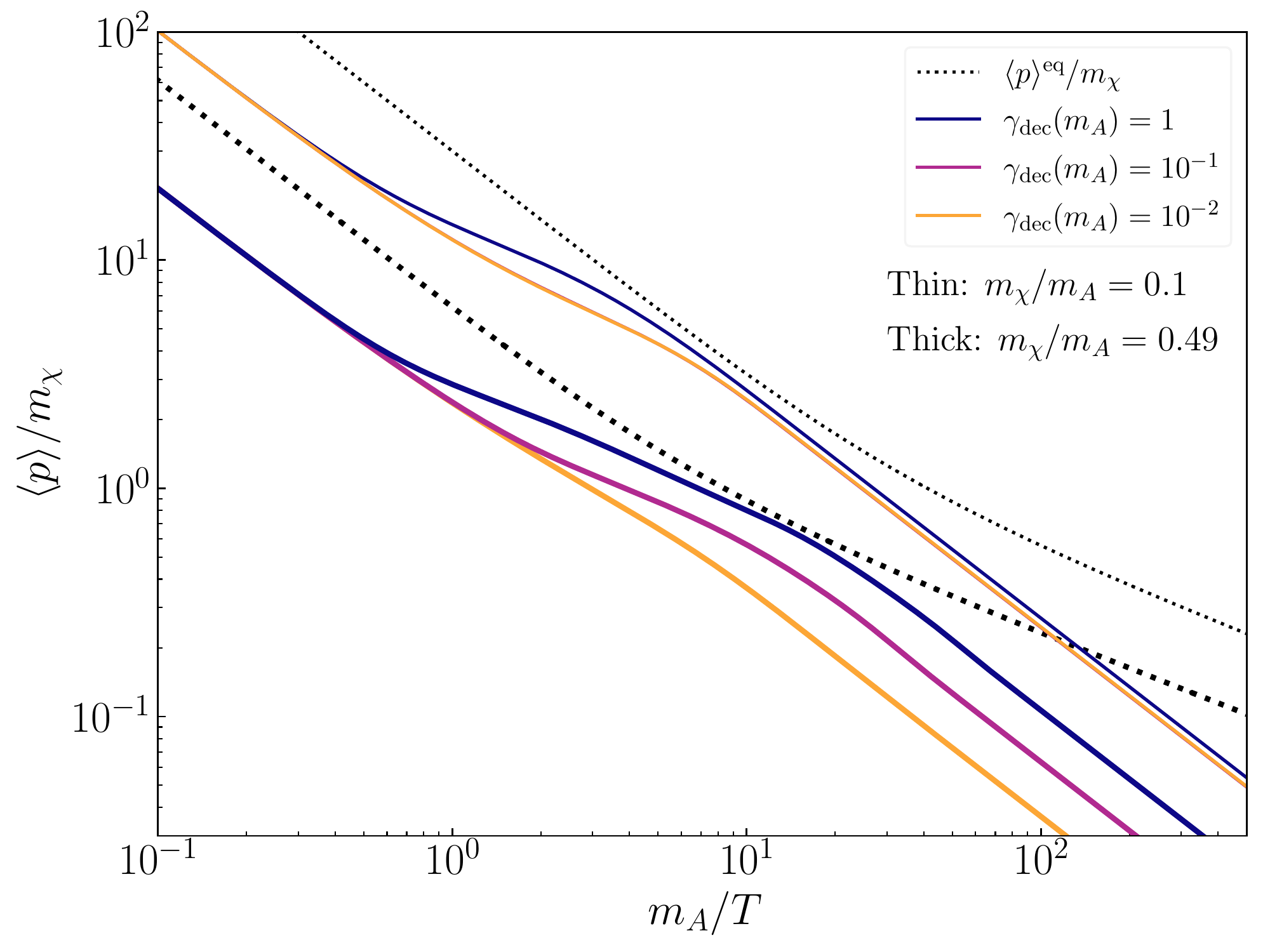}
\caption{\textbf{Left}: Late-time phase-space distribution of dark matter $\chi$ produced via \mbox{$A\leftrightarrow\chi+\bar{\chi}$} in which the mediator $A$ is assumed to be in thermal equilibrium with SM sector throughout the production. 
The blue, purple and orange colors correspond to different decay rates, and the black color correspond to the distribution in thermal equilibrium.
In addition, the thin and thick curves correspond to $m_\chi/m_A=0.1$ and $0.49$, respectively.
\textbf{Right}:  
The evolution of the averaged physical momentum of $\chi$ associated with each case on the left.}\label{fg:numden1}
\end{figure}

All in all, the observation from FIG.~\ref{fg:numden} suggests that not only it is necessary to include the annihilation of dark matter when the interaction rate is close to the transition point between freeze-in and freeze-out,
in this region, it is also necessary to use the phase-space-distribution approach rather than the conventional number-density approach for a more reliable and rigorous estimation of the relic abundance if the mass ratio is large.

We also present in FIG.~\ref{fg:numden1} the phase-space distribution of $\chi$ (left panel) at late time ($m_A/T=100$) and the evolution of the average dark-matter momentum (right panel). 
One can see that a smaller mass ratio $m_\chi/m_A$ generally yields a wider distribution with higher momentum as dark-matter particles inherit more energy from the mass of particle $A$.  
On the other hand, for the same mass ratio, a larger decay width yields a narrower distribution with higher momentum as dark-matter particles extract more energy from the thermal bath through particle $A$. 
The plots in the right column also show that the average momentum of dark matter cannot exceed that if it were in the thermal equilibrium with the thermal bath.

\FloatBarrier

%%%%%%%%%%%%%%%%%%%%%%%%%%%%%%%%
\subsection{{Out-of-equilibrium decay}}\label{subsec:out_of_eq_dec}
%%%%%%%%%%%%%%%%%%%%%%%%%%%%%%%%
%%%%%%%%%%%%%%%%%Haolin Merge%%%%%%%%%%%%%%
In this subsection, we focus on the scenario where $\chi$ is produced from the out-of-equilibrium decay of $A$. 
In contrast to the in-equilibrium decay discussed in the last subsection where $f_A$ follows the thermal distribution, we need in principle to consider the process $\psi + \psi \leftrightarrow A + A$ (as we have mentioned before, $\psi$ is a SM species), and
solve for the coupled Boltzmann equations for both $f_A$ and $f_\chi$.

\begin{figure}
\includegraphics[width=0.49\textwidth]{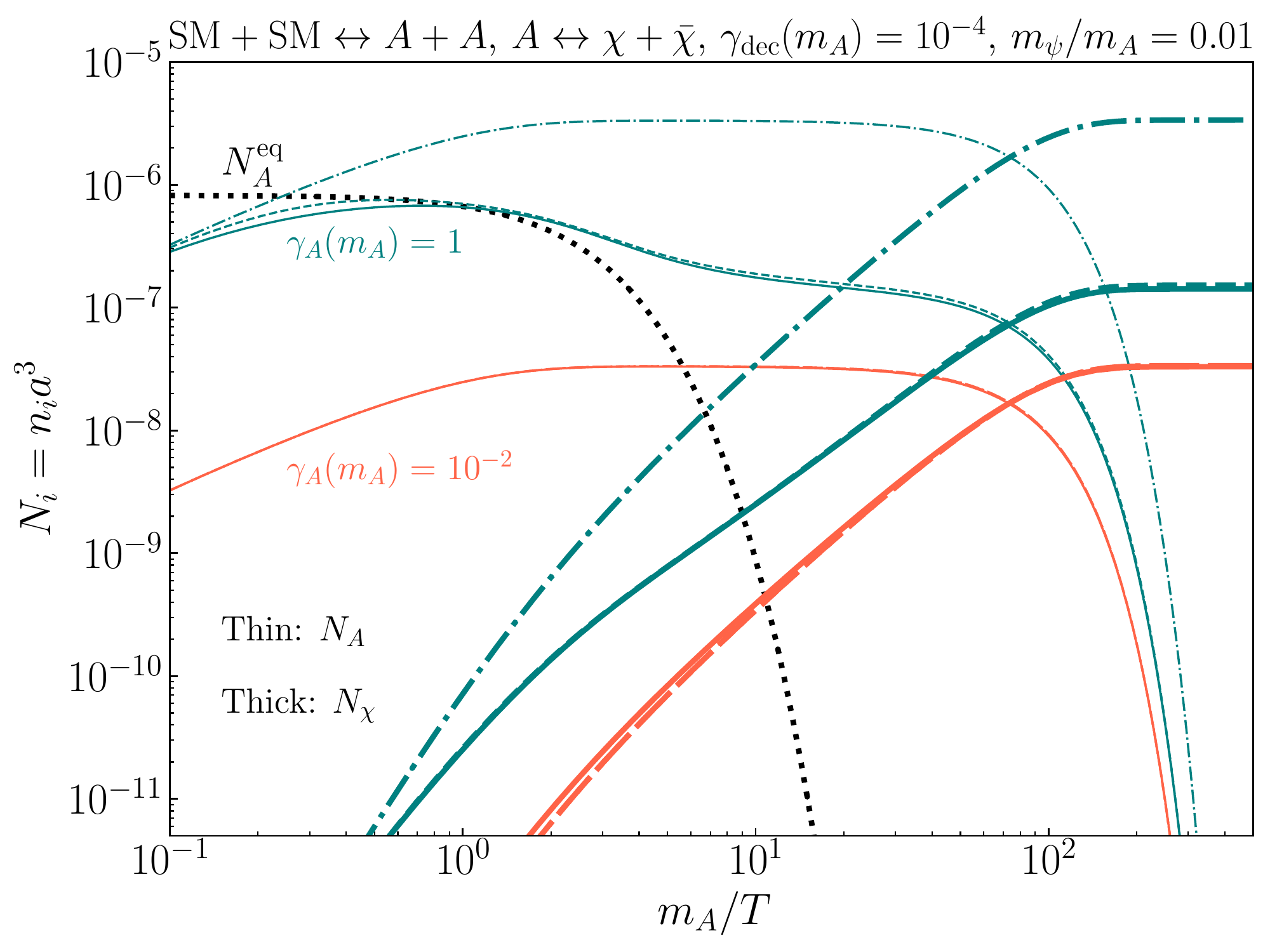}
\includegraphics[width=0.49\textwidth]{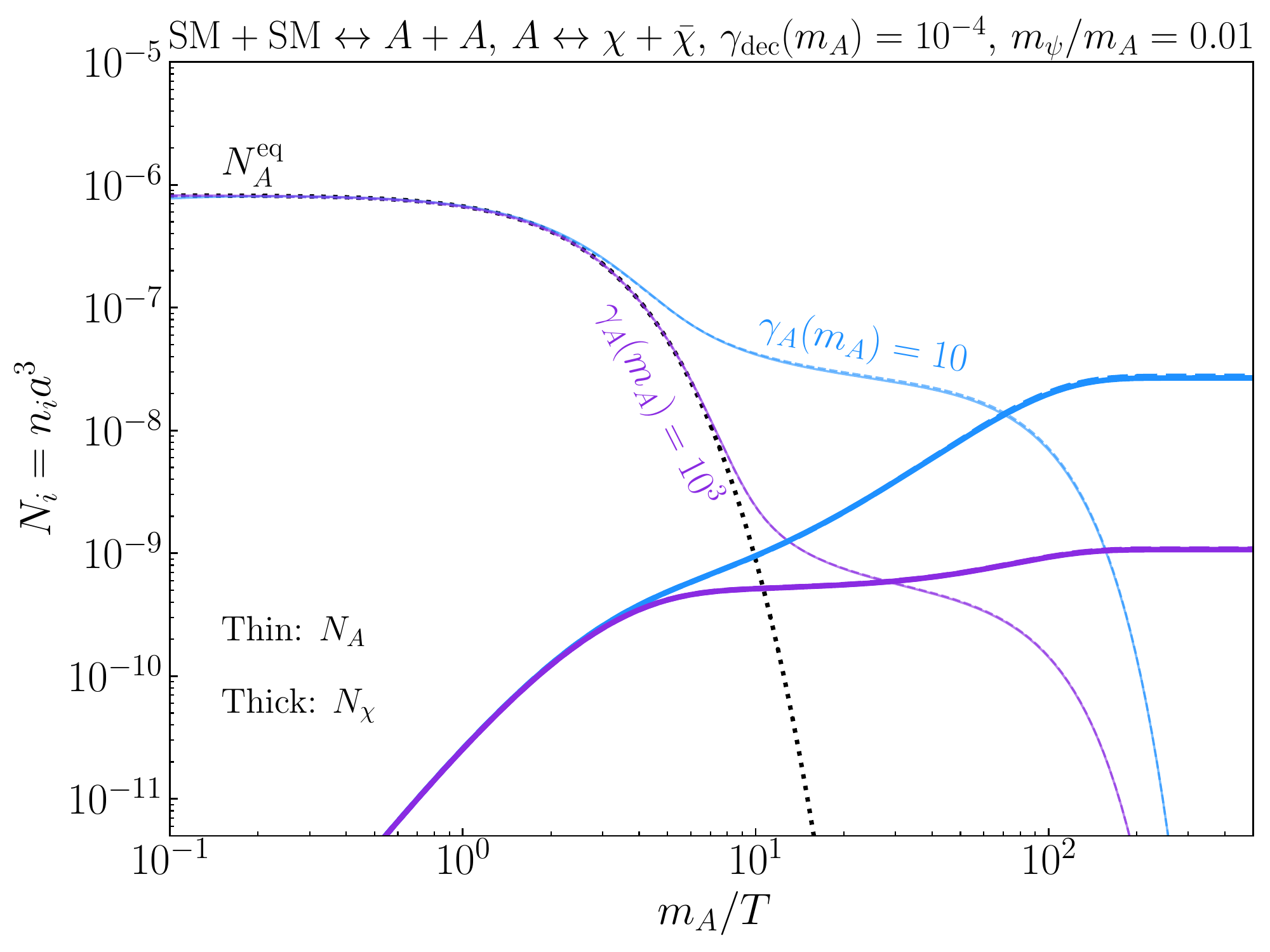}
\caption{Freeze-in production of $\chi$ from $A\leftrightarrow\chi+\bar{\chi}$, 
while the mediator $A$ is produced through freeze-in (left panel) or freeze-out (right panel) via $\psi + \psi \leftrightarrow A + A$.
The black dotted curves show the comoving number density of $A$ when it is in thermal equilibrium with the thermal bath, 
whereas the orange, teal, blue and purple curves represent the actual evolution of $N_A$ (thin curves) and $N_\chi$ (thick).
Like FIG.~\ref{fg:numden}, solid curves are results of Eqs.~\eqref{eq:Boltzmann_fA} and \eqref{eq:Boltzmann_fx}, dashed curves are results of Eqs.~\eqref{eq:Boltzmann_n_A_dec} and \eqref{eq:Boltzmann_n_chi_dec}, and dash-dotted curves are results of Eqs.~\eqref{eq:Boltzmann_n_A_dec} and \eqref{eq:Boltzmann_n_chi_dec} without including the inverse process $A+A\to \psi+\psi$. }\label{fg:Afreezeinvsfreezeoutn}
\end{figure}

Depending on the coupling strength between $A$ and $\psi$, $A$ can be generated either through freeze-out or through freeze-in. 
To compare the difference between these two cases, we fix the decay width of $A$ at $\gamma_{\rm dec}(m_A)=10^{-4}$  and the mass ratio $m_\psi/m_A=0.01$, and take four benchmarks for $ \gamma_A=n_A^{\rm eq} \langle\sigma_{\rm ann} v\rangle/H$. 
Two of them --- $ \gamma_A(m_A)=10^{-2}$ and $1$ are for freeze-in production of $A$, while the other two --- $\gamma_A(m_A)=10$ and $10^{3}$ for freeze-out production of $A$.
We then present our results in the left and the right panels of FIG.\,\ref{fg:Afreezeinvsfreezeoutn} for freeze-in and freeze-out production of $A$, respectively.

In FIG.\,\ref{fg:Afreezeinvsfreezeoutn}, the orange, teal, blue and purple colors represent different values of $ \gamma_A$, and the thin and thick curves represent the comoving number density of $A$ and $\chi$ respectively. 
Similar to the ones in the in-equilibrium-decay case in FIG.~\ref{fg:numden}, the dashed and the solid curves represent the solutions with the full collision term from the number-density and distribution equations respectively.
However, the dash-dotted curves now represent the solutions from using the number density directly without the inverse process $A+A\to \psi+\psi$.
The results from excluding the inverse-decay process are not presented, since we find that, for the out-of-equilibrium decay, the inclusion of the inverse-decay process does not significantly alter the relic abundance of dark matter. 
In both plots, we also show the number density of $A$ if it is in thermal equilibrium with the thermal bath with the dotted black curves. Comparing the teal and orange curves in the left panel, one finds that the back reaction process has significant effect only when $A$ is produced through the freeze-in mechanism with relatively large annihilation rate as shown by the teal curves. 
This is because, for the freeze-in production of $A$, a smaller $\gamma_A$ renders the number density of $A$ too small to gain significant back reaction rate in the entire period of dark matter production.
On the other hand, in both plots, one finds that the results from solving number-density equations and distribution equations are quite similar, and thus one can safely estimate the dark-matter relic abundance using the number-density equations directly.

\begin{figure}
\centering
\includegraphics[width=0.33\textwidth]{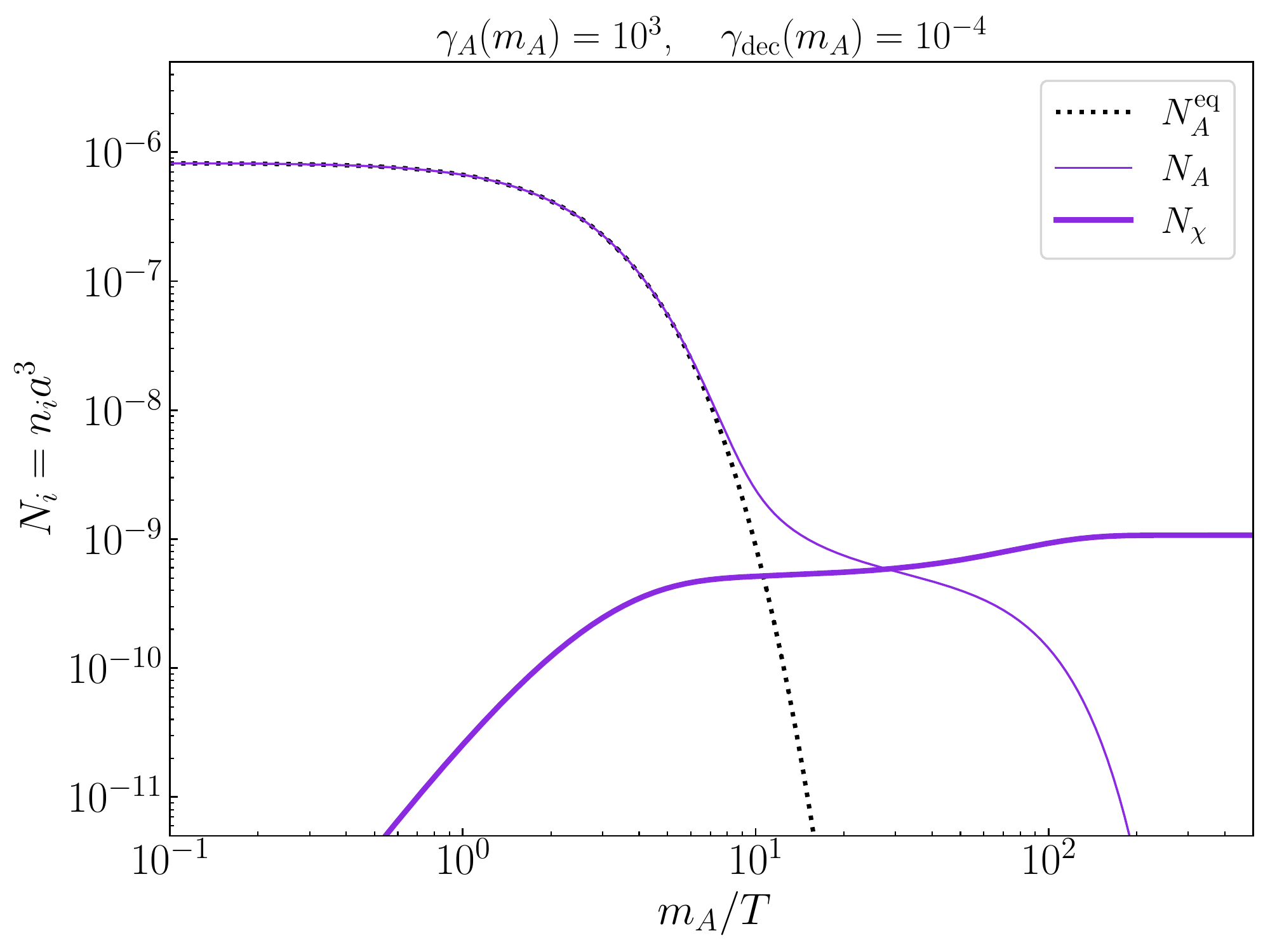}\includegraphics[width=0.33\textwidth]{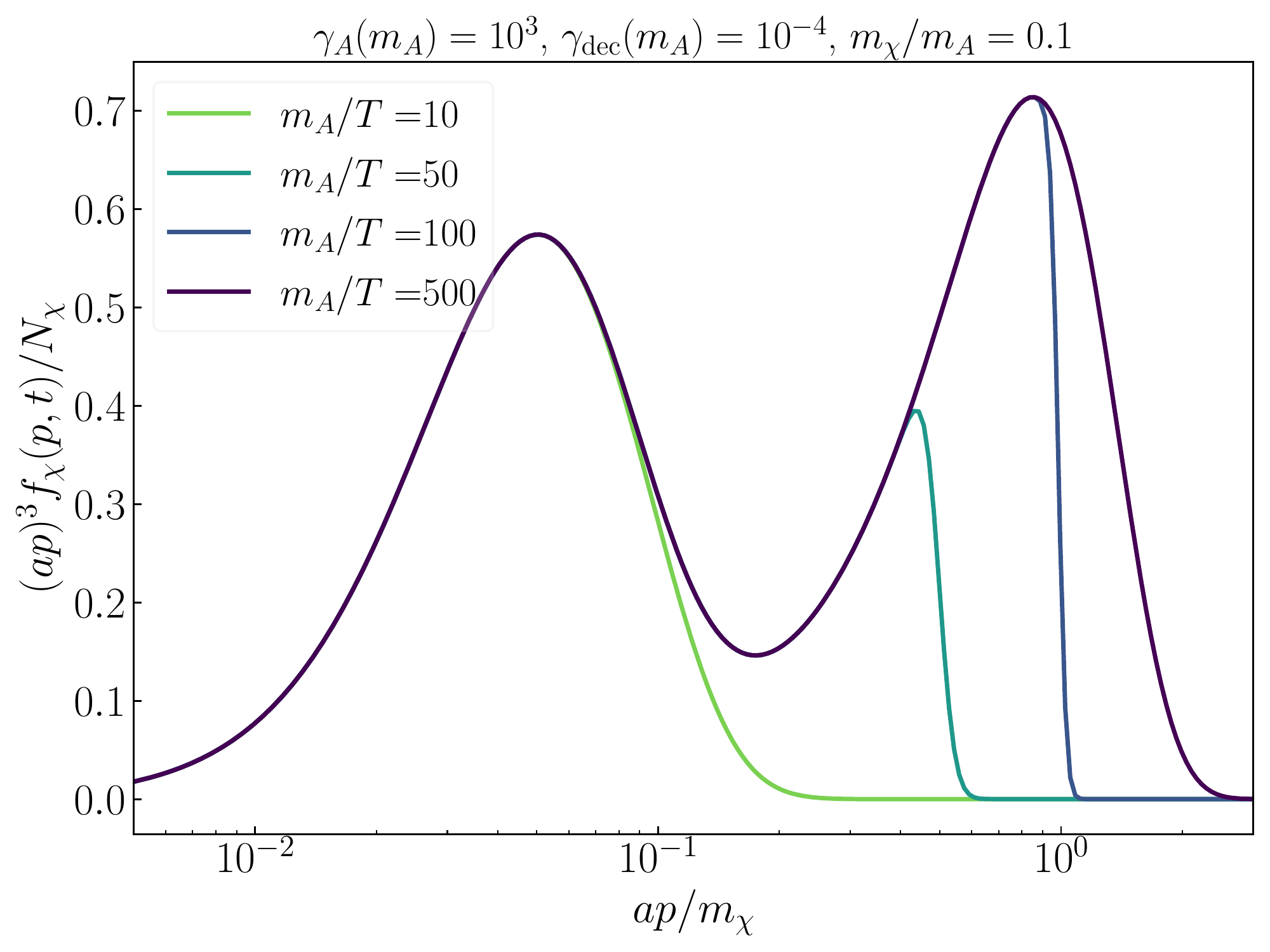}\includegraphics[width=0.33\textwidth]{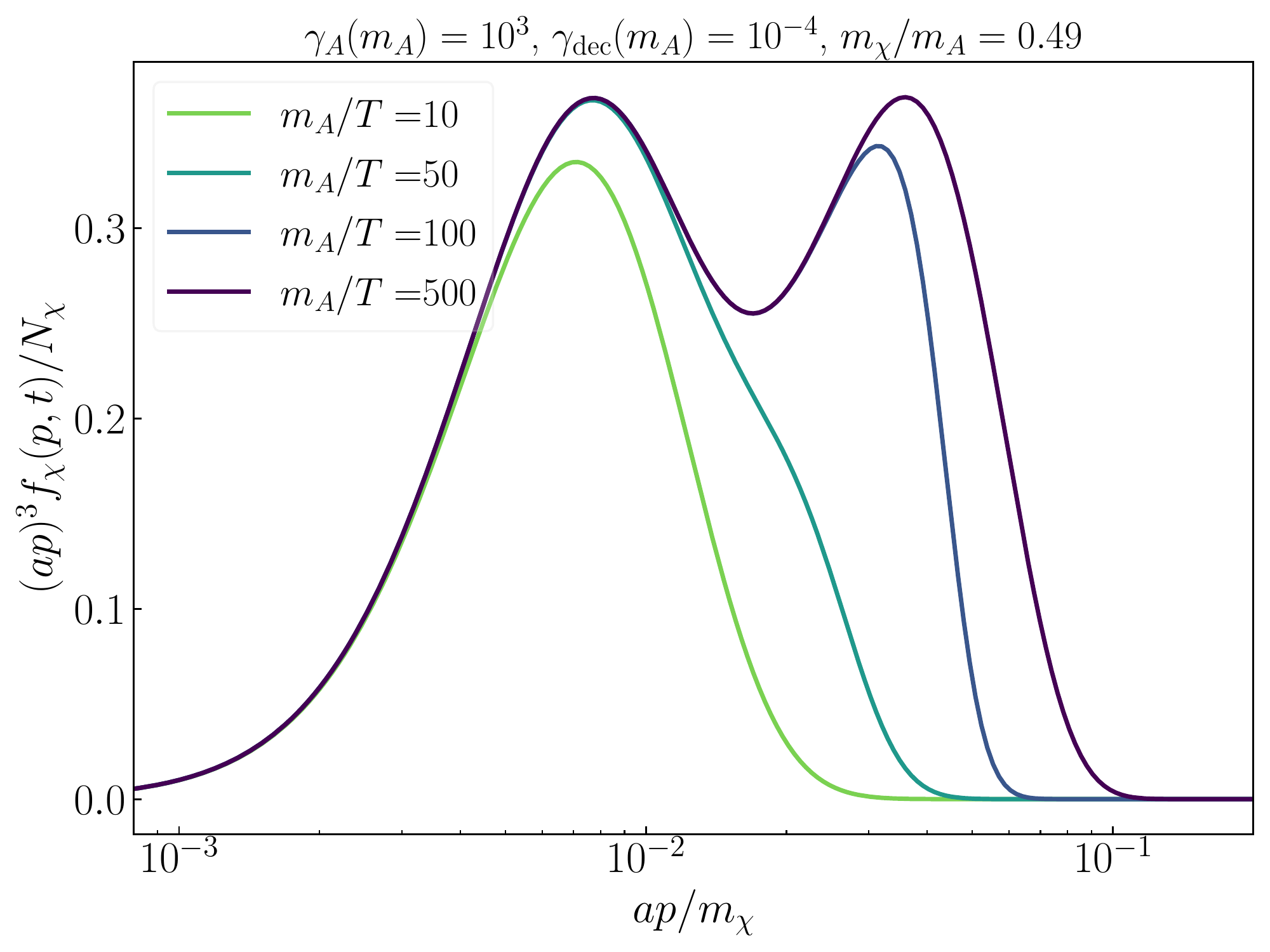}\\
\includegraphics[width=0.33\textwidth]{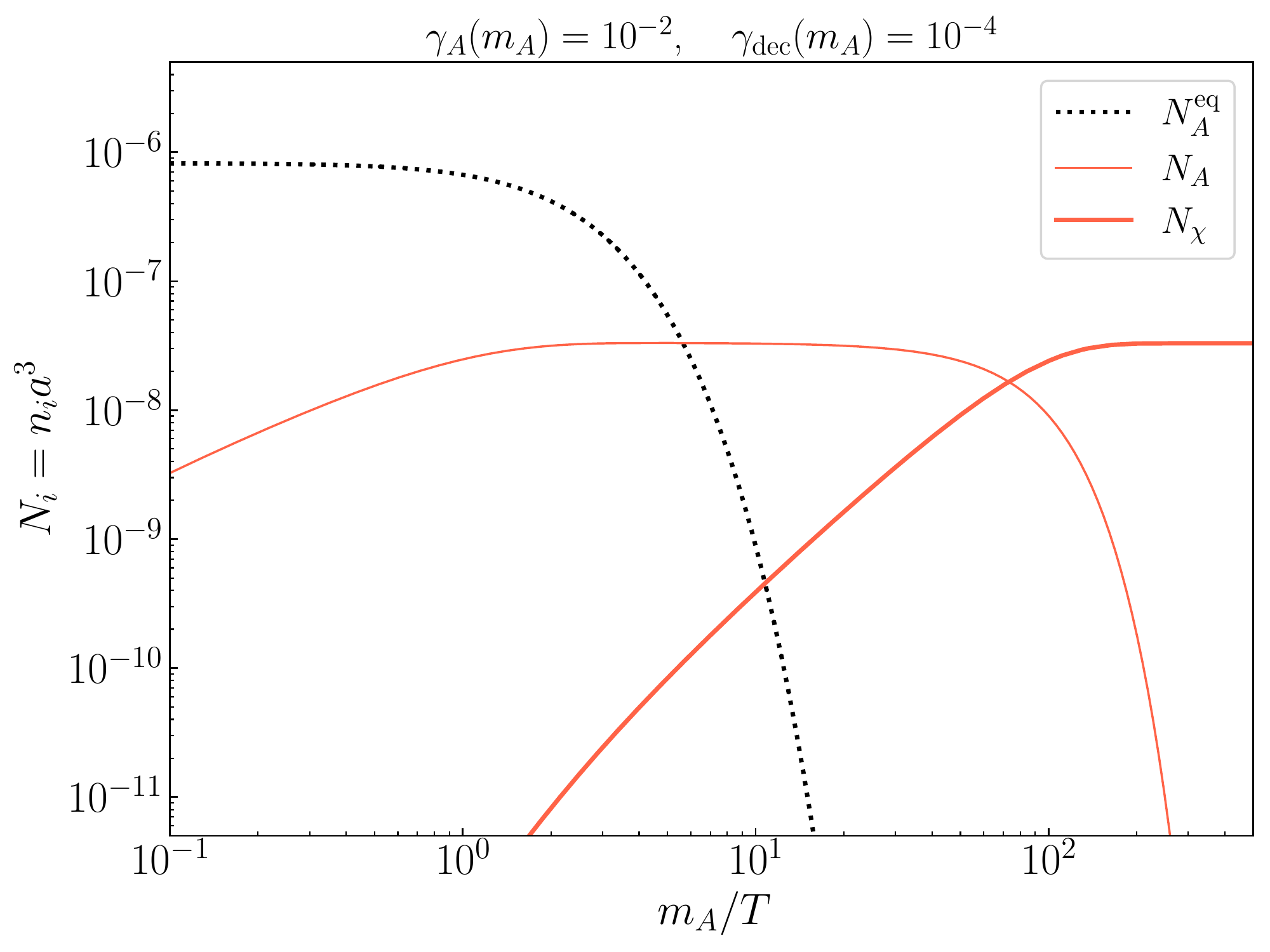}\includegraphics[width=0.33\textwidth]{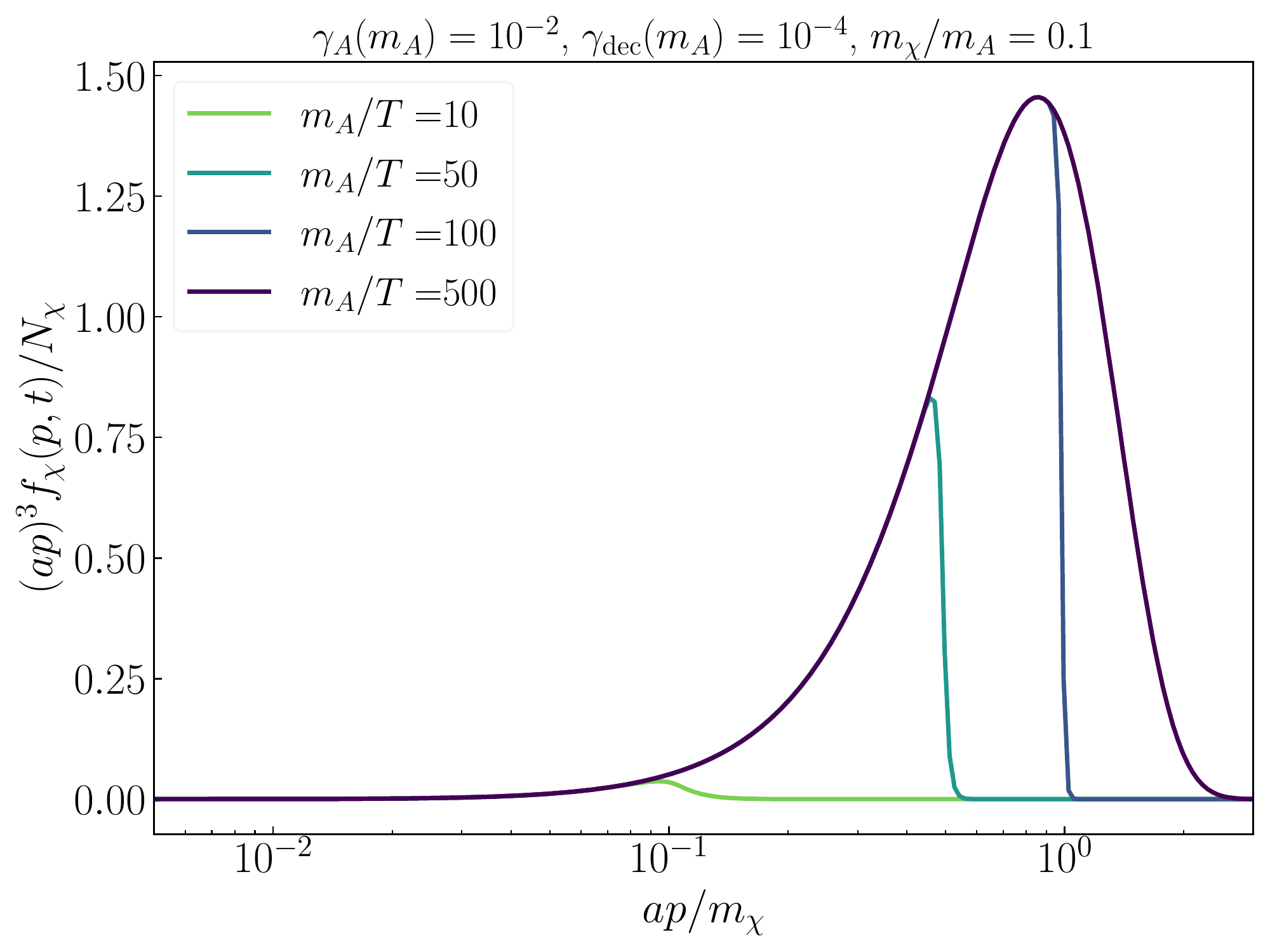}\includegraphics[width=0.33\textwidth]{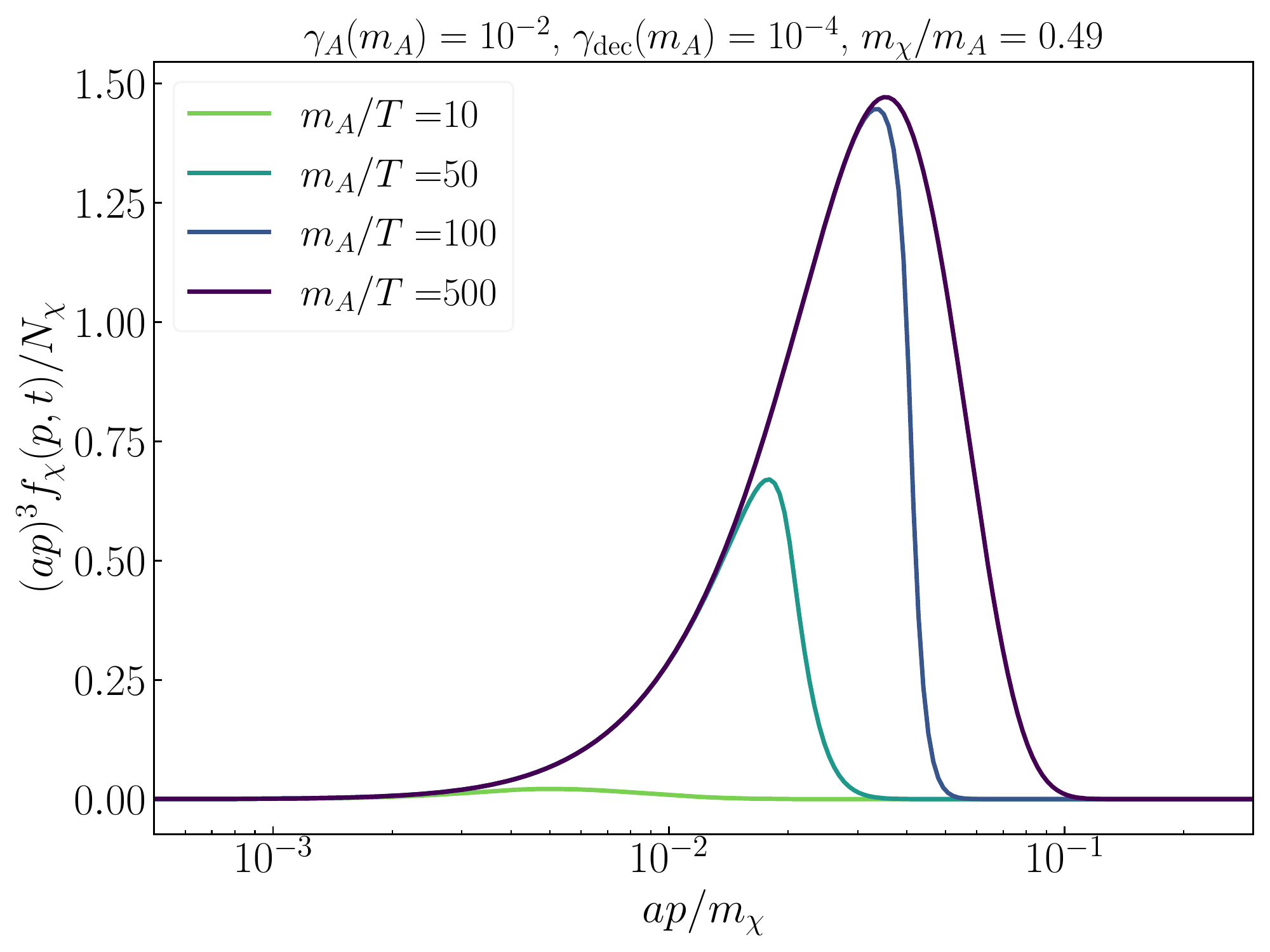}
\caption{Freeze-in through $A\leftrightarrow\chi+\bar{\chi}$ with different mass ratios $m_\chi/m_A$.
In the left panels,
we take two examples from FIG.~\ref{fg:Afreezeinvsfreezeoutn} for freeze-in (orange) and freeze-out (purple), respectively.
The middle and right panels show snapshots of the phase-space distribution of $\chi$ at different instants during the production, 
where two mass ratios are chosen with $m_\chi/m_A=0.1$ in the middle panels and $m_\chi/m_A=0.49$ in the right panels.
Notice that the mass ratio has no distinguishable effect on the comoving number densities, and thus we only present one set of curves for each choice of $\gamma_A(m_A)$.}\label{fg:Afreezeinvsfreezeout}
\end{figure}

\begin{figure}
\includegraphics[width=0.49\textwidth]{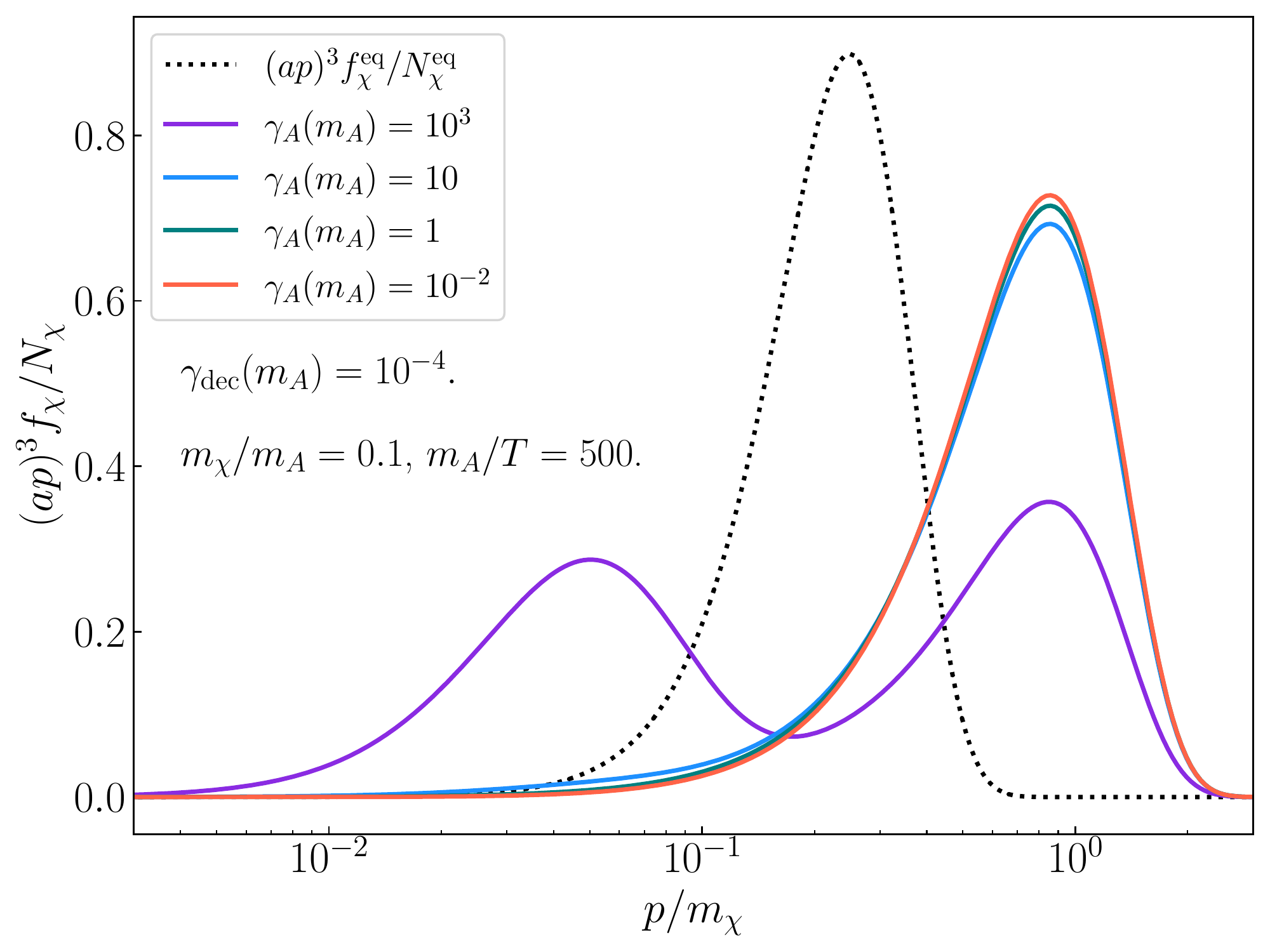} \includegraphics[width=0.49\textwidth]{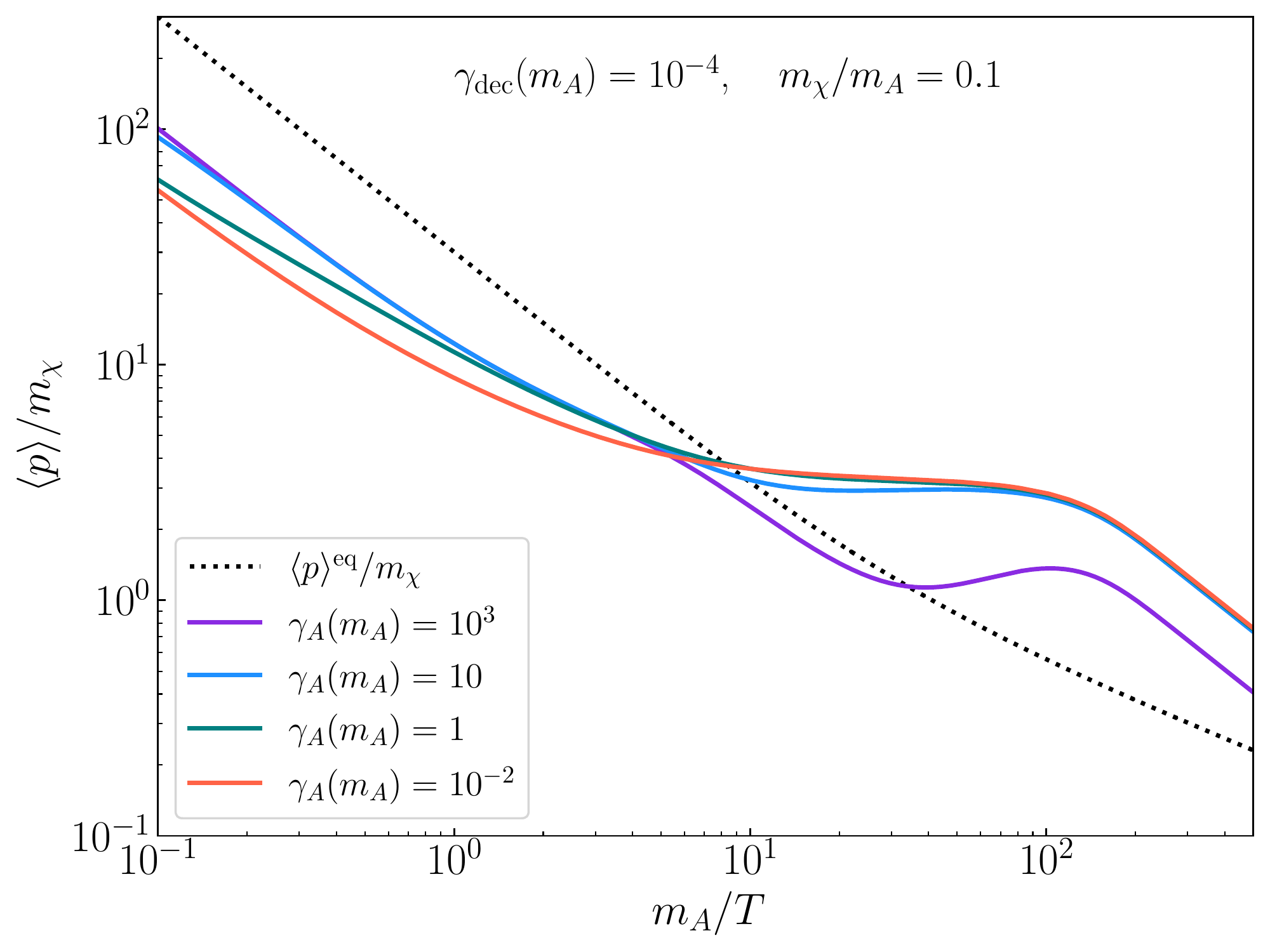}\\
\includegraphics[width=0.49\textwidth]{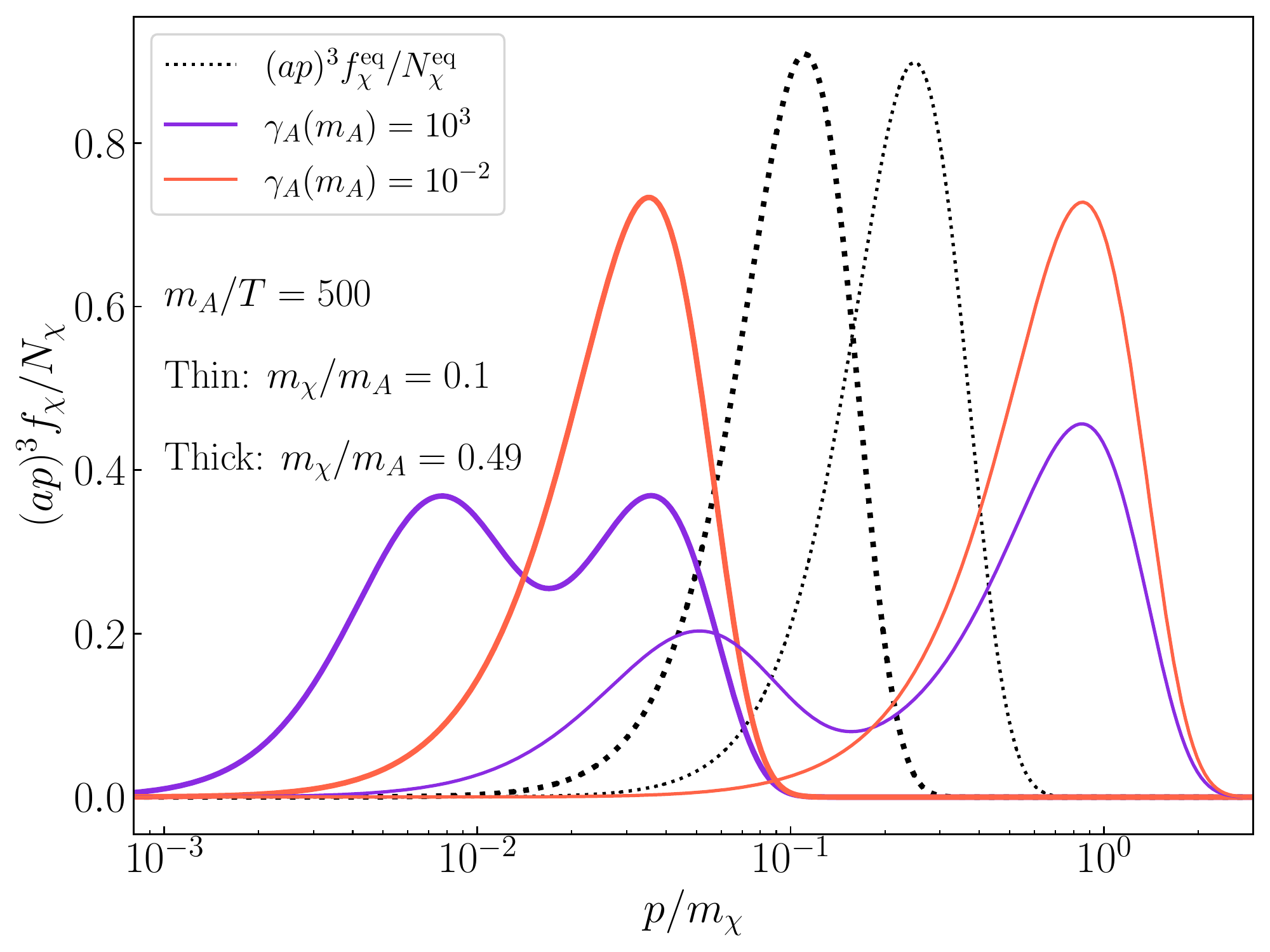} \includegraphics[width=0.49\textwidth]{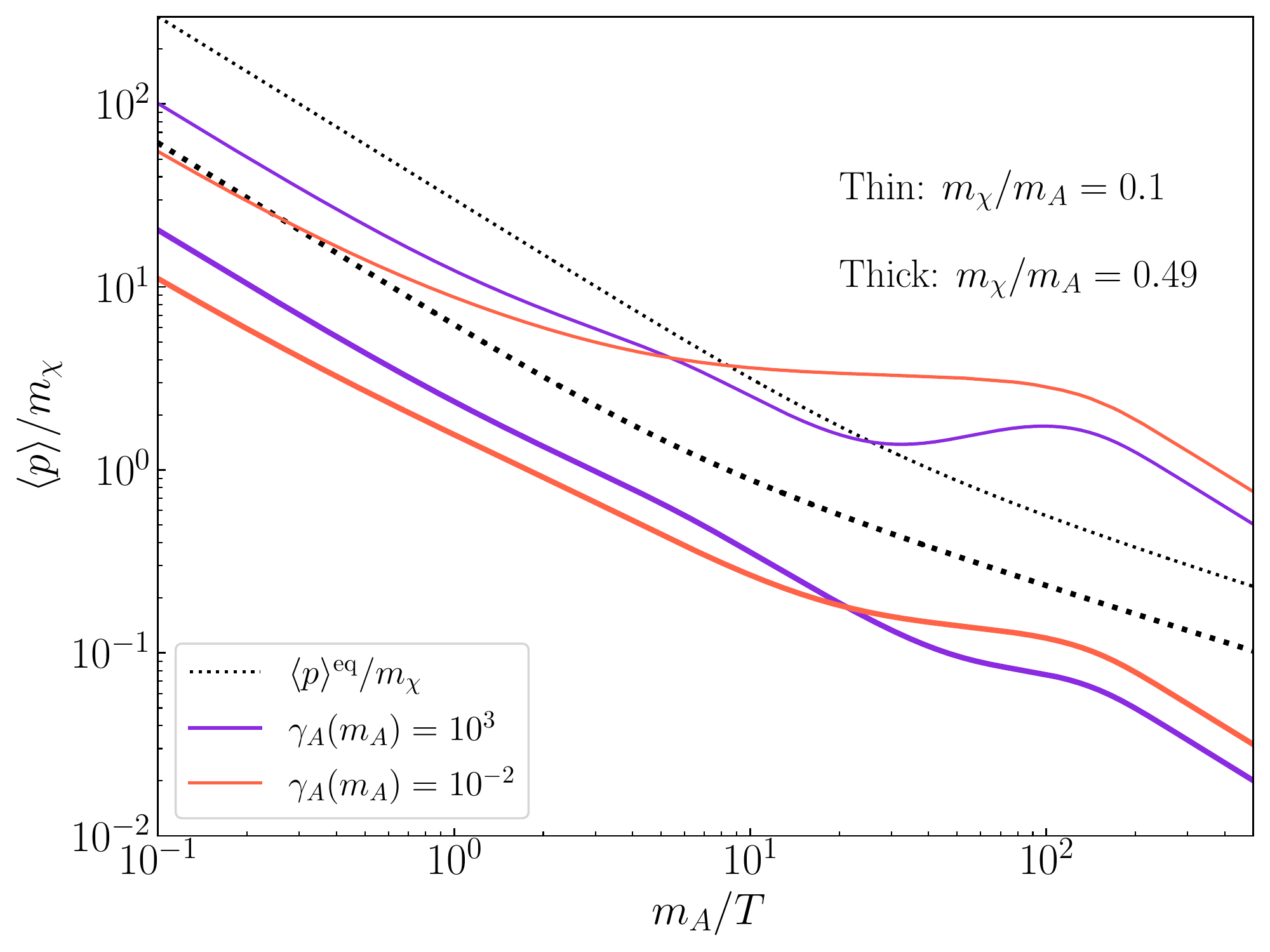}
\caption{Late-time phase-space distributions (\textbf{Left}) and the evolution of physical momentum (\textbf{Right}).  
In the upper panels, we fix $m_\chi/m_A=0.1$ and compare the results from different annihilation rates $\gamma_A(m_A)$.
In the lower panels, we choose the two benchmark annihilation rates from FIG.~\ref{fg:Afreezeinvsfreezeout}, and compare the results from two mass ratios $m_\chi/m_A=0.1$ (thin curves) and $m_\chi/m_A=0.49$ (thick curves), for which the decay products are relativistic and non-relativistic, respectively. 
}\label{fg:Afreezeinvsfreezeout2}
\end{figure}

In FIG.\,\ref{fg:Afreezeinvsfreezeout}, we further illustrate the evolution of phase-space distributions of dark matter for different $A$ production mechanisms and the mass ratios $m_\chi/m_A$. 
We find that for both the freeze-in and freeze-out production of $A$, the mass ratio barely changes the evolution of number densities for the fixed $\gamma_A$ and $\gamma_{\rm dec}$, and we present the numerical result of number densities in the first column. 
However, different mass ratios do have significant effects on the phase-space distributions of dark matter as one can see from the second and the third columns. 
The four plots in these two columns demonstrate the evolution of dark matter phase-space distributions with curves of different colors. 
The right two plots in the first row with $A$ produced through freeze-out present two-peak structures in the late-time distributions, and one can clearly see the growth of the first peak before $A$ freeze-out and the second peak due to the decay of $A$ after it freezes out. 
One can also find that a larger mass ratio tends to bring the two peak closer.
This is because a larger mass ratio leads to a smaller momentum of dark matter in the out-of-equilibrium decay of $A$ which brings the second peak to a lower value close to the first peak. 
On the other hand, in the second row one finds that no two-peak structure is observable.
This is because during the early time of the freeze-in process, the number density of the $A$ is too small to produce a significant amount of dark matter, and the majority of dark matter is produced after the completion of the freeze-in production of $A$, leading to a simple unimodal structure in the distribution.

In FIG.\,\ref{fg:Afreezeinvsfreezeout2}, we show how the dark-matter phase-space distribution is affected by varying the production rate of $A$.
In the first row, we plot the late time dark matter phase-space distributions with different values of annihilation rate $\gamma_A(m_A)$ in the left panel. 
As one can see, increasing $\gamma_A(m_A)$ leads to the appearance of the two-peak feature in the dark matter phase-space distribution as it increases the fraction of $\chi$ produced before the late out-of-equilibrium decay of $A$.
We also plot the evolution of dark-matter average momentum for these cases in the right panel.
We find that, a smaller production rate of $A$ leads to a lower average momentum at early times during the freeze-in stage, which is consistent with our observation in Sec.~\ref{sec:tworegimes}, but a higher average momentum at late times after $A$ decays.
Such observation still holds even when the mass ratio is relatively large as shown in the bottom right panel.
This is simply because, when $\gamma_A(m_A)$ is smaller, a larger fraction of dark-matter particles is produced through the late-time decay of $A$ after the freeze-in production of $A$ is complete, and this fraction of particles experiences less redshift.
The difference between small and large mass ratios can be observed from the second row --- the average momentum for the large mass ratio would not surpass $\expt{p}^{\rm eq}$ 
during the entire evolution, whereas this is not necessarily true for the small mass ratio as more energy in the mass of $A$ is released as kinetic energy.

\subsubsection{Effects of elastic scatterings on multi-modal distribution}

\begin{figure}
\includegraphics[width=0.49\textwidth]{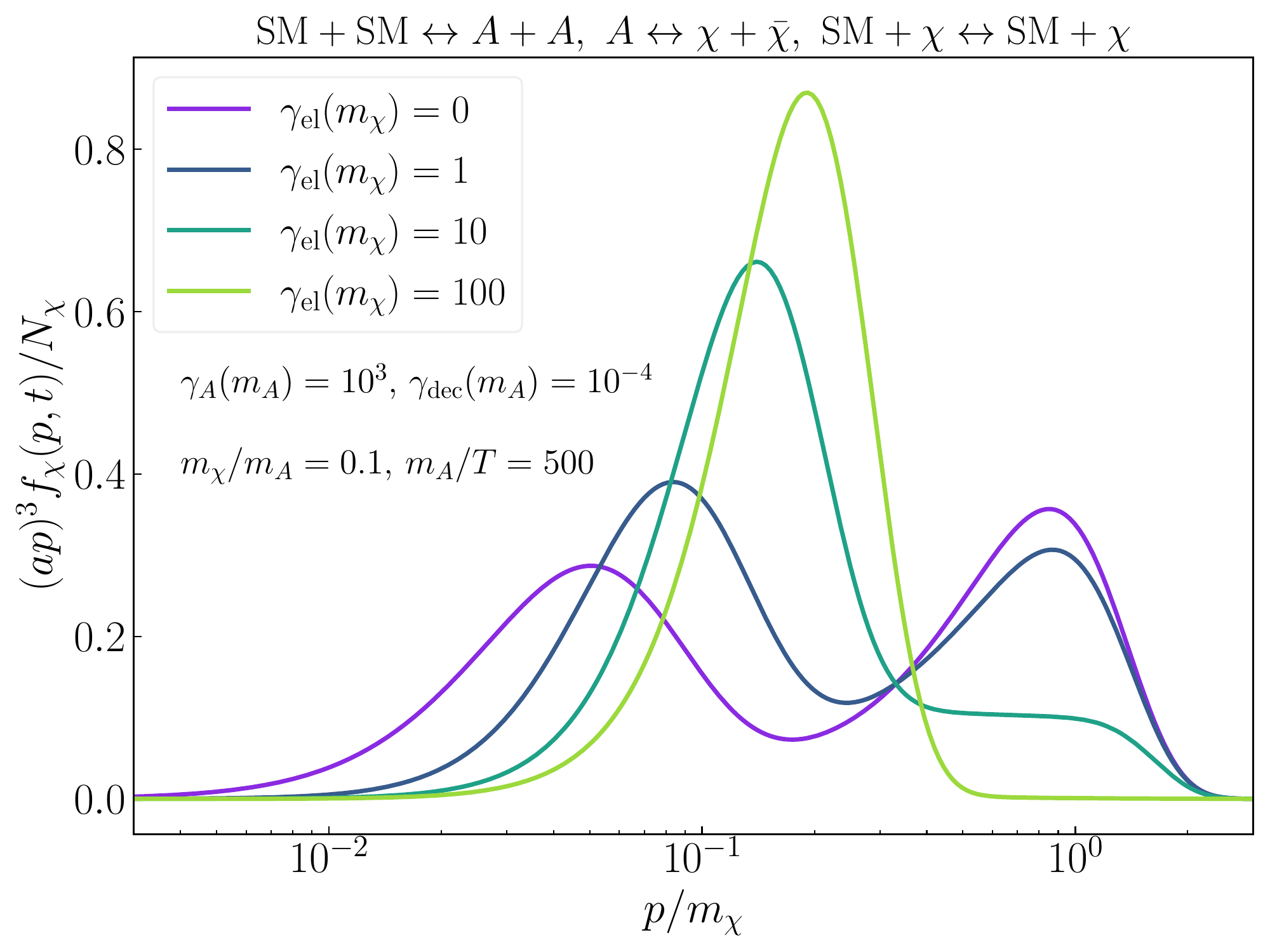} \includegraphics[width=0.49\textwidth]{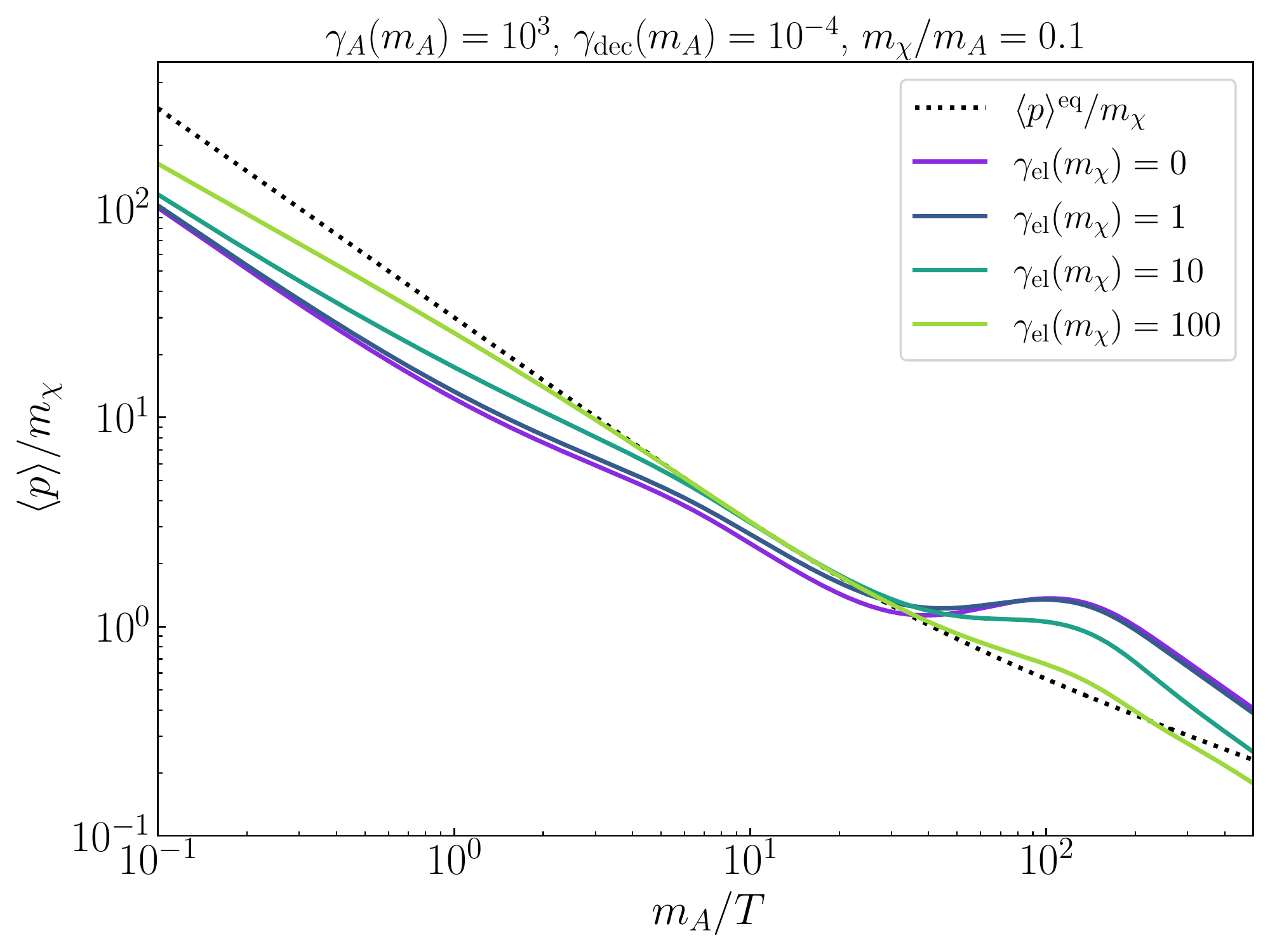}\\
\includegraphics[width=0.49\textwidth]{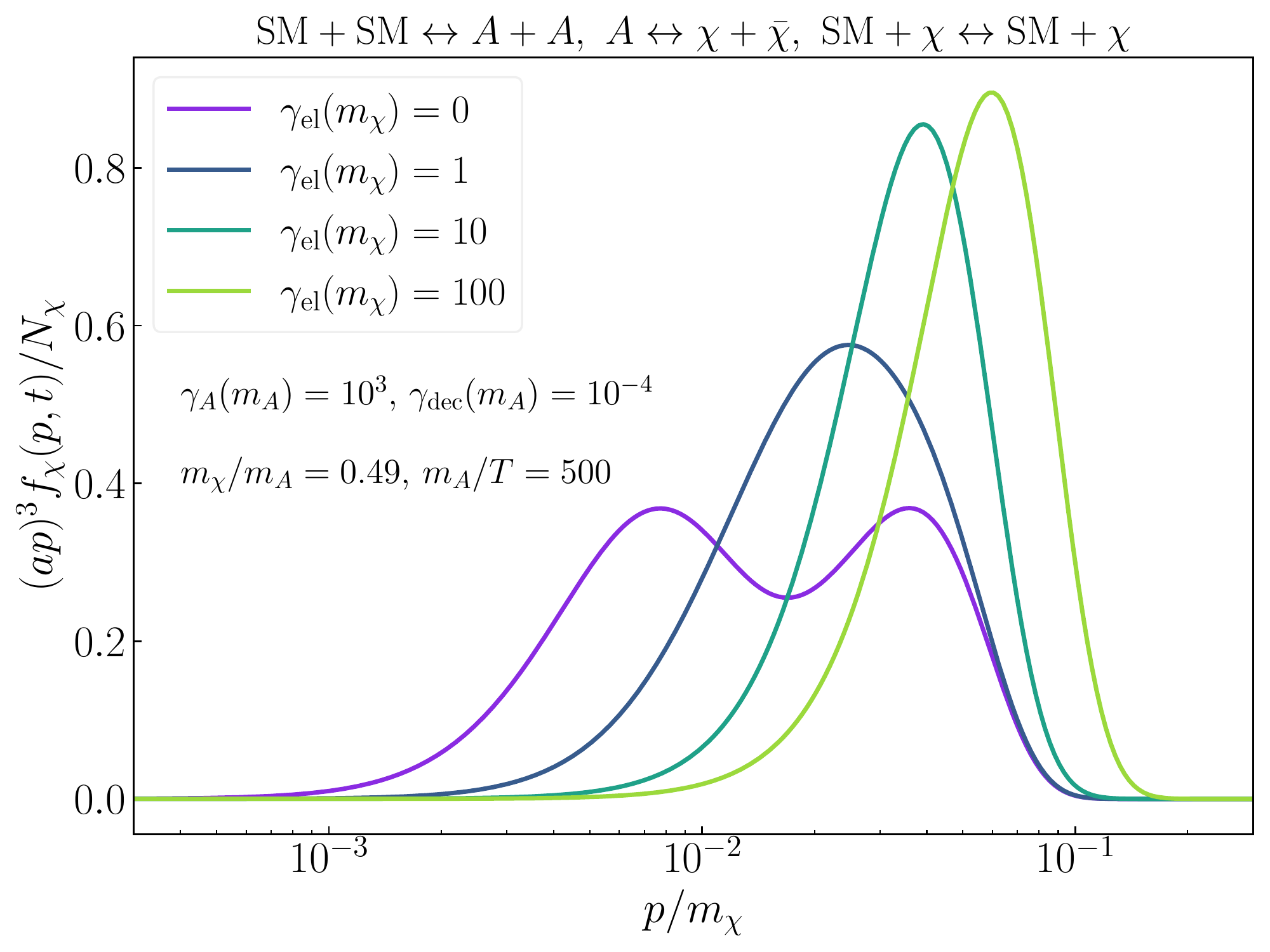} \includegraphics[width=0.49\textwidth]{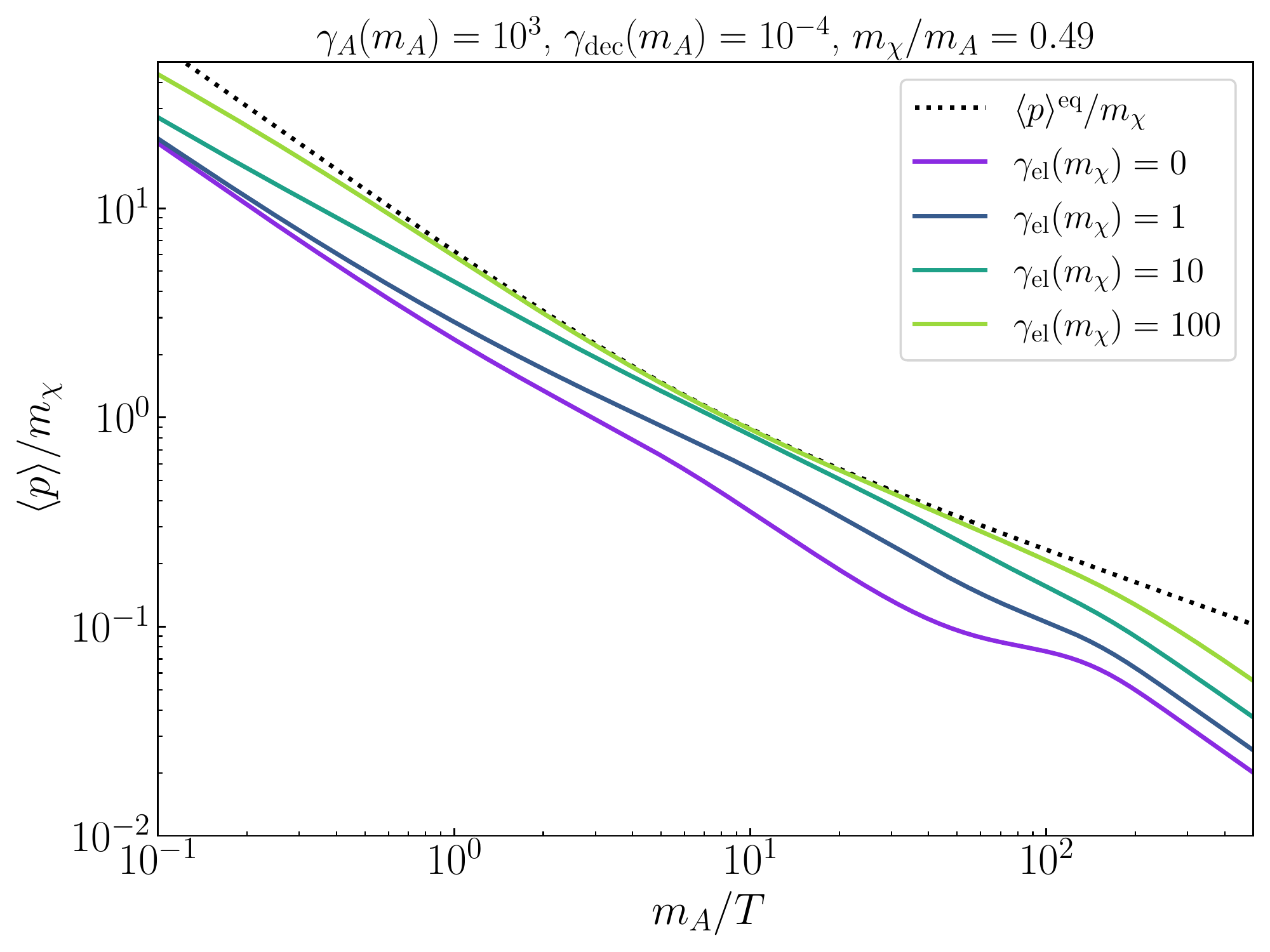}
\caption{Effects of elastic scattering process on the dark-matter phase-space distribution and the average physical momentum.
The upper and lower panels correspond to \mbox{$m_\chi/m_A=0.1$ and $0.49$}, respectively.}\label{fg:Afreezeinvsfreezeout3}
\end{figure}

In the above analysis, we have not taken into account the elastic scattering between dark matter and other particles, and thus the dark matter phase-space distribution is solely determined by the decay of $A$ and the production mechanism of $A$. 
In FIG.\,\ref{fg:Afreezeinvsfreezeout3}, we show the effects on the dark-matter phase-space distribution when the elastic-scattering with a SM particle $\chi+{\rm SM}\leftrightarrow\chi+{\rm SM}$ is taken into account. 
Like in Sec.~\ref{sec:tworegimes}, we will simply tune the elastic scattering amplitude without concerning possibility of introducing a new production channel.
In other words, we are merely treating the $\chi+{\rm SM}\leftrightarrow\chi+{\rm SM}$ process as a tool to calculate a well defined elastic-scattering cross section.
Since multiple processes are involved in these examples, for convenience, we characterize the elastic-scattering rate by $\gamma_{\rm el}\equiv n_\psi^{\rm eq}\langle \sigma_{\rm el}v\rangle /H$ evaluated at $T=m_\chi$.
The plots in the first and the second rows show examples with two benchmark mass ratios $m_\chi/m_A = 0.1$ and $0.49$, respectively. 
In each row, the left panel shows the late-time phase-space distribution of dark matter for different values of $\gamma_{\rm el}(m_\chi)$, 
and the right panel shows the evolution of the average momentum. 

As expected, the elastic scattering tends to distort the distribution towards the thermal distribution such that the non-thermal features might be erased.
For the examples in the upper panels, since the mass ratio is small, the decay products of $A$ has a typical momentum larger than that if they are in equilibrium with the thermal bath.
Therefore, while the early contribution from the freeze-in production of $\chi$ is overall colder than the thermal bath, the late contribution from the late decay of $A$ is in general warmer than it.
The elastic scattering then tends to bring them together as one can see in the top left panel.
As a result, in the top right panel, we find that the elastic scattering can even dissipate the kinetic energy of dark-matter particles since the final momentum becomes smaller when $\gamma_{\rm el}(m_\chi)$ gets larger.
On the other hand, the examples in the lower panels have a larger mass ratio which makes the decay products always colder than the thermal bath.
For these cases, elastic scattering can only push them to larger momentum, and, as a result, it facilitates the extraction of kinetic energy just like what we have seen in Sec.~\ref{sec:tworegimes}.

\FloatBarrier

\section{Conclusion}\label{sec:conclusion}

The freeze-in and freeze-out production mechanisms have been studied for years with the emphasis on producing the correct dark-matter relic abundance.
Most of these studies are based on solving Boltzmann equations of the number densities of the particles involved only, and, approximations such as neglecting the back reaction from the particles produced via freeze-in are often used.
Although such approximations are reasonable in many cases, it is not clear at which point the assumptions behind these approximations might not hold, and how much the error could be.

In this work, we revisit the freeze-in and freeze-out production of dark mater at a more fundamental level of the phase-space distribution $f(p,t)$.
Using the $2\to2$ and the $1\to2$ processes for illustration, we quantitatively investigate the evolution of the dark-matter number density
and compare the results of the relic abundance obtained by solving the number-density Boltzmann equations with or without the back reaction from the dark-matter particles or the decaying mediator.
We then explicitly show when and to what extent the traditional number-density approaches could hold or fail.

Beyond the relic abundance, we also study the evolution of the dark-matter phase-space distribution, and discuss how the average momentum and the shape of the distribution change as we adjust the interaction strength or modify the thermal history of the relevant particles.
In addition, we also investigate the effects of the elastic scattering, focusing on how it modifies the average momentum, the shape of the distribution, and even the possibility of erasing the multi-modal features.

Our findings are listed below.
For the vanilla $2\to2$ process, we find that:
\begin{itemize}
\item  {For freeze-in, the relic abundance of dark matter obtained by the solutions of the phase-space distribution and that obtained by the number-density approach show noticeable difference in the transition regime between freeze-in and freeze-out.
For the case of constant annihilation amplitude studied in this paper, the difference is around 10\% when the annihilation strength is close to the transition point between freeze-in and freeze-out. 
The difference could be even as large as a factor of $\mathcal{O}(10)$ if one neglects the re-annihilation of dark-matter particles.}

\item In both the $2\to2$ freeze-in and freeze-out scenarios, the resulting phase-space distributions of dark matter at late times are all colder than the thermal ones, and in general the freeze-in mechanism tends to generate colder phase-space distribution than the freeze-out mechanism. 

\item The inclusion of the elastic scattering between dark matter and the thermal bath particles tends to heat up dark matter and narrow the width of the phase-space distribution.
\end{itemize}

On the other hand, for dark matter produced through $1\to2$ decay:

\begin{itemize}
\item {The production mechanism of the decaying particle $A$ and whether the decay occurs in or out of equilibrium are especially important for the final distribution of dark matter. 
In the case where $A$ decays while in equilibrium with the thermal bath, the inclusion of inverse decay is important when the decay width of the mediator is relatively large.
Otherwise, the relic-abundance estimate can be off by orders of magnitude depending on the mass ratio between the products and the decaying particle.
Even when the inverse decay is included, if the mass ratio is large, the number-density approach can still lead to a relatively large error ($\sim$50\%) for moderately large decay width.
On the other hand, the inverse decay is negligible when dark matter is produced through the out-of-equilibrium decay of $A$.
However, in this case, the annihilation of $A$ may have a non-negligible impact on the relic-abundance estimation in the transition regime between freeze-in and freeze-out of the mediator --- neglecting such process can also lead to an error as large as one order of magnitude.}

\item The phase-space distribution of dark matter produced through in-equilibrium decay is unimodal, and the shape depends on both the decay rate and the mass ratio between dark matter and the decaying particle.
For dark matter produced through out-of-equilibrium decay, the production mechanism for the decaying particle can also alter the shape of the phase-space distribution. 
Particularly, when the decaying particle is produced through freeze-out, a multi-modal distribution can occur.

\item The inclusion of the dark matter elastic scattering in the decay production mechanism tends to distort the phase-space distribution towards the thermal one, and thus can blur the multi-modal pattern in the phase-space distribution, or even make such pattern disappear.
\end{itemize}

With our numerical framework and solutions of dark matter phase-space distributions, not only can we analyze the difference in the yield of dark matter between the number-density and the distribution methods, 
the information of the phase-space distribution can also be used to constrain dark-matter models through its implications in structure formation.
The data from Lyman-alpha forest~\cite{Murgia:2018now}, perturbations in arcs of strongly lensed galaxies~\cite{Vegetti:2018dly}, number counts of Milky-Way satellites~\cite{Newton:2020cog} and gaps in tidal streams from satellite galaxies~\cite{Banik:2019smi} can all serve as promising observables to constrain the dark matter phase-space distribution. 
Studies on nonthermal dark-matter phase-space distributions have also provoked interest in relevant N-body simulations \cite{Stucker:2021vyx}.
Our work provides the framework to obtain the crucial ingredients in the program connecting concrete dark-matter models to predictions of the aforementioned studies, and thus facilitates a more comprehensive analysis for different dark-matter models in the future, which also provides extra motivation for efforts in these directions.

\acknowledgments
The work is supported by the National Natural Science Foundation of China (NSFC) under Grants No.\ 12022514 and No.\ 11875003 and CAS Project for Young Scientists in Basic Research YSBR-006. J.\ H.\ Y.\  is also supported by the National Natural Science Foundation of China (NSFC) under Grant No.\ 12047503 and the National Key Research and Development Program of China under Grant No. 2020YFC2201501. 
F.\ H.\ is supported by the International Postdoctoral Exchange Fellowship Program, the National Natural Science Foundation of China under grants No.\ 12025507, 11690022, 11947302, and is also supported by the Strategic Priority Research Program and Key Research Program of Frontier Science of the Chinese Academy of Sciences under Grants No.\ XDB21010200, XDB23010000, ZDBS-LY-7003.

%%%%%%%%%%%%%%%%%%%%%%%%%%%
\appendix
%%%%%%%%%%%%%%%%%%%%%%%%%%%
\section{Derivation of collision terms in Boltzmann equation}\label{sec:app}

The {techniques for evaluating the} collision-term integral has been extensively studied, {especially for neutrino physics since the 1990s\,
\cite{Hannestad:1995rs,Dolgov:1997mb,Mangano:2001iu,Oldengott:2014qra,Du:2021idh}}
as it is essential for a precise prediction of
the effective number of neutrino species in the early universe.
Based on these previous efforts, we briefly show the our strategy for numerically calculating the collision terms needed in this work in this appendix.
{For simplicity, we shall ignore the dependence on the internal degrees of freedom of each particle species.}

\subsection{Collision term of \texorpdfstring{$1\to2$}{1->2} decay}

We devote this subsection to the $1\to 2+\chi$ processes. The collision term of $f_{\chi}$ for the decay process can be written as
\beqn
C(t,p_\chi)& = &\frac{1}{2E_\chi}\int \frac{d^3p_1}{(2\pi)^3 2E_1} \int\frac{d^3p_2}{(2\pi)^3 2E_2} (2\pi)^4 \delta^4(p_1-p_2-p_\chi)\nn\\
&~&\times~\overline{\abs{\mathcal{M}_{1\to 2+\chi}}^2} \left( f_1(1\mp f_2)(1\mp f_\chi) - f_2 f_\chi (1\mp f_1) \right) .\nn\\
& = &\frac{1}{32\pi^2 E_\chi}\int \frac{d^3p_1}{E_1 E_2^s} \delta(E_1-E_2^s-E_\chi)\overline{\abs{\mathcal{M}_{1\to 2+\chi}}^2}\nn\\
&~&\times\left( f_1(1\mp f_2)(1\mp f_\chi) - f_2 f_\chi (1\mp f_1) \right),
\eeqn
where we use the fact that $\overline{|M_{1\to 2+\chi}|^2}=\overline{|M_{2+\chi\to 1}|^2}$ and integrate over $p_2$ in the second line. $E_2^s =\sqrt{m_2^2+(\vec{p}_1-\vec{p}_\chi)^2}$ is fixed by the delta function $\delta^3(\vec{p}_1-\vec{p}_2-\vec{p}_\chi)$.

To integrate the remaining $\delta(E_1-E_2^s-E_\chi)$, we notice that:
\beqn
\delta(E_1-E_2^s-E_\chi)&=&\delta\left(\sqrt{p_1^2+m_1^2}-E_\chi - \sqrt{m_2^2+p_1^2+p_\chi^2-p_1 p_\chi \cos\theta}\right)\nn\\
&=&\frac{E_2^s}{p_1p_\chi}\delta(\cos\theta-\cos\theta^s) ,
\eeqn
where $\theta$ is the angle between $\vec{p}_1$ and $\vec{p}_\chi$, and $\theta^s$ is the root of $\theta$ inside the $\delta$-function. In the meantime, the existence of the root implies the upper and lower bounds on the magnitude of $\vec{p}_1$, leading to:
\beq
C(t,p_\chi) = \frac{1}{16\pi E_\chi p_\chi}\int_{E_1^{\rm min}}^{E_1^{\rm max}} dE_1  \overline{\abs{\mathcal{M}_{1\to 2+\chi}}^2} \left( f_1(1\mp f_2)(1\mp f_\chi) - f_2 f_\chi (1\mp f_1) \right), \label{eq:collsion_decay_general}
\eeq
where $E_1^{\rm max/min}=\sqrt{(p_1^{\rm max/min})^2+m_1^2}$ and
\beqn
p_1^{\rm max}&=&\frac{(m_1^2+m_\chi^2-m_2^2)p_\chi+2m_1E_\chi p^*}{2m_\chi^2},\\
p_1^{\rm min}&=&
\begin{cases}
\displaystyle\frac{(m_1^2+m_\chi^2-m_2^2)p_\chi-2m_1E_\chi p^*}{2m_\chi^2} &\text{if~~} p_\chi>p^*,\\
\displaystyle\frac{-(m_1^2+m_\chi^2-m_2^2)p_\chi+2m_1E_\chi p^*}{2m_\chi^2} &\text{if~~} p_\chi\leq p^*,
\end{cases}
\eeqn
with $p^*$ being the momentum of decay products in the center of mass frame:
\beq
p^*\equiv \frac{\sqrt{[m_1^2-(m_2+m_\chi)^2][m_1^2-(m_2-m_\chi)^2}]}{2m_1}\,\,.
\eeq

In the case where the amplitude is independent of $p_1$, ignoring the Bose enhancement and Pauli blocking effects, Eq. (\ref{eq:collsion_decay_general}) can be further simplified:
\beqn
C(t,p_\chi) &=&
\begin{cases}
\displaystyle\frac{ T \left( e^{-E_\chi/T}-  f_\chi \right)}{16\pi E_\chi p_\chi}\overline{\abs{\mathcal{M}_{1\to 2+\chi}}^2} \left( e^{-E_1^{\rm min}/T}-e^{-E_1^{\rm max}/T} \right) &\text{if~~} f_{1,2}=e^{-E_{1,2}/T},\nn\\[20pt]
 \begin{aligned}
 &\pm \frac{T}{16\pi E_\chi p_\chi}\overline{\abs{\mathcal{M}_{1\to 2+\chi}}^2} \ \nn\\ 
 &\times\ln \left( \frac{1\mp e^{-E_1^{\rm max}/T}}{1\mp e^{-E_1^{\rm min}/T}} \right) -f_\chi\left(F_2(\Delta E_1^{\rm max})-F_2(\Delta E_1^{\rm min})\right)  
\end{aligned}
&\text{if~~} f_{1,2}=\frac{1}{e^{E_{1,2}/T}\mp 1},
\end{cases}
\\
\eeqn
where \mbox{$\Delta E^{\rm max/min}_1 = E_1^{\rm max/min}-E_\chi$ and $F_2(E_2)=\int_{m_1}^{E_2}f_2(\bar{E}_2)d\bar{E}_2$} is the accumulative distribution function of $f_2$.

Similarly, the collision term of $f_1$ for the decay process is
\beqn
C(t,p_1)& = &-\frac{1}{2E_1} \int\frac{d^3p_2}{(2\pi)^3 2E_2}\int \frac{d^3p_\chi}{(2\pi)^3 2E_\chi}  (2\pi)^4 \delta^4(p_1-p_2-p_\chi)\nn\\
&~&\times~\overline{\abs{\mathcal{M}_{1\to 2+\chi}}^2} \left( f_1(1\mp f_2)(1\mp f_\chi) - f_2 f_\chi (1\mp f_1) \right) .\nn\\
& = & -\frac{1}{16\pi E_1 p_1}\int_{E_2^{\rm min}}^{E_2^{\rm max}} dE_2  \overline{\abs{\mathcal{M}_{1\to 2+\chi}}^2} \left( f_1(1\mp f_2)(1\mp f_\chi) - f_2 f_\chi (1\mp f_1) \right),\nn\\
\eeqn
where $E_2^{\rm max/min}=\sqrt{(p_2^{\rm max/min})^2+m_2^2}$ and
\beqn
p_2^{\rm max}&=&\frac{(m_1^2+m_\chi^2-m_2^2)p_1+2m_1E_1 p^*}{2m_1^2},\\
p_2^{\rm min}&=&
\abs{\frac{(m_1^2+m_\chi^2-m_2^2)p_1-2m_1E_1 p^*}{2m_1^2} }.
\eeqn

%%%%%%%%%%%%%%%%%%%%%%%%%%%%%%%%%%%%%%%%%%%%%%%%%%%%%%%%%%%%
\subsection{Collision term of \texorpdfstring{$2\to2$} annihilation}\label{sec:col_2to2_ann}
%%%%%%%%%%%%%%%%%%%%%%%%%%%%%%%%%%%%%%%%%%%%%%%%%%%%%%%%%%%%

In this section, we consider the process $1+2\leftrightarrow 3+\chi$. 
We follow the calculations in Ref.~\cite{Bae:2017dpt} and generalize the results therein by including the inverse process and not assuming that the particle $3$ is in thermal equilibrium.

The collision term can be explicitly written as follows
\beqn\label{eq:colint}
C(t,p_\chi)&=&\frac{1}{2E_\chi}\int \prod_{i=1}^{3} d\pi_i (2\pi)^4 \delta^4(p_1+p_2-p_3-p_\chi)\nn\\
&~&\times~\Big[\overline{\abs{\mathcal{M}_{1+2\to 3+\chi}}^2}f_1^{\rm eq}f_2^{\rm eq}(1\mp f_3)(1\mp f_\chi)\nn\\
&~&~~~~~-\overline{\abs{\mathcal{M}_{ 3+\chi\to 1+2}}^2}f_3f_\chi(1\mp f_1^{\rm eq})(1\mp f_2^{\rm eq})\Big],
\eeqn
{in which two symmetries of the collision integral can be exploited to simplify the calculation with certain assumption:}
\begin{itemize}
    \item The integration over the three momenta is only a function of the magnitude of $\vec{p}_\chi$ presenting a rotational invariance, which enables the alignment of $\vec{p}_\chi$ and the z-axis of the coordinate, thus $\vec{p}_\chi=(0,0,p_\chi)$.
    \item If one neglects the Pauli blocking or Bose enhancement effect and approximates the product of the distribution functions $f_1^{\rm eq}f_2^{\rm eq}(1\mp f_3)(1\mp f_\chi)$ in the second line of Eq.~\eqref{eq:colint} by $f_3^{\rm eq}f_\chi^{\rm eq}$ using the detailed balance condition, then the integration over the momenta $d\pi_1$ and $d\pi_2$ is Lorentz invariant, thus one can always perform the integration in the center of mass frame. 
\end{itemize}

We adopt the above assumptions and proceed with the integration over $d\pi_1d\pi_2$:
\beqn
&~&\int d\pi_1 d\pi_2 (2\pi)^4\delta^4(p_1+p_2-p_3-p_\chi)\dots\nn\\
&=&\frac{1}{16\pi^2}\int\frac{d^3p_1}{E_1E_2}\delta (E_1+E_2-E_3-E_\chi)\dots\nn\\
&=&\frac{1}{16\pi}\int\frac{dt dE_1}{p_\chi E_2}\delta (E_1+E_2-E_3-E_\chi)\dots,
\eeqn
where from the second line to the third line we have used the relation {$dp_1d\cos\theta_{1}={E_1}/({2p_1^2p_\chi})~dt dE_1$}, with 
$t=(p_1-p_\chi)^2=(p_2-p_3)^2$, and $E_2=\sqrt{\abs{\vec{p}_3+\vec{p}_\chi-\vec{p}_1}^2+m_2^2}$.

Since the integration over $d\pi_1d\pi_2$ is Lorentz invariant, we exploit the convenience of the center of mass frame and obtain
\beq
\delta (E_1+E_2-E_3-E_\chi)=\frac{\delta\left(E_1-\frac{\sqrt{s}}{1+E_2^s/E_1^s}\right)}{1+E_1^s/E_2^s}
\eeq
where
\beq
E_1^s = \frac{s+m_1^2-m_2^2}{2\sqrt{s}},\quad E_2^s = \frac{s+m_2^2-m_1^2}{2\sqrt{s}}.
\eeq
Integrating over $dE_1$ yields
\beqn
\int d\pi_1 d\pi_2 (2\pi)^4\delta^4(p_1+p_2-p_3-p_\chi)\dots&=&\int_{t^{\rm min}}^{t^{\rm max}}\frac{dt}{16\pi\sqrt{s}p_{\chi}^{*}}\dots,\label{eq:int_delta}
\eeqn
with
\beq
t^{\rm max/min}=\left[\frac{m_1^2-m_2^2-m_\chi^2+m_3^2}{2\sqrt{s}} \right]^2-(p_1^*\mp p_3^*)^2,
\eeq
and  the momenta in the center-of-mass frame are
\beqn
p_3^*&=&p_{\chi}^*=\frac{\sqrt{[s-(m_3+m_\chi)^2][s-(m_3-m_\chi)^2}]}{2\sqrt{s}},\\
p_1^*&=&p_2^*=\frac{\sqrt{[s-(m_1+m_2)^2][s-(m_1-m_2)^2}]}{2\sqrt{s}}.
\eeqn
The resulting collision term can then be written as
\beqn
C(t,p_\chi)&= &\frac{1}{512\pi^3 E_\chi p_\chi}\int_{(m_3+m_\chi)^2}^{\infty} \frac{ds}{\sqrt{s}p_{\chi}^*} \int_{E_3^{\rm min}}^{E_3^{\rm max}} dE_3\nn\\
&~&
\times \int^{t^{\rm max}}_{t^{\rm min}} dt~\Big[e^{-(E_\chi+E_3)/T}-f_\chi(p_\chi) f_3(p_3)\Big]\overline{\abs{\mathcal{M}_{1+2\to 3+\chi}}^2},
\eeqn
in which we replace the $f^{\rm eq}_{3,\chi}$ with the Maxwell-Boltzmann distribution, and $E_3^{\rm max/min}=\sqrt{(p^2_3)^{\rm max/min }+m_3^2}$ is the maximum/minimum value of $E_3$ when $p_\chi$ and $s$ is fixed, where $p_3^{\rm max/min}$ can be expressed as a function of $s$:
\beqn
p_3^{\rm max/min}=
\begin{cases}
\displaystyle \frac{\pm (s-m_3^2-m_\chi^2)p_\chi+2E_\chi\sqrt{s}p_\chi^*}{2m_\chi^2},\,\,\,\,\, &\text{if~~} p_\chi\leq \frac{\sqrt{s}}{m_3} p_\chi^*\,.\nn\\
\displaystyle \frac{(s-m_3^2-m_\chi^2)p_\chi \pm 2E_\chi\sqrt{s}p_\chi^*}{2m_\chi^2}, &\text{if~~} p_\chi > \frac{\sqrt{s}}{m_3} p_\chi^*\,.
\end{cases}\\
\eeqn

One can separate the collision term into forward and inverse parts denoted as $C = \overrightarrow{C}+\overleftarrow{C}$. Since only the factor $f^{({\rm eq})}_3(p_3)$ in the integrand is dependent on $E_3$ for both the forward and the inverse processes, one can integrate over $dE_3$ over the distribution function $f_3$ and obtain the simplified collision terms for both forward and inverse process as: 
\beqn
\overrightarrow{C}(t,p_\chi)&\simeq&\frac{Te^{-E_\chi/T}}{512\pi^3 E_\chi p_\chi}\int_{(m_3+m_\chi)^2}^{\infty}\frac{ds}{p_{\chi}^*(s)\sqrt{s}}
\left(e^{-E_3^{\rm min}(s)/T}-e^{-E_3^{\rm max}(s)/T}\right)\nn\\
&~&\times\int^{t^{\rm max}}_{t^{\rm min}} dt~\overline{\abs{\mathcal{M}_{1+2\to 3+\chi}}^2},\\
\overleftarrow{C}(t,p_\chi)&\simeq&\frac{-f_\chi(p_\chi)}{512\pi^3 E_\chi p_\chi}\int_{(m_3+m_\chi)^2}^{\infty} \frac{ds}{p_{\chi}^*(s)\sqrt{s}} \left(F_3(E_3^{\rm max})-F_3(E_3^{\rm min})\right)\nn\\
&~&
\times \int^{t^{\rm max}}_{t^{\rm min}} dt~ \overline{\abs{\mathcal{M}_{3+\chi\to 1+2}}^2},
\eeqn
where $F_{3}(E_3)=\int_{m_3}^{E_3}d\bar{E}_3f(\bar{E_3})$ is the accumulative distribution function for the distribution $f_3(E_3)$. In principle, both the forward and the inverse processes should be considered in the collision term.
However, in the pure freeze-in scenario, the inverse  process is often neglected as its rate is much smaller than the forward one.
Therefore, for the pure freeze-in scenario, when numerically solving the Boltzmann equation, one often simply takes $C\approx \overrightarrow{C}$.

%%%%%%%%%%%%%%%%%%%%%%%%%%%%%%%%%%%%%%%%%%%%%%%%%%%%%%%%%%%%
\subsection{Collision term of \texorpdfstring{$2\to2$}{2->2} elastic scattering}
The collision term of elastic scattering is

\begin{equation}\label{eq:elasint}
	C_{\rm el}(t,p_\chi)=\frac{1}{2E_\chi}\int \prod_{i=1}^{3} d\pi_i (2\pi)^4 \delta^4(p_1+p_2-p_3-p_\chi) \overline{\abs{\mathcal{M}_{\rm el}}^2} \left( f_1^{\rm eq}f_2^{\chi}-f_3^{\rm eq} f_\chi\right),
\end{equation}
where we have neglected the Bose enhancement and the Pauli blocking effects.
For the inverse process, the similar center of mass frame trick can be implemented as in the annihilation case because $f_3^{eq}f_\chi$ is independent of $p_1$ and $p_2$. However, for the forward process,  $f_1^{\rm eq}f^\chi_2$ cannot be pulled out of the integration of $d\pi_1d\pi_2$ due to the unknown form of $ f_2=f_\chi $. Thus, we will focus on the treatment of this part in this section. Here we follow the calculations in Ref.~\cite{Yunis:2020woq} and generalize the result for arbitrary $m_\chi=m_2$ and $m_1=m_3$.

Substituting
\begin{equation}
	\int \frac{d^3 p_3}{2 E_3}= \int d^3 p_3 d E_3 \delta( E_3^2-\vec{p_3} ^2-m_3^2) \Theta(E_3),
\end{equation}
and
\begin{equation}
	\left\{\begin{aligned}
		\vec{p}_\chi&=p_\chi(0,0,1) \\
		\vec{p}_1&=p_1(0, \sin \theta_1, \cos \theta_1) \\
		\vec{p}_2&=p_2(\sin \beta \sin \theta_2, \cos \beta \sin \theta_2, \cos \theta_2)
	\end{aligned}\right.,
\end{equation}
into Eq.\,\eqref{eq:elasint}, one obtains
\begin{equation}
	\begin{aligned}
		C_{\rm el}(t,p_\chi)&= \frac{1}{8(2 \pi)^{4} E_{\chi}} \int d p_1  d\cos \theta_1  d p_2 d\cos \theta_2 \frac{p_1^{2}}{E_{1}} \frac{p_2^2}{E_{2}} \Theta\left(E_{1}+E_{2}-E_{\chi}\right) \\
		 &\quad \times 		 \overline{\abs{\mathcal{M_{\rm el}}}^2} \left( f_1^{\rm eq}f_2^{\chi}-f_3^{\rm eq} f_\chi\right) \int_{0}^{2 \pi} d \beta \delta_{D}\left(g\left(...\right)\right),
	\end{aligned}
\end{equation}
where
\begin{equation}
	\begin{aligned}
		g\left(...\right) &\equiv (E_{1}+E_{2}-E_{\chi})^2- (\vec{p}_{1}+\vec{p}_{2}-\vec{p}_{\chi})^2 - m_3^2 \\
		&= 2 \bar{m}^{2}+2 E_{1} E_{2}-2 E_{2} E_{\chi}-2 E_{1} E_{\chi}+ 2p_1 p_\chi \cos \theta_1 +2p_2 p_\chi \cos \theta_2 \\
		&\quad - 2p_1 p_2 (\cos \theta_1\cos \theta_2+ \sin \theta_1\sin \theta_2 \cos \beta).
	\end{aligned}
\end{equation}
and
$ \bar{m}^{2}\equiv (m_1^2+m_2^2+m_\chi^2-m_3^2)/2 $.

The $\delta_D(g)$ integral can be performed quite straightforwardly with solution 
\begin{equation}
\label{eq:tutbar}
	\begin{aligned}
		\cos \beta^* &=(p_1 p_2 \sin \theta_1 \sin \theta_2)^{-1} \left( \frac{\bar{t}}{2} +E_1E_2-E_2E_\chi  +p_2 \cos \theta_2( p_\chi -p_1 \cos \theta_1)\right) .
	\end{aligned}
\end{equation}
The $t,u,\bar{t}$ above are defined as:
\begin{equation}
	\left\{\begin{aligned}
		t &\equiv (p_1 -p_\chi)^2=m_1^2+m_\chi^2- 2E_1E_\chi +2 p_1p_\chi \cos \theta_1 \\
		u&\equiv (p_2 -p_\chi)^2=m_2^2+m_\chi^2- 2E_2E_\chi +2 p_2p_\chi \cos \theta_2 \\
		\bar{t}&\equiv t+m_2^2-m_3^2
	\end{aligned}\right. .
\end{equation}
In the meantime, the existence of the solution $\cos\beta^*$ can be converted to a $\Theta$ function $\Theta\left(\abs{{\partial g}/{\partial \beta}} ^2\right)$, which then translates to the integration bounds on $\cos\theta_2$.
Here, we express $ \abs{ {\partial g}/{\partial \beta} }^2 $ in the polynomial of $ \cos \theta_2 $:
\begin{equation}
	\begin{aligned}
		\abs{\frac{\partial g}{\partial \beta}} ^2 &= a \cos^2 \theta_2 + b \cos \theta_2 +c \\
		a &= - 4 p_2^2 \abs{\vec{p_1}-\vec{p_\chi}}^2 \\
		b &= -8p_2 \left( \frac{\bar{t}}{2} +E_1E_2-E_2E_\chi \right)  ( p_\chi -p_1 \cos \theta_1) \\
		c &= 4 p_1^2 p_2^2 \sin^2 \theta_1 - 4 \left( \frac{\bar{t}}{2} +E_1E_2-E_2E_\chi \right)^2\\
		\Delta &= b^2-4 a c = 64 p_1^2 p_2^2 \sin^2 \theta_1 \bar{t} \left\lbrace - \frac{t}{\bar{t}} E_2^2 - E_2(E_1-E_\chi) -\left[ \frac{\bar{t}}{4} + \frac{ m_2^2 \abs{\vec{p_1}-\vec{p_\chi}}^2}{ \bar{t}}  \right] \right\rbrace .
	\end{aligned}
\end{equation}
Note that $a\leq 0$.\footnote{For $a=0$, we have $\vec{p}_1=\vec{p}_\chi$ or $p_2=0$, which fixes $\cos\theta_1=1$, $b=0$ and $\abs{\frac{\partial g}{\partial \beta}} ^2=c\leq 0$, yielding a vanishing collision term.} We further express the two roots of the polynomial below:
\beq
x_1=\frac{-b+\sqrt{\Delta}}{2a},\quad x_2=\frac{-b-\sqrt{\Delta}}{2a},
\eeq
the integration over $\cos\theta_2$ then reads:\footnote{We have numerically check that $x_1\geq -1$ and $x_2\leq1$ in our parameter setup.}
\beq
\int d\cos\theta_2\dots=
\begin{cases}
\int_{x_1}^{x_2}d\cos\theta_2\dots, &  a<0, \Delta>0, \\[10pt]
0, & {\rm else.}
\end{cases}
\eeq
For the case $a<0$ and $\Delta>0$, then the collision integral becomes:
\begin{equation}
	C_{\rm el}(t,p_\chi)= \frac{1}{16 (2 \pi)^{4} E_{\chi} p^2_{\chi}} \int d E_1 dt \int_{E_2^-}^{E_2^+} d E_2 \int_{u(x_1)}^{u(x_2)}  d u \frac{\overline{\abs{\mathcal{M}_{\rm el}}^2}} {\sqrt{\abs{ {\partial g}/{\partial \beta} }^2 }}  \times  \left( f_1^{\rm eq}f_2^{\chi}-f_3^{\rm eq} f_\chi\right), 
\end{equation}
where we have used $d p_1 d \cos \theta_1 = {E_1}/({2 p_1^2 p_\chi}) d E_1 dt$, $d p_2 d \cos \theta_2 = {E_2}/({2 p_2^2 p_\chi}) d E_2 du$, and the integration bounds on $E_2$ is constraint by $\Delta>0$.
Note that since $\Delta$ is a quartic function of $E_2$ whose roots can be written as:
\begin{equation}\label{key}
	R_{\pm} = \frac{\bar{t}}{2 t} \left[ E_\chi -E_1 \pm \abs{\vec{p_1}-\vec{p_\chi}} \sqrt{1- \frac{4 m_2^2 t}{\bar{t}^2}}  \right]. 
\end{equation}
A crucial observation is
\beq
R_{\pm}^2=m_2^2+\left((E_\chi-E_1)\sqrt{1- \frac{4 m_2^2 t}{\bar{t}^2}}\pm \abs{\vec{p_1}-\vec{p_\chi}}\right)^2\geq m_2^2,
\eeq
indicating $|R_{\pm}|>m_2$.
Therefore we have the following bounds on integration over $E_2$:
\beq
\int dE_2\dots=
\begin{cases}
\int_{\max (m_2,R_{\pm},E_\chi-E_1)}^{\infty}dE_2\dots & \text{if }t<0,\text{and }1-4m^2t/\bar{t}^2> 0,\\[10pt]
\int_{R_{-}}^{R_{+}}dE_2\dots &\text{if }t>0,\text{and }1-4m^2t/\bar{t}^2> 0,\\[10pt]
0, &{\rm else}.
\end{cases}
\eeq
If the scattering amplitude is a constant as is the case in the large centor-of-mass energy limit, the integration over $du$ can be done analytically:
\begin{eqnarray}
C_{\rm el}(t,p_\chi)&=& \frac{\overline{\abs{\mathcal{M}_{\rm el}}^2}}{16 (2 \pi)^{4} E_{\chi} p^2_{\chi}} \int_{m_1}^{\infty} d E_1 \int^{t_{\rm max}}_{t_{\rm min}}dt \int_{E_2^-}^{E_2^+} d E_2 \nn\\
&&\times \frac{1}{\sqrt{(E_\chi-E_1)^2-t}}    \left( f_1^{\rm eq}f_2^{\chi}-f_3^{\rm eq} f_\chi\right), 
\end{eqnarray}
where $t^{\rm max/min}$ is obtained by setting $\cos\theta_1$ to $\pm 1$ in Eq.~\eqref{eq:tutbar}.

\section{Boltzmann equation: From distribution to number density}\label{sec:app2}

From the Boltzmann equation for the distribution
\beq
\frac{\partial f}{\partial t} - Hp\frac{\partial f}{\partial p}=C[f]\,,
\eeq
we can derive the Boltzmann equation for the number density and the thermally averaged cross section.
Integrating both sides over the corresponding phase space, the left-hand-side of the equation becomes
\beq
\int\frac{d^3 p}{(2\pi)^3} \left( \frac{\partial f}{\partial t} - Hp\frac{\partial f}{\partial p}\right)  = \frac{dn}{dt}+3Hn\,.
\eeq

\subsection{\texorpdfstring{$2\to2$}{} annihilation}
For the process $ 1+2\rightarrow 3+\chi $, the right-hand side of the Boltzmann equation for the number density is
\begin{equation}\label{numc1}
	\int\frac{d^3 p_\chi}{(2\pi)^3}C[f_\chi] = \int d\pi_1 d\pi_2 d\pi_3 d\pi_\chi (2\pi)^4 \delta^{(4)}(p_1+p_2-p_3-p_\chi) \overline{\abs{\mathcal{M}_{1+2\rightarrow 3+\chi}}^2} \left( f_1 f_2 -f_3 f_\chi \right).
\end{equation}
Since $ f_\chi $ is unkonwn a priori, this integral cannot be done theoretically. In the formalism of the Boltzmann equation for number density, all the particles are supposed to keep in kinetic equilibrium with the thermal bath, such that $f_\chi\sim e^{-(E-\mu_\chi)/T} \sim f_\chi^{\rm eq} n_\chi/n_\chi^{\rm eq}$.

Define the thermally averaged cross section as
\begin{equation}\label{key}
	\left\langle \sigma v\right\rangle \equiv \frac{1}{n_3^{\rm eq} n_\chi^{\rm eq}}\int d\pi_1 d\pi_2 d\pi_3 d\pi_\chi (2\pi)^4 \delta^{(4)}(p_1+p_2-p_3-p_\chi) \overline{\abs{\mathcal{M}_{1+2\rightarrow 3+\chi}}^2}  f_1^{\rm eq} f_2 ^{\rm eq},
\end{equation}
Eq.~\eqref{numc1} becomes
\begin{equation}\label{key}
	\int\frac{d^3 p_\chi}{(2\pi)^3}C[f_\chi] = \left\langle \sigma v\right\rangle \left(n_3^{\rm eq} n_\chi^{\rm eq} \frac{n_1 n_2}{n_1^{\rm eq} n_2^{\rm eq}} - n_3 n_\chi \right) \,.
\end{equation}

If particle 1 and 2 are in thermal equilibrium and particle is also dark matter, we will get the well-known number density Boltzmann equation
\begin{equation}\label{key}
	\frac{dn_\chi}{dt}+3Hn_\chi =  \left\langle \sigma v\right\rangle \left[ \left( n_\chi^{\rm eq}\right) ^2  - n_\chi^2 \right] \,.
\end{equation}

\subsection{Decay process \texorpdfstring{$1 \rightarrow 2 + 3$}{1->2+3}}
For the process $ 1 \rightarrow 2 + 3 $, the collision term for number density is
\begin{equation}\label{key}
	\int\frac{d^3 p}{(2\pi)^3}C[f] = \int d\pi_1 d\pi_2 d\pi_3 (2\pi)^4 \delta^{(4)}(p_1-p_2-p_3) \overline{\abs{\mathcal{M}_{1\rightarrow 2 + 3}}^2} \left( f_1 -f_2 f_3  \right).
\end{equation}
Similarly, define
\begin{equation}\label{key}
	\Gamma_1 \equiv \frac{1}{n_1^{\rm eq}} \int d\pi_1 d\pi_2 d\pi_3 (2\pi)^4 \delta^{(4)}(p_1-p_2-p_3) \overline{\abs{\mathcal{M}_{1\rightarrow 2 + 3}}^2}  f_1^{\rm eq},
\end{equation}
the collision term becomes
\begin{equation}\label{key}
	\int\frac{d^3 p}{(2\pi)^3}C[f] = \Gamma_1 \left( n_1 -  n_1^{\rm eq} \frac{n_2 n_3}{n_2^{\rm eq} n_3^{\rm eq}} \right) .
\end{equation}

Therefore, if particle 2 and 3 are in thermal equilibrium, the Boltzmann equation of particle 1 is
\begin{equation}\label{key}
	\frac{dn_1}{dt}+3Hn_1 = -\Gamma_1 \left( n_1 -  n_1^{\rm eq} \right) ,
\end{equation}
if particle 1 is in thermal equilibrium and particle 2 and 3 are dark matter $ \chi $, the Boltzmann equation of $ \chi $ is
\begin{equation}\label{key}
	\frac{dn_\chi}{dt}+3Hn_\chi =  \Gamma_1 \frac{n_1^{\rm eq}}{ \left( n_\chi^{\rm eq}\right) ^2} \left[ \left( n_\chi^{\rm eq}\right) ^2  - n_\chi^2 \right]\,.
\end{equation}

\bibliographystyle{JHEP}
\bibliography{ref}

\end{document}